\documentclass[listof=totoc,bibliography=totoc,index=totoc,BCOR=10mm,captions=tableheading,figuresright,fleqn,titlepage,oneside,12pt,paper=a4,pagesize]{scrbook}
\usepackage{lmodern}

\usepackage[latin9]{inputenc}
\usepackage{geometry}
\usepackage{color}
\definecolor{note_fontcolor}{rgb}{0.800781, 0.800781, 0.800781}
\usepackage[naustrian,english,british]{babel}
\usepackage{verbatim}
\usepackage{varioref}
\usepackage{float}
\usepackage{rotfloat}
\usepackage{units}
\usepackage{amsmath}
\usepackage{amssymb}
\usepackage{graphicx}
\usepackage{nomencl}

\providecommand{\makenomenclature}{\makeglossary}
\makenomenclature
\usepackage[unicode=true,
 bookmarks=true,bookmarksnumbered=true,bookmarksopen=true,bookmarksopenlevel=2,
 breaklinks=true,pdfborder={0 0 0},pdfborderstyle={},backref=false,colorlinks=true]
 {hyperref}
\hypersetup{pdftitle={Quantum Motion},
 pdfauthor={Johannes Mesa Pascasio},
 pdfsubject={Dissertation},
 pdfpagelayout=OneColumn, pdfnewwindow=true, pdfstartview=XYZ, plainpages=false}

\makeatletter

\providecommand{\tabularnewline}{\\}
\newenvironment{lyxgreyedout}
  {\textcolor{note_fontcolor}\bgroup\ignorespaces}
  {\ignorespacesafterend\egroup}

\usepackage{iftex}
\ifPDFTeX
\else
   \ifXeTeX
      \usepackage{polyglossia}
      \setdefaultlanguage{english}
      \setotherlanguage{german}
      \usepackage{xltxtra}
      \usepackage[varg]{txfonts}        
   \else
      \usepackage{luatextra}
   \fi
   \defaultfontfeatures{Ligatures=TeX}
\fi

\usepackage{nicefrac}
\usepackage[headsepline]{scrlayer-scrpage}

\setkomafont{captionlabel}{\bfseries}
\setcapindent{1em}

\let\mySection\section\renewcommand{\section}{\suppressfloats[t]\mySection}
\let\mySubsection\subsection\renewcommand{\subsection}{\suppressfloats[t]\mySubsection}

\usepackage[labelformat=simple]{subfig}

\usepackage{graphicx}
\DeclareGraphicsExtensions{.png,.pdf,.eps}
\graphicspath{{pictures/}}

\usepackage{microtype} 
\usepackage{csquotes}
\date{} 

\usepackage[style=alphabetic,defernumbers,natbib=true,backend=biber]{biblatex}  
\DeclareNameAlias{default}{last-first}  
\renewbibmacro{in:}{}  

\DeclareFieldFormat[article]{volume}{\textbf{#1}\space}
\DeclareFieldFormat{pages}{#1}

\renewbibmacro*{issue+date}{%
    \iffieldundef{issue}{}{{No.}\printfield{issue}\setunit*{\addspace}}%
    \printtext[parens]{\usebibmacro{date}}%
  \newunit}

\DeclareFieldFormat{edition}%
                   {\ifinteger{#1}%
                    {\mkbibordedition{#1}\addthinspace{}ed.}%
                    {#1\isdot}}

\DeclareFieldFormat{sentencecase}{\MakeSentenceCase{#1}}

\renewbibmacro*{title}{%
  \ifthenelse{\iffieldundef{title}\AND\iffieldundef{subtitle}}
    {}
    {\ifthenelse{\ifentrytype{article}\OR\ifentrytype{inbook}%
      \OR\ifentrytype{incollection}\OR\ifentrytype{inproceedings}%
      \OR\ifentrytype{inreference}}
      {\printtext[title]{%
        \printfield[sentencecase]{title}%
        \setunit{\subtitlepunct}%
        \printfield[sentencecase]{subtitle}}}%
      {\printtext[title]{%
        \printfield[titlecase]{title}%
        \setunit{\subtitlepunct}%
        \printfield[titlecase]{subtitle}}}%
     \newunit}%
  \printfield{titleaddon}}

\DeclareSourcemap{
  \maps[datatype=bibtex]{
    \map{
      \step[fieldsource=author, match={'t Hooft}, final]
      \step[fieldset=sortname, fieldvalue={tHooft}]
      \step[fieldset=shortauthor, fieldvalue={Hooft}]
      }
    \map{
      \step[fieldsource=author, match={El Naschie}, final]
      \step[fieldset=shortauthor, fieldvalue={ElNaschie}]
      }
    \map{
      \step[fieldsource=author, match={De Raedt}, final]
      \step[fieldset=shortauthor, fieldvalue={DeRaedt}]
      }
  }
}

\defbibenvironment{nolabelbib}
  {\list
     {}
     {\setlength{\leftmargin}{\bibhang}%
      \setlength{\itemindent}{-\leftmargin}%
      \setlength{\itemsep}{\bibitemsep}%
      \setlength{\parsep}{\bibparsep}}}
  {\endlist}
  {\item}

\@ifundefined{showcaptionsetup}{}{%
 \PassOptionsToPackage{caption=false}{subfig}}
\usepackage{subfig}
\makeatother

\begin{document}
\pagestyle{empty}
\selectlanguage{naustrian}%

\subject{Dissertation}

\title{{\Huge{}Current-based Simulation Models of Quantum Motion}}

\subtitle{\bigskip{}
}

\author{\textsf{\textbf{Johannes Mesa Pascasio}}}
\selectlanguage{english}%

\publishers{\textsf{\bigskip{}
}\\
\textsf{Supervisor\medskip{}
}\\
\textsf{Prof. Manfried Faber\vspace{1cm}
}\\
\textsf{Referees\medskip{}
}\\
\textsf{Prof. Maurice A. de Gosson\smallskip{}
}\\
\textsf{Prof. Basil Hiley}}

\date{\textsf{TU Wien, Rigorosum am 26.\ April 2017}}

\maketitle
\cleardoublepage{}

\selectlanguage{british}%
\pagenumbering{roman}
\makeatletter
\if@twoside
   \setcounter{page}{2}
\else
   \setcounter{page}{2}
\fi
\makeatother

\pagestyle{plain}

\selectlanguage{naustrian}%

\chapter*{Kurzfassung}

\addcontentsline{toc}{chapter}{Kurzfassung} 

In den letzten Jahren wurden Teilchenbewegungen auf makroskopischer
Ebene beobachtet, die bislang nur aus der Quantenmechanik bekannt
waren. Obgleich es sich bei solchen Experimenten, wie sie von der
Gruppe um Couder und Fort ausgeführt wurden, um rein klassische Physik
handelt, gelingt in einer Analogiebetrachtung eine neuartige Beschreibung
mikroskopischer Phänomene.

In dieser Arbeit wird mit Hilfe rein klassischer Mittel ein Modell
des Bouncer\textendash Walker Systems eines elementaren Teilchens
konstruiert, das zugleich die alte Idee de~Broglies, des Welle-Teilchen
Dualismus, widerspiegelt. Dieses Modell beinhaltet einerseits eine
mögliche Erklärung des Energieaustausches zwischen diesen separierten
Bewegungen und somit eine Begründung für die Energiequantelung wie
ursprünglich von Max Planck postuliert. Andererseits erlaubt das Modell
die präzise Ausführung der bohmschen Bewegungen in perfekter Übereinstimmung
mit der Quantenmechanik.

Zur Berechnung quantenmechanischer Teilchenbahnen im Ein- oder Mehrspaltsystem
eignet sich die ballistische Diffusionsgleichung, eine spezielle Form
der Diffusionsgleichung mit zeitabhängiger Diffusivität. Dies macht
es möglich, wie hier gezeigt werden soll, den Zerfall eines gaußschen
Wellenpakets auf elegante Weise zu simulieren.

Mit diesen Instrumenten wird in dieser Arbeit schließlich eine Rechenvorschrift
zur Behandlung der auftretenden Ströme entwickelt, die äquivalent
zur de~Broglie\textendash Bohm Theorie bleibt. Damit lassen sich
Talbot-Muster und die Talbot-Distanz für beliebige Mehrspaltsysteme
elegant reproduzieren.

Bei großen Unterschieden der Intensitäten in Doppelspaltexperimenten
wird der Strahl geringer Intensität nach außen gedrückt und trotz
anfänglich senkrechter Bewegung aus dem Spalt der Schirm seitlich
von der Austrittsstelle getroffen. In dieser Arbeit wird die seitliche
Anordnung des Schirms als mögliche alternative Messmethode untersucht.

Schließlich werden die mathematischen Simulationsverfahren, deren
Limitierungen und mögliche Erweiterungen vorgestellt. Entkoppelt von
der Diffusion lässt sich die Wirkung und somit die Phase als eine
neue Quantität einer Gaußverteilung berechnen. Für ein Mehrspaltsystem
genügt es in Folge, die Phasen zu kombinieren um die korrekte Intensitätsverteilung
sowie die zugehörigen Wahrscheinlichkeitsströme zu erhalten. Die Entkopplung
erlaubt überdies die Berechnung variabler Spaltbreiten sowie Phasenverschiebung
auf einfache Weise.

\begin{lyxgreyedout}
\global\long\def\VEC#1{\mathbf{#1}}
\global\long\def\d{\,\mathrm{d}}
 \global\long\def\e{\mathrm{e}}
\global\long\def\i{\mathrm{i}}
 \global\long\def\meant#1{\left<#1\right>}
 \global\long\def\meanx#1{\overline{#1}}
 \global\long\def\p{\partial}

\global\long\def\intopinfty{\intop}
\global\long\def\vardelta{\mathbf{\boldsymbol{\delta}}}
\global\long\def\i{\mathrm{i}}
\end{lyxgreyedout}
\selectlanguage{british}%

\selectlanguage{english}%

\chapter*{Abstract}

\addcontentsline{toc}{chapter}{Abstract} 

In recent years particles' trajectories have been observed at a macroscopic
level which had been associated with nothing but quantum mechanical
theory before, even though these experiments carried out by Couder
and Fort's group are purely classical physics. By analogical considerations
a new kind of description of microscopic phenomena is possible.

With purely classical tools a model for a bouncer-walker system of
an elementary particle will be derived in this work which reflects
the old idea of de~Broglie's particle-wave duality. This model contains,
on the one hand, a possible explanation of the work-energy exchange
between the two separated motions, thereby providing an energy quantisation
as originally postulated by Max Planck. On the other hand, the system
perfectly obeys the Bohmian-type law of motion in full accordance
with quantum mechanics.

For the calculation of elementary particles' trajectories a ballistic
diffusion equation will be derived which is a special case of a diffusion
equation with a time-dependent diffusivity. Therewith the simulation
of spreading of an elementary Gaussian is made easy as will be shown
herein. 

With these tools one also accounts for Born's rule for multi-slit
systems and develops a set of current rules that directly leads to
a new formulation of the guiding equation equivalent to the original
one of the de~Broglie\textendash Bohm theory. As will be shown in
this thesis, this tool reproduces Talbot patterns and Talbot distance
for an arbitrary multi-slit system. 

Moreover, the sweeper effect is shown to arise when the intensity
relation of two beams of a double-slit experiment exhibit a big difference.
Then, the low-intensity beam is pushed aside in a sense that its initial
propagation straight out of the slit is bent towards the side. A sideways
screen as an alternative measurement method is proposed.

At last, mathematical simulation tools as well as their limitations
and possible extensions are provided. Decoupled from the diffusion
part the action and thus also the phase can be calculated as a new
quantity of each single Gaussian. Then, for a multi-slit system a
simple combination of these phases yields the correct intensity distributions
including the complete interference patterns as well as the associated
probability currents. The decoupling further allows for calculation
of setups comprising variable slit widths as well as phase shifting.
\selectlanguage{british}%

\selectlanguage{british}

\pagestyle{plain}
\tableofcontents
\thispagestyle{plain}

\pagestyle{scrheadings}

\selectlanguage{english}%

\chapter*{Introduction}

\addcontentsline{toc}{chapter}{Introduction}
\pagestyle{plain}

\selectlanguage{british}%
\pagenumbering{arabic}
\makeatletter
\if@twoside
   \setcounter{page}{1}
\else
   \setcounter{page}{1}
\fi
\makeatother

\selectlanguage{english}%
Fundamental quantum phenomena are the basis of modern technologies
like information theory or cryptography, for example. Even in well
settled fields like semiconductor physics a deeper understanding is
necessary, e.g., to scale circuits down, make them more reliable,
or even to replace them by newer technologies like quantum circuits.
Other topics in focus are quantum interference and quantum coherence,
but also nonlinear optical phenomena which are intensively discussed
nowadays. There are numerous possible applications of these results:
Quantum circuits, sensors, molecules, or, more related to this thesis'
field, the coherent control of atomic motion or secure communication
with entangled photons.

A macroscopic body's motion is well understood in classical physics.
At a microscopic level, a particle's motion is not yet completely
understood. One may inquire into the trajectories of these particles,
as, for example, Bohm did before~\cite{Bohm.1993undivided}, though
the solutions therefore are widely spread in literature. Recently,
averaged trajectories of photons have been reported~\cite{Kocsis.2011observing,Bliokh.2013photon}
which obey also the rules of quantum mechanical theory.

A few years ago a French group around Yves Couder and Emmanuel Fort
discovered the existence of quantum-type rules at macroscopic level
in practical experiments using oil droplets bouncing on a vibrating
oil bath~\cite{Couder.2006single-particle,Protiere.2006,Eddi.2009,Fort.2010path-memory}.
Investigations on those experiments showed that there is a kind of
particle-wave duality similar to the explanation of de~Broglie~\cite{DeBroglie.1960book,Couder.2012probabilities,Harris.2013wavelike}.

One aspect of this work is to adapt de~Broglie's particle-wave duality
to microscopic level in order to investigate the energy exchange so
that a model for the interaction between the particle and the wave
can be specified. For simplicity, the description is restricted to
one-dimensional, nonrelativistic cases. At the beginning, a classical
particle behaves as a damped oscillator which also carries out a random
motion that is superposed to the oscillatory motion. The necessary
conditions to keep both motions alive will be shown and the energy
balance will be derived. Accordingly, these conditions are shown to
enforce natural motions, i.e.~moving particles, and even accelerated
ones being prevented of radiating. This adaptation has been published
in reference~\cite{Groessing.2011explan,Schwabl.2012quantum}.

Another aspect concerns the motion of those particles: While in literature~\cite{Holland.1982electromagnetism,Sanz.2012trajectory}
the quantum mechanical equations are solved, herein a different description
of a particle's motion will be developed which also simplifies the
numerical calculations concerning the random motion carried out by
the particle, its averaged motion and also the trajectories of particles,
their velocities and probability currents. By setting initial probability
distributions right after a single slit, or even multiple slits, the
conditions of the decay of these distributions can be studied. It
will be shown that the decay obeys a ballistic diffusion which leads
to an explanation of interference effects, velocities and densities,
and even to calculatory rules for the probability distributions, and
the associated probability currents. If the intensities of two beams
of a double-slit experiment provide big differences then the low-intensity
beam is swept aside, away from the high-intensity beam. It is suggested
to record the intensities at a screen perpendicular to the double-slit
setup. Parts of this aspect have already been shown in~\cite{Groessing.2010emergence,Groessing.2012doubleslit,Mesa.2013variable,Fussy.2014multislit,Groessing.2014attenuation}.

These two aspects are of course closely tied to each other since the
particle's motion inherits the waves' motion that itself influences
the particle again, which has also been referred to in further publications
of our group~\cite{Groessing.2010entropy,Groessing.2011dice,Groessing.2012vaxjo,Grossing.2012quantum,Mesa.2012classical,Schwabl.2012quantum,Groessing.2013joaopessoa,Groessing.2013dice,Groessing.2014emqm13-book}.

A last aspect of the present work is the description of the numerical
means to simulate a particle's motion. The mathematical background
for the derivation and calibration of simulation tools is provided.
Furthermore, the limitations one is confronted with when using these
tools will be shown. 

The theoretical framework presented herein has been worked out in
tight cooperation with all members of our group. Accordingly, this
thesis contains parts of our already published content, yet enhanced
and completed to give a picture of what has been developed so far.
The numerical treatment of the second part as well as the whole simulation
procedure including programming has been worked out by myself. The
practicability of the derived framework is demonstrated herein with
numerous images obtained by these simulations.\selectlanguage{british}%

\pagestyle{scrheadings}

\chapter{The fluid droplet picture adapted to quantum mechanics\label{sec:2.fluid-droplet}}
\begin{quote}
Inspired by the experiments of Couder and Fort's group who show that
a macroscopic particle may both regularly oscillate in time with its
characteristic frequency \emph{and }propagate irregularly in space,
we distinguish between these two types of motion and call the former
\textsl{bouncer} and the latter \textsl{walker}. We discuss this two-fold
perspective of an individual particle and discuss an analogous sub-quantum
model simultaneously characterized by regular periodic and stochastic
motions, both of which must, however, be comparable on the level of
the work-energy expended during a certain amount of time. We shall
calculate the respective work-energies for each aspect separately,
afterwards they will be compared during the same time-span.

We assume that phenomena of standard quantum mechanics like Planck's
energy relation or the Schrödinger equation can be assessed as the
property of the vacuum combined with diffusion processes reflecting
also a stochastic nature. Thus we obtain the quantum mechanical results
as an averaged behaviour of sub-quantum processes.~\cite{Groessing.2011explan,Schwabl.2012quantum}
\end{quote}

\section{The macroscopic fluid droplet\label{sec:The-macroscopic-fluid}}

Masses and waves are well-known constitutive elements of classical
physics. The idea of the wave-particle duality had no equivalent on
a macroscopic scale for a long time until a small group of French
physicists around Yves Couder~\cite{Couder.2005,Couder.2006single-particle,Protiere.2006,Eddi.2009,Couder.2010walking,Fort.2010path-memory,Couder.2012probabilities}
published experiments providing bouncing masses and waves coupled
tight, on the one hand, but being different objects on the other hand.
More curious is the fact that both, the waves and the bouncing masses
comprised of the same substance in those experiments,i.e.\ silicon
oil.

Consider a coffee machine comprising a filter containing the coffee
powder and a glass pot where the finished coffee is collected at last.
Every now and then a brown coffee droplet goes down from the filter
into the glass pot, falls on the surface of the same liquid and disappears
rapidly. Sometimes the droplet bounces back from that surface for
two or three times thereby leaving some waves on the surface. Everybody
knows that, however, the French scientists asked themselves how they
could keep the droplets bouncing for a longer time. Therefore, they
replaced the static pot by a vertically vibrating bath and the coffee
by silicon oil because of its higher viscosity. Surprisingly, the
droplet kept bouncing without disappearing for a long time, even in
the order of hours or days.

Due to the vertical vibration of the bath, the characteristic acceleration
thereby being generated causes the droplet to bounce on the surface
periodically. As the droplet collides with the interface, it remains
separated by a continuous air film. Before this air film can break,
the droplet lifts off again. At each successive bounce, the droplet
forms a surface wave which is thereby attenuated so that the force
acting on the droplet guides it towards the next surface point and
so on. Couder and his group then showed that by controlling the vibrating
bath the droplets can be guided along artificial paths reminiscent
of quantum mechanics.

If those macroscopic experiments are able to reproduce~\textendash{}
to a certain extent~\textendash{} quantum mechanical experiments
like diffraction of a single object or double-slit interference, then
at least it should be worth to investigate this mechanism peculiarly
with regard to quantum mechanical similarities. This has also been
suggested by other authors, for example by Brady and Anderson~\cite{Brady.2014why}
or Richardson \emph{et al}.\ \cite{Richardson.2014analogy}. In other
words, consider the bouncing mass to be an elementary particle like
an electron or a neutron whose intrinsic oscillation generates and
affects the wave-like landscape around itself. Of course, this wave-like
landscape has to be built up in an underlying sub-structure of the
vacuum, the sub-quantum medium, combined with diffusion processes.

According to Couder's experiments we distinguish between these two
types of motion and call them \textsl{bouncer} and \textsl{walker},
respectively. We discuss this two-fold perspective of an individual
particle and, after individual inspection, these two tools will be
compared, or coupled, respectively. We are interested in reproducing
the energy exchange and conservation between these two types of motion
with respect to well-known quantum mechanical principles. We assume
that phenomena of standard quantum mechanics like Planck's energy
relation or the Schrödinger equation can be assessed as the emergent
property of an underlying sub-structure of the vacuum combined with
diffusion processes reflecting also the stochastic parts of the zero-point
field, i.e.\ the zero-point fluctuations~\cite{Cetto.2012quantization,Groessing.2008vacuum,Groessing.2009origin,Groessing.2010emergence,Cetto.2014emerging-quantum}.
With respect to an analogous sub-quantum model, then, this means that
the zitterbewegung\index{zitterbewegung} is simultaneously characterized
by regular periodic and stochastic motions, both of which must, however,
be comparable on the level of the work-energy expended during a certain
amount of time. We shall calculate the respective work-energies for
each aspect separately, afterwards they will be compared during the
same time-span. This will lead to requirements which have to be fulfilled
by a such modelled quantum mechanical system. Thus we obtain the quantum
mechanical results as an averaged behaviour of sub-quantum processes.

\section{A classical oscillator: The bouncer\label{sec:aceq.sub.3}}

We assume a system comprising  two subsystems, the first one is a
harmonic oscillator and the second one undergoes a Brownian-type motion~\cite{Groessing.2011explan,Schwabl.2012quantum}.
To recall the above-mentioned picture, one could consider Couder's
droplet: The droplet bounces on a wave, thus the droplet itself represents
the harmonic oscillator which, at the same time, moves along together
with a wave driven by the oscillations. Here, in a first step, we
focus on the harmonic oscillation of the first subsystem.

We can write down the following Newtonian equation of a classical
forced oscillator with friction (see any good textbook, e.g.~\cite{Demtroeder.2005exp1})
with one degree of freedom 
\begin{equation}
m\ddot{x}+m\omega_{0}^{2}x+2\gamma m\dot{x}=F_{0}\cos\omega t.\label{eq:aceq.2.1}
\end{equation}
Equation~(\ref{eq:aceq.2.1}) describes a forced oscillation of a
mass $m$\nomenclature[mass]{$m$}{mass} swinging around a centre
point along $x(t)$. The resonant angular frequency is $\omega_{0}$\nomenclature[omega0]{$\omega_0$}{angular frequency of the undamped system}\nomenclature[omega]{$\omega$}{angular frequency of the damped system}
for the case $m$ would swing freely. Due to the damping/friction
$\gamma$\nomenclature[gamma]{$\gamma$}{friction or damping coefficient}
of the swinging particle, for periodic motion there is a need for
a locally independent driving force $F_{0}\cos\omega t$.

The general solution of the inhomogeneous equation~(\ref{eq:aceq.2.1})
comprises a general solution of the homogeneous equation (the left
hand side of Eq.~(\ref{eq:aceq.2.1})) plus a special solution of
the inhomogeneous equation. Accordingly, the general solution must
be of form
\begin{equation}
x(t)=r_{1}\e^{-\gamma t}\cos(\omega_{1}t+\varphi_{1})+r\cos(\omega t+\varphi).
\end{equation}
After short calculation $\omega_{1}=\sqrt{\omega_{0}^{2}-\gamma^{2}}$
appears as the frequency of the free damped oscillation. 

However, for $t\gg\gamma^{-1}$ the amplitude $r_{1}\e^{-\gamma t}$
of the first term vanishes, thus $\gamma^{-1}$ plays the role of
a relaxation time. The second term remains and represents a stationary
solution of Eq.~(\ref{eq:aceq.2.1}), 
\begin{equation}
x(t)=r\cos(\omega t+\varphi).\label{eq:aceq.2.2}
\end{equation}
As we suppose the oscillator to be in a steady state we are only interested
in the stationary solution~(\ref{eq:aceq.2.2}) further on. By substitution
of~(\ref{eq:aceq.2.2}) into~(\ref{eq:aceq.2.1}) we find after
some calculations for the phase shift between the forced oscillation\index{forced oscillation}
and the forcing oscillation that 
\begin{equation}
\tan\varphi=-\frac{2\gamma\omega}{\omega_{0}^{2}-\omega^{2}}\,,\label{eq:aceq.2.3}
\end{equation}
 and for the amplitude of the forced oscillation 
\begin{equation}
r(\omega)=\frac{F_{0}/m}{\sqrt{(\omega_{0}^{2}-\omega^{2})^{2}+(2\gamma\omega)^{2}}}\,.\label{eq:aceq.2.4}
\end{equation}

Next, we derive the work-energy $W_{\mathrm{bouncer}}$ of the stationary
system for $t\gg\gamma^{-1}$ and $\omega=\omega_{0}$. From~(\ref{eq:aceq.2.2})
we find for the kinetic energy of the harmonic oscillator that 
\begin{equation}
E_{\mathrm{kin}}=\frac{1}{2}m\dot{x}^{2}=\frac{1}{2}m\omega^{2}r^{2}\sin^{2}(\omega t+\varphi)\label{eq:diss.Ekin}
\end{equation}
and from~(\ref{eq:aceq.2.1}) for the potential energy
\begin{equation}
E_{\mathrm{pot}}=\intop_{0}^{x}m\omega_{0}^{2}x\d x=\frac{1}{2}m\omega_{0}^{2}x^{2}=\frac{1}{2}m\omega_{0}^{2}r^{2}\cos^{2}(\omega t+\varphi).\label{eq:diss.Epot}
\end{equation}
Therefore, the sum of the kinetic and the potential energy reads
\begin{equation}
\begin{aligned}E & =E_{\mathrm{kin}}(t)+E_{\mathrm{pot}}(t)=\frac{1}{2}m\dot{x}^{2}+\frac{1}{2}m\omega_{0}^{2}x^{2}\\
 & =\vphantom{\intop_{0}^{x}}\frac{1}{2}mr^{2}\left[\omega^{2}\sin^{2}(\omega t+\varphi)+\omega_{0}^{2}\cos^{2}(\omega t+\varphi)\right]\\
 & =\frac{1}{2}mr^{2}\left[\omega^{2}+\left(\omega_{0}^{2}-\omega^{2}\right)\cos^{2}(\omega t+\varphi)\right]
\end{aligned}
\label{eq:diss.1.8}
\end{equation}
Generally, energy $E$ oscillates for $\omega\neq\omega_{0}$ whereas
for $\omega=\omega_{0}$ Eq.~(\ref{eq:diss.1.8}) reduces to
\begin{equation}
E\Bigr|_{\omega\to\omega_{0}}=\frac{1}{2}m\omega_{0}^{2}r^{2}=\mathrm{const.}\label{eq:aceq.2.14}
\end{equation}
This means, the damped, forced system turns out to be stationary if
it is driven at the resonance frequency $\omega=\omega_{0}$ of the
free undamped oscillator\index{oscillator!resonance frequency}.

For $\omega=\omega_{0}$ we obtain the work-energy multiplying Eq.~(\ref{eq:aceq.2.1})
with $\dot{x}$ 
\begin{equation}
m\ddot{x}\dot{x}+m\omega_{0}^{2}x\dot{x}=-2\gamma m\dot{x}^{2}+F_{0}\cos(\omega_{0}t)\dot{x}\label{eq:aceq.2.5}
\end{equation}
which can also be written as 
\begin{equation}
\frac{\mathrm{d}}{\mathrm{d}t}\left(\frac{1}{2}m\dot{x}^{2}+\frac{1}{2}m\omega_{0}^{2}x^{2}\right)=-2\gamma m\dot{x}^{2}+F_{0}\cos(\omega_{0}t)\dot{x}.\label{eq:aceq.2.6}
\end{equation}
In parentheses on the left hand side one easily recognizes the sum
of the kinetic and the potential energy of Eq.~(\ref{eq:diss.1.8})
which is constant for a stationary solution~(\ref{eq:aceq.2.2}),
which is why the l.h.s.\  of Eq.~(\ref{eq:aceq.2.6}) equals zero.
As Eq.~(\ref{eq:aceq.2.6}) provides the power balance of the forced
oscillator, we identify the damping of $-2\gamma m\dot{x}^{2}$ as
the expended power going off the oscillator to the bath, whereas,
in turn, $F_{0}\cos(\omega_{0}t)\dot{x}$ represents the power which
is regained from the energy bath and applied back to the system. We
conclude that the driving force and the friction force have to cancel
each other 
\begin{align}
F_{0}\cos(\omega_{0}t) & =2\gamma m\dot{x}=-2\gamma m\omega_{0}r\sin(\omega_{0}t+\varphi).
\end{align}
We get 
\begin{equation}
F_{0}=2\gamma m\omega_{0}r,\qquad\varphi=-\frac{\pi}{2}
\end{equation}
and thus
\begin{equation}
x(t)=r\sin(\omega_{0}t),\qquad\dot{x}(t)=\omega_{0}r\cos(\omega_{0}t),\qquad r=\frac{F_{0}}{2\gamma m\omega_{0}}\,.\label{eq:aceq.2.12}
\end{equation}
One can write down the net work-energy that is taken up by the bouncer
during each period $\tau=\frac{2\pi}{\omega_{0}}$ as 
\begin{equation}
\begin{aligned}W_{\mathrm{bouncer}} & =\intop_{0}^{\tau}F_{0}\cos(\omega_{0}t)\dot{x}\d t=\intop_{0}^{\tau}2\gamma m\dot{x}^{2}\d t\\
 & =2\gamma m\omega_{0}^{2}r^{2}\intop_{0}^{\tau}\cos^{2}(\omega_{0}t)\d t=\gamma m\omega_{0}^{2}r^{2}\tau=2\pi\gamma m\omega_{0}r^{2}.
\end{aligned}
\label{eq:aceq.2.7}
\end{equation}

Let us recall that $W_{\mathrm{bouncer}}$ is the energy floating
in one period from the energy bath via the oscillator to friction
energy. In addition, we have the constant energy as mentioned in connection
with Eq.~(\ref{eq:aceq.2.6}) of the oscillator. Further on, we call
this constant energy $E_{\mathrm{bouncer}}$, which is the energy~(\ref{eq:aceq.2.14})
of the linear harmonic oscillator whose mean energies are given by
Eqs.~(\ref{eq:diss.Ekin}) and (\ref{eq:diss.Epot}) together with~(\ref{eq:aceq.2.12})
by\footnote{We shall use different symbols for mean values over space $\meanx x$,
and mean values over time $\meant x$, if not otherwise noted (see
e.g.~\cite{Schwabl.2006en}).} 
\begin{equation}
\meant{E_{\mathrm{kin}}}=\meant{E_{\mathrm{pot}}}=\frac{1}{4}m\omega_{0}^{2}r^{2},\qquad E=\meant{E_{\mathrm{kin}}}+\meant{E_{\mathrm{pot}}}=\frac{m\omega_{0}^{2}r^{2}}{2}.
\end{equation}

In our work \cite{Groessing.2011explan} we are concerned about the
lowest energy of a harmonic oscillator. From quantum mechanics we
know that this lowest energy is $\hbar\omega_{0}/2$. In order to
bring this classical harmonic oscillator in a quantum mechanical context,
we request
\begin{equation}
E_{\mathrm{bouncer}}:=E=\frac{m\omega_{0}^{2}r^{2}}{2}=\frac{\hbar\omega_{0}}{2},\label{eq:aceq.120}
\end{equation}
where the symbol $E_{\mathrm{bouncer}}$ assigns the energy to the
bouncer system. From Eq.~(\ref{eq:aceq.120}) we immediately find
that
\begin{equation}
mr^{2}\omega_{0}=\hbar.\label{eq:aceq.2.18}
\end{equation}
For the work-energy $W_{\mathrm{bouncer}}$ of Eq.~(\ref{eq:aceq.2.7})
we get 
\begin{equation}
W_{\mathrm{bouncer}}=2\pi\gamma\hbar.\label{eq:aceq.2.19}
\end{equation}

In the generalized case of an $N$\textendash dimensional space, we
can separate each dimension with the use of Eq.~(\ref{eq:aceq.2.2})
and get for the oscillators' amplitudes, 
\begin{equation}
\begin{array}{rcl}
x_{1}(t) & = & r_{x_{1}}\cos(\omega_{0}t+\phi_{x_{1}}),\\
 & \vdots\\
x_{N}(t) & = & r_{x_{N}}\cos(\omega_{0}t+\phi_{x_{N}}).
\end{array}\label{eq:aceq.2.20}
\end{equation}
We obtain the work-energy during each single period $\tau$ by integrating
over the $N$ components~(\ref{eq:aceq.2.20}) and get 
\begin{align}
W_{\mathrm{bouncer}} & =\intop_{\tau}2\gamma m(\dot{r}_{x_{1}}^{2}+\cdots+\dot{r}_{x_{N}}^{2})\d t=N\gamma m\omega_{0}^{2}r^{2}\tau\label{eq:aceq.2.21}
\end{align}
with
\begin{equation}
r^{2}=r_{x_{1}}^{2}=\ldots=r_{x_{N}}^{2}.
\end{equation}
Assuming again (\ref{eq:aceq.2.18}), we can write down for any number
$N$ of dimensions that 
\begin{equation}
W_{\mathrm{bouncer}}=2\pi N\gamma\hbar.\label{eq:aceq.2.22}
\end{equation}

\section{Brownian motion of a particle: The walker\label{sec:sub.2}}

After having discussed the first subsystem which fulfilled a harmonic
oscillation, we focus on the second subsystem which obeys a Brownian
motion, embedded in an environment comprising an energy bath with
a white noise driving force (cf.~\cite{Coffey.2004}). The latter
oblige the subsystem to undergo rapid and random movements due to
statistical independent kicks of random magnitude and direction.

The Brownian motion\index{Brownian motion!walker} of a thus characterized
particle, which we call a walker, is then described by a Langevin\index{Langevin equation}
stochastic differential equation with velocity $u=\dot{x}$, a time-dependent
stochastic force $f(t)$, and friction coefficient $\zeta$ (c.f.~\cite[chapter 8.1]{Schwabl.2006en}
and \cite{Uhlenbeck.1930theory,Coffey.2004}), 
\begin{equation}
m\dot{u}=-m\zeta u+f(t),\qquad t\ge0\label{eq:Langevin}
\end{equation}
which describes stochastic processes which we investigate for $t\ge0$
only. Since the force $f(t)$ is stochastic \textendash{} and hence
is the velocity stochastic \textendash{} one has as usual for the
averages 
\begin{equation}
\meant{f(t)}=0,\quad\meant{f(t)f(t')}=\phi(t-t'),\label{eq:aceq.3.2}
\end{equation}
where $\phi(\tau)$ differs noticeably from zero only for intervals
$\tau<\tau_{\mathrm{c}}$. The correlation time $\tau_{\mathrm{c}}$
denotes the time during which the fluctuations of the stochastic force
remain correlated\footnote{Under the precondition that the collisions of the particles undergoing
a Brownian motion are completely uncorrelated, the correlation time
is roughly equal to the duration of a collision. \cite{Schwabl.2006en}}. We are only interested in the Brownian-type motion of the particle,
therefore we restrict ourselves to $\tau\gg\tau_{\mathrm{c}}$ that
further allows us to  introduce a coefficient $\lambda$ that measures
the strength of the mean square deviation of the stochastic force,
such that 
\begin{equation}
\phi(\tau)=\lambda\delta(\tau).\label{eq:aceq.3.7}
\end{equation}

One solves the Langevin equation with the help of the retarded Green's
function $G(t)$
\begin{equation}
\dot{G}+\zeta G=\delta(t),\qquad G(t)=\Theta(t)\e^{-\zeta t}
\end{equation}
with the Heaviside step function $\Theta(t)=\int_{-\infty}^{t}\delta(\tau)\d\tau$.
Letting 
\begin{equation}
u(t=0)=u_{0}\label{eq:diss.u0}
\end{equation}
 be the initial value of the velocity, one obtains 
\begin{equation}
\begin{aligned}u(t) & =u_{0}\e^{-\zeta t}+\intop_{0}^{\infty}\d\tau\,G(t-\tau)f(\tau)/m\\
 & =u_{0}\e^{-\zeta t}+\e^{-\zeta t}\intop_{0}^{t}\d\tau\,\e^{\zeta\tau}f(\tau)/m.
\end{aligned}
\label{eq:diss.2.26}
\end{equation}
Using this solution and the assumptions~(\ref{eq:aceq.3.2}) we find
for the mean value of the velocity\footnote{The mean value $\meant{}$ can be understood either as an average
over time or an average over an ensemble at a fixed time, $\meant{x(t)}=\int_{-\infty}^{\infty}xP(x,t)\d x$.
For a stationary process the mean value is constant because of $P(x,t)=P(x)$
(see e.g.~\cite{Schwabl.2006en,Burgdorfer.2007statistical}).}
\begin{equation}
\meant{u(t)}=\underbrace{\meant{u_{0}}}_{=u_{0}}\e^{-\zeta t}+\intop_{0}^{t}\d\tau\,\e^{-\zeta(t-\tau)}\underbrace{\meant{f(\tau)}}_{=0}/m=u_{0}\e^{-\zeta t},
\end{equation}
where no average is involved over the friction terms because they
are constant describing the averaged interaction of the system with
the bath. Therefore, one does not consider the average value of $u(t)$,
but instead that of its square, 
\begin{equation}
\begin{aligned}\meant{u^{2}(t)} & =\meant{u_{0}^{2}}\e^{-2\zeta t}+2\e^{-2\zeta t}\intop_{0}^{t}\d\tau\,\e^{\zeta\tau}\underbrace{\meant{u_{0}f(\tau)}}_{=0}\frac{1}{m}\\
 & \hphantom{=u_{0}^{2}\e^{-2\zeta t}}+\e^{-2\zeta t}\intop_{0}^{t}\d\tau\,\intop_{0}^{t}\d\tau'\thinspace\e^{\zeta(\tau+\tau')}\meant{f(\tau')f(\tau)}\frac{1}{m^{2}}\\
 & =u_{0}^{2}\e^{-2\zeta t}+\e^{-2\zeta t}\intop_{0}^{t}\d\tau\,\intop_{0}^{t}\d\tau'\thinspace\e^{\zeta(\tau+\tau')}\phi(\tau-\tau')\frac{1}{m^{2}}\\
 & =\vphantom{\intop_{0}^{t}}u_{0}^{2}\e^{-2\zeta t}+\frac{\lambda}{2\zeta m^{2}}\left(1-\e^{-2\zeta t}\right)\quad\stackrel{t\gg\zeta^{-1}}{\longrightarrow}\quad\frac{\lambda}{2\zeta m^{2}}
\end{aligned}
\label{eq:aceq.3.8}
\end{equation}
where the velocity $u_{0}$ at $t=0$ is independent of and hence
uncorrelated with the random force $f(t)$ and hence $\meant{u_{0}f(\tau)}=\meant{u_{0}}\meant{f(\tau)}=0$.
For $t\gg\zeta^{-1}$, the term containing $u_{0}$ becomes negligible,
i.e.\ $\zeta^{-1}$ then plays the role of a relaxation time. We
require that our particle attains thermal equilibrium~\cite{Groessing.2008vacuum,Groessing.2009origin}
after long times so that due to the \textit{equipartition theorem
on the sub-quantum level}~\footnote{We assume the equipartition theorem to be the same and hence borrowed
from classical statistical mechanics.} the average value of the kinetic energy becomes 
\begin{equation}
\meant{E_{\mathrm{kin}}}=\frac{1}{2}m\,\meant{u^{2}(t)}\quad\stackrel{t\gg\zeta^{-1}}{\longrightarrow}\quad\frac{\lambda}{4\zeta m}=:E_{\mathrm{zp}},\label{eq:aceq.3.9}
\end{equation}
with $E_{\mathrm{zp}}$ being the average kinetic energy of the zero-point
field. One can define the $E_{\mathrm{zp}}$ per degree of freedom
as~\footnote{As we are probably at a length scale where the thermodynamical laws
have not yet proven valid, we stick to formally using $E_{\mathrm{zp}}$.
Surely, Eq.~(\ref{eq:aceq.3.3a}) is the sub-quantum analogon to
the thermodynamical expression $k_{\mathrm{B}}T/2$, however, as for
today we neither know $T_{0}$ nor the constant $k$ \textendash{}
unless it should turn out as identical to $k_{\mathrm{B}}$.} 
\begin{equation}
E_{\mathrm{zp}}:=\frac{kT_{0}}{2}\label{eq:aceq.3.3a}
\end{equation}
with $k$ being a constant equivalent to Boltzmann's constant $k_{\mathrm{B}}$,
and $T_{0}$ denotes the vacuum temperature in our scenario in close
analogy to the usual thermodynamical formalism.

Next we derive the velocity correlation function
\begin{equation}
\begin{aligned}\meant{u(t)u(t')} & =\meant{u_{0}^{2}}\e^{-\zeta(t+t')}+\e^{-\zeta(t+t')}\intop_{0}^{t}\d\tau\,\intop_{0}^{t'}\d\tau'\thinspace\e^{\zeta(\tau+\tau')}\meant{f(\tau)f(\tau')}\frac{1}{m^{2}}\\
 & \qquad+\e^{-\zeta(t+t')}\left(\intop_{0}^{t}\d\tau\,\e^{\zeta\tau}\underbrace{\meant{u_{0}f(\tau)}}_{=0}+\intop_{0}^{t'}\d\tau\,\e^{\zeta\tau}\underbrace{\meant{u_{0}f(\tau)}}_{=0}\right)\frac{1}{m}\\
 & =u_{0}^{2}\e^{-\zeta(t+t')}+\e^{-\zeta(t+t')}\intop_{0}^{t}\d\tau\,\intop_{0}^{t'}\d\tau'\thinspace\e^{\zeta(\tau+\tau')}\delta(\tau-\tau')\frac{\lambda}{m^{2}}\\
 & =\frac{\lambda}{2\zeta m^{2}}\e^{-\zeta|t-t'|}+\left(u_{0}^{2}-\frac{\lambda}{2\zeta m^{2}}\right)\e^{-\zeta(t+t')}.
\end{aligned}
\label{eq:diss.2.30}
\end{equation}
For $t,t'\gg\zeta^{-1}$ one can neglect the last term in~(\ref{eq:diss.2.30}).
Then, one obtains the mean square displacement of $x(t)$ by integrating~(\ref{eq:diss.2.30})
twice, assuming $x(0)=0$, which yields~\footnote{We stress that even if we use the same character $x$ as for the oscillating
particle, now the meaning is different: $x(t)$ in section~\ref{sec:aceq.sub.3}
signified a deterministic harmonic displacement of mass point $m$
in the case of an oscillating particle (bouncer), whereas $x(t)$
now means a stochastic random walk variable for the particle that
carries out a Brownian motion of the walker.}
\begin{equation}
\meant{x^{2}(t)}=\intop_{0}^{t}\d\tau\intop_{0}^{t}\d\tau'\frac{\lambda}{2\zeta m^{2}}\e^{-\zeta|\tau-\tau'|}\quad\stackrel{t\gg\zeta^{-1}}{\longrightarrow}\quad\frac{\lambda}{\zeta^{2}m^{2}}t=2Dt,\label{eq:aceq.3.11}
\end{equation}
 with the diffusion constant 
\begin{equation}
D=\frac{\lambda}{2\zeta^{2}m^{2}}=\frac{2E_{\mathrm{zp}}}{\zeta m}.\label{eq:aceq.3.12}
\end{equation}

Next, we calculate the work-energy $W_{\mathrm{walker}}$ of the stationary
system. We remind ourselves that we have to do with a steady-state
system. Due to the friction $\zeta$, there exists a flow of (kinetic)
energy into the environment. Consequently, there must also exist a
work-energy flow back into our system of interest. Therefore, we calculate
the averaged power by multiplying Eq.~(\ref{eq:Langevin}) by $u=\dot{x}$
and obtain an averaged power-balance equation
\begin{equation}
m\meant{\ddot{x}\dot{x}}=-m\zeta\meant{\dot{x}^{2}}+\meant{f(t)\dot{x}}.\label{eq:diss.1.36}
\end{equation}
In contrast to Eq.~(\ref{eq:aceq.2.5}) we are dealing with stochastic
variables and thus we are fine with averaged values for the power-balance.
Even though, we assume in close analogy to Eq.~(\ref{eq:aceq.2.5}),
that the average system's energy being constant due to a stationary
state of the system. Therefore, the terms on the right hand side of
Eq.~(\ref{eq:diss.1.36}) providing the power balance must cancel.
This yields for the duration of time $\tau$ the net work-energy of
the walker 
\begin{equation}
W_{\mathrm{walker}}={\color{blue}{\color{black}\intop_{\tau}m\zeta\meant{\dot{x}^{2}}\d t=\intop_{\tau}m\zeta\meant{u^{2}}\d t.}}\label{eq:aceq.3.13}
\end{equation}
Here, we want to ensure that the work-energy we shall obtain is comparable
with Eq.~(\ref{eq:aceq.2.19}). Therefore, we choose the basic time
interval $\tau=2\pi/\omega_{0}$ of the walker-system the same as
in Eq.~(\ref{eq:aceq.2.12}) of the bouncer-system. Furthermore,
as we are dealing with a walker-system that obeys a stochastic motion,
we have to work with mean values to make all fluctuating contributions
negligible due to averaging over these statistical variations.

Inserting~(\ref{eq:aceq.3.9}) into~(\ref{eq:aceq.3.13}), we obtain
\begin{equation}
W_{\mathrm{walker}}=\tau m\zeta\,\meant{u^{2}(t)}=2\tau\zeta E_{\mathrm{zp}}.\label{eq:aceq.3.14}
\end{equation}
The work-energy for the particle undergoing Brownian motion can thus
be written as 
\begin{equation}
W_{\mathrm{walker}}=\frac{4\pi}{\omega_{0}}\zeta E_{\mathrm{zp}}.\label{eq:aceq.3.15}
\end{equation}

Turning now to the $N$-dimensional case, the average squared velocity
of a particle is 
\begin{equation}
\meant{u^{2}}=\meant{u_{x_{1}}^{2}}+\cdots+\meant{u_{x_{N}}^{2}},\label{eq:aceq.3.16}
\end{equation}
 with equal probability for each direction, 
\begin{equation}
\meant{u_{x_{1}}^{2}}=\cdots=\meant{u_{x_{N}}^{2}}=\frac{1}{N}\meant{u^{2}}.\label{eq:aceq.3.17}
\end{equation}
 Accordingly, the average kinetic energy of a moving particle with
$N$ degrees of freedom becomes 
\begin{equation}
E_{\mathrm{zp}}^{(N)}=\frac{1}{2}m\meant{u^{2}}=NE_{\mathrm{zp}}\label{eq:aceq.3.18}
\end{equation}
 and thus 
\begin{equation}
\meant{u^{2}(t)}=2N\,\frac{E_{\mathrm{zp}}}{m}.\label{eq:aceq.3.19}
\end{equation}

Again, we note that Eq.~(\ref{eq:aceq.3.19}) describes an energy
equipartition which, however, here relates to the sub-quantum level,
i.e.\ to the vacuum temperature $T_{0}$. It should thus not be confused
with the equipartition theorem as discussed, e.g.\ with respect to
blackbody radiation and the Planck spectrum.

With the analogical explanation as for the one-dimensional case, we
find for the work-energy of the walker in $N$-dimensional space 
\begin{align}
W_{\mathrm{walker}} & =m\zeta\intop_{\tau}\left[\meant{u_{x_{1}}^{2}(t)}+\cdots+\meant{u_{x_{N}}^{2}(t)}\right]\d t=m\zeta\intop_{\tau}\meant{u^{2}(t)}\d t.\label{eq:aceq.3.20}
\end{align}
 Inserting (\ref{eq:aceq.3.19}), we obtain 
\begin{equation}
W_{\mathrm{walker}}=\tau m\zeta\meant{u^{2}(t)}=2\tau\zeta NE_{\mathrm{zp}},\label{eq:aceq.3.21}
\end{equation}
which is $N$ times the value of the one-dimensional case in Eq.~(\ref{eq:aceq.3.15}).
Therefore, the work-energy for the particle undergoing Brownian motion
can be written as 
\begin{equation}
W_{\mathrm{walker}}=\frac{N4\pi}{\omega_{0}}\zeta E_{\mathrm{zp}},\label{eq:aceq.3.22}
\end{equation}
for the general case of $N$ degrees of freedom.

\section{The walking bouncer\label{sec:walking}}

Our model of a single-particle quantum system comprises a bouncer-system
and a walker-system. So far, we have analysed these two systems independently.
Now we construct an energy exchange mechanism for our model where
we assume a continuous energy flow from the bath to the oscillator,
and \textit{vice versa}. Accordingly, the walker gains its energy
from the heat bath via the oscillations of the bouncer\textendash bath
system in $N$ dimensions: The bouncer pumps energy to and from the
heat bath via the friction $\gamma$. 

In the centre of mass frame, the system is characterized by a single
degree of freedom. However, in the $N$-dimensional reference frame
of the laboratory, the oscillation is not fixed \textit{a priori}.
Rather, possible exchanges of energy will be equally distributed in
a stochastic manner. Concerning the latter, the flow of energy is
on average distributed evenly via the friction $\gamma$ in all $N$
dimensions of the laboratory frame. It can thus also be considered
as the stochastic source of the particle moving in $N$ dimensions,
each described by the Langevin equation~(\ref{eq:Langevin}).

Therefore, we recognize \textit{friction} in both cases, as represented
by $\gamma$ and $\zeta$, respectively, to generally describe the
coupling between the oscillator (or particle in motion) on the one
hand, and the bath on the other hand. Moreover, and most importantly,
during that flow, the averaged coupling of the bouncer can be assumed
to be exactly identical with the coupling of the walker. For this
reason we directly compare the results of Eqs.~(\ref{eq:aceq.2.22})
and (\ref{eq:aceq.3.22}), 
\begin{equation}
W_{\mathrm{bouncer}}=W_{\mathrm{walker}},\label{eq:aceq.4.1}
\end{equation}
 providing 
\begin{equation}
2\pi N\gamma\hbar=\frac{N4\pi}{\omega_{0}}\zeta E_{\mathrm{zp}}.\label{eq:aceq.4.2}
\end{equation}

Now, our single-particle quantum model consists of two parts, each
of which possesses a certain energy, which we expressed by Eqs.~(\ref{eq:aceq.120})
and (\ref{eq:aceq.3.9}), respectively. Even described by two different
mechanisms, the bouncer-system and the walker-system are still two
different aspects of our assumed single-particle quantum model. Therefore,
the energy $E$ of each system must be the same, being the minimum
energy of the single particle. We derived the energies of the sub-systems
as
\begin{equation}
E_{\mathrm{bouncer}}=\frac{m\omega_{0}^{2}r^{2}}{2}=\frac{\hbar\omega_{0}}{2}\label{eq:diss.1.49}
\end{equation}
being the energy of the bouncer and as
\begin{equation}
E_{\mathrm{zp}}=\frac{kT_{0}}{2}\label{eq:diss.1.50}
\end{equation}
being energy of the walker, respectively. Comparing these two equations
yields
\begin{equation}
\frac{\hbar\omega_{0}}{2}=\frac{kT_{0}}{2}\label{eq:hw=00003Dkt}
\end{equation}
and hence the zero-point energy in terms of $\hbar\omega$ reads as
\begin{equation}
E_{\mathrm{zp}}=\frac{\hbar\omega_{0}}{2}\thinspace.\label{eq:diss.1.52}
\end{equation}
Substituting this result into Eq.~(\ref{eq:aceq.4.2}) leads directly
to 
\begin{equation}
\gamma=\zeta\label{eq:gamma=00003Dzeta}
\end{equation}
which means the bouncer and the walker are coupled with the same strength
to the ZPF bath, i.e.\ the friction coefficient for both the bouncer
and the walker is identical.

For a quantitative derivation of the friction coefficients of both
the bouncer and the walker, we introduce the action function $S(\VEC x,t)$
such that the total energy of the whole system is given by 
\begin{equation}
E_{\mathrm{tot}}(\VEC x,t)=-\frac{\partial S(\VEC x,t)}{\partial t}\,.\label{eq:3.2.2}
\end{equation}
We need to specify that a quantum system's total energy consists of
the energy of the system of interest (i.e., the particle with frequency
$\omega_{0}$), and of some term representing energy throughput related
to the surrounding vacuum, i.e.\ effectively some function $F$ of
the heat flow $\Delta Q$,~\cite{Groessing.2008vacuum}
\begin{equation}
E_{\mathrm{tot}}(\VEC x,t)=E(\omega_{0},\VEC x,t)+F\left[\Delta Q(\VEC x,t)\right].\label{eq:3.1.6}
\end{equation}
The first term in Eq.~(\ref{eq:3.1.6}) corresponds to a particle's
energy. The second term, being equivalent to some kinetic energy,
can be recast with the aid of a fluctuating momentum term, $\delta\VEC p$,
of the particle with momentum $\VEC p$, by 
\begin{equation}
F\left[\Delta Q(\VEC x,t)\right]=\frac{(\delta\mathbf{p})^{2}}{2m}\,.\label{eq:diss.1.53}
\end{equation}
We consider as usual the momentum $\mathbf{p}$ of the particle as
given by 
\begin{equation}
\VEC p(\VEC x,t)=\nabla S(\VEC x,t)=m\VEC v,\label{eq:p=00003DnablaS=00003Dmv}
\end{equation}
noting, however, that this will not be the effective particle momentum
yet, due to the additional momentum coming from the heat flow, described
by momentum fluctuation of Eq.~(\ref{eq:diss.1.53}) as 
\begin{equation}
\delta\VEC p=\delta(\nabla S)=\nabla(\delta S):=m\VEC u,\label{eq:dp=00003DnabladS=00003Dmu}
\end{equation}
where velocity $\VEC u$ is assumed to be the same as in the Langevin
equation~(\ref{eq:Langevin}). Our task is now to find an adequate
expression for $\delta\VEC p$ from our central assumption, i.e.,
from an underlying nonequilibrium thermodynamics. To begin, we remember
the distinction between ``heat'' as disordered internal energy on
one hand, and mechanical work on the other: heat as disordered energy
cannot be transformed into useful work by any means. According to
Boltzmann, if a particle trajectory is changed by some supply of heat
$\Delta Q$ to the system, this heat will be spent either for the
increase of disordered internal energy, or as ordered work furnished
by the system against some constraint mechanism,~\cite{Boltzmann.1866uber}
\begin{equation}
\Delta Q=\Delta E_{\mathrm{internal}}+\Delta W_{\mathrm{constraints}}.\label{eq:3.2.6}
\end{equation}
Now, in order to proceed in our quest to obtain an expression for
the momentum fluctuation~(\ref{eq:dp=00003DnabladS=00003Dmu}) from
our thermodynamical approach, we can again rely on a formula originally
derived by Ludwig Boltzmann. As mentioned above, Boltzmann considered
the change of a trajectory by the application of heat $\Delta Q$
to the system. Considering a very slow transformation, i.e., as opposed
to a sudden jump, Boltzmann derived a formula which is easily applied
to the special case where the motion of the system of interest is
oscillating with some period $\tau=2\pi/\omega_{0}$. Boltzmann's
formula for periodic systems~(\ref{eq:boltz.40}) relates the applied
heat $\Delta Q$ to a change in the action function~(\ref{eq:boltz.39})
$S=\int(E_{\mathrm{kin}}-V)\d t$, i.e., $\delta S=\delta\int E_{\mathrm{kin}}\d t$,
providing\footnote{The period $\tau$ is assumed to remain constant during a change $\Delta Q$.}
\begin{equation}
\Delta Q=2\omega_{0}\delta S=2\omega_{0}\left[\delta S(\tau)-\delta S(0)\right].\label{eq:diss.1.57}
\end{equation}
The gradient reads as 
\begin{equation}
\nabla Q=2\omega_{0}\nabla(\delta S),\label{eq:diss.1.58}
\end{equation}
with abbreviation $\nabla(\Delta Q)=:\nabla Q$, which leads by using~(\ref{eq:dp=00003DnabladS=00003Dmu})
to
\begin{equation}
m\mathbf{u}=\frac{\nabla Q}{2\omega_{0}}\,.\label{eq:aceq.4.8a}
\end{equation}
 As the friction force in Eq.~(\ref{eq:Langevin}) is equal to the
gradient of the heat flux, 
\begin{equation}
m\zeta\mathbf{u}=\nabla Q,\label{eq:aceq.4.9}
\end{equation}
 comparison of~(\ref{eq:aceq.4.8a}) and~(\ref{eq:aceq.4.9}) together
with~(\ref{eq:gamma=00003Dzeta}) provides
\begin{equation}
\zeta=\gamma=2\omega_{0}.\label{eq:aceq.4.10}
\end{equation}
Note that with Eqs.~(\ref{eq:diss.1.52}) and (\ref{eq:aceq.4.10})
one obtains in any one dimension the expression for the diffusion
constant~(\ref{eq:aceq.3.12})\index{diffusion constant} as 
\begin{equation}
D=\frac{2E_{\mathrm{zp}}}{\zeta m}=\frac{\hbar}{2m}\,,\label{eq:aceq.4.12}
\end{equation}
 which is exactly the usual expression for $D$ in the context of
quantum mechanics.

\section{Conclusions and perspectives}

In this chapter a new type of objects has been presented obeying the
laws of Newtonian mechanics which can exhibit simultaneously particle
and wave properties. As a prerequisite, classical non-equilibrium
thermodynamics has been assumed, i.e. a mechanism of stationary energy
flow, which enables a work-energy exchange between an oscillating
bouncer and a stochastically driven walker. It has been shown that
such an exchange can be derived with two classical differential equations,
the Newtonian equation and the Langevin equation, together describing
the two-fold perspective of a single particle called the walking bouncer.
Each of these equations contains a friction factor, which has been
shown to be equal for both equations, on the one hand, and responsible
for the coupling and hence the characteristic feature of the transfer,
on the other hand.

Both equations used, the Newtonian and the Langevin equation, are
classical equations leading naturally to classical solution. To build
a connection to the quantum regime, the minimum energy of a quantum
oscillator has been used to introduce energy quantisation. Once applied
this step, all used attributes turned out to be equal to the ones
known from quantum mechanics, especially the diffusion constant.

The given picture leaves  open which part, the bouncer or the walker,
is the sender of the energy transfer and which one the receiver of
the exchanged energy, respectively. Certainly, one could surely find
an answer for macroscopic particles when taking a close look at Couder's
experiments. However, one should not expect to determine thereby an
adequate answer for the mechanism translated into the language of
quantum mechanics as the model presented herein should rather be considered
a toy-model hopefully giving one a clue to find a precise mathematical
description of the whole system underlying quantum mechanics. In this
sense, the derived walking bouncer should be recognized as a model
for further discussions on how an object could act as a particle and
a wave simultaneously, thereby replacing the old fashioned picture
of an object that could either act as a particle or a wave, dependent
on particular circumstances.

In the following chapters we will implicitly make use of such type
of a particle, even though the zitterbewegung, modelled by the stochastic
movement of the walker, will silently disappear in the mathematical
description due to averaging processes. This also means we shall leave
here the level of stochastic description and turn towards a phenomenological
approach of a particle's behaviour, i.e.~the decay of a Gaussian
distribution.

\chapter{Probability distributions and velocities\label{sec:3.probability-distributions}}
\begin{quote}
Following the same idea as in section~\ref{sec:The-macroscopic-fluid},
it should be worth to investigate the mechanism of the microscopic
picture of the fluid droplet in particular with regard to underlying
processes. Also, Bohmian theory gives an answer to this question,
however, with regard to the underlying diffusion processes, we shall
in fact find a different answer: In this chapter we shall provide
a completely different view of the diffusion process which emerges
out of uncorrelated Gaussian position distributions as well as momentum
distributions, with the spreading of the resulting wave packet being
characterized as a ballistic diffusion. 

By introducing a slit setup which will serve as the main environment
to our investigations further on, it is sufficient to analyse one-dimensional
distributions only. In a further step, we shall derive the ballistic
diffusion equation which allows us the complete description of the
spreading wave packet.
\end{quote}

\section{Outline}

Based on the results derived in chapter~\ref{sec:2.fluid-droplet},
we move now towards a phenomenological approach of a particle's behaviour.
Therefore, we assume the particles to emerge from a source one by
one propagating through a slit, and finally hitting a screen becoming
visible, or being measured there. According to our discussion in chapter~\ref{sec:2.fluid-droplet},
we model a system in which each single particle obeys the random motion
of Brownian-type. We draw conclusion from the measurement patterns
of such experiments that in the average of a sufficiently big number
of single events we can assume smooth trajectories thereby describing
the influence of an underlying diffusion process. In order to keep
things simple, we always assume an aperture with whose edges the particle's
interaction is negligible. Furthermore, we restrict our investigations
to Gaussian-shaped probability distributions only, as this is a function
class widely used in physical theory, which reproduces all the quantum
measurements considered herein. As a result, such a smooth diffusion
process will turn out to be a ballistic diffusion\footnote{The term ballistic diffusion will be defined in chapter~\ref{sec:3.3.The-derivation-of-Dt}
by Eq.~\eqref{eq:ballisticDE}.}.

\section{The constituting setup\label{sec:The-constituting-setup}}

For the following, it will be helpful to let ourselves be guided by
the picture provided by the walkers\footnote{Although the fluid droplet model includes both, a bouncer and a walker,
we consider it a single system due to the tight coupling. We prefer
to point out the walker facet of the duality as this aspect suits
better to ones understanding of the propagating particle.} introduced in chapter~\ref{sec:sub.2}. For also with a walker,
one is confronted with a rapidly oscillating object, which itself
is guided by an environment that also contributes some fluctuating
momentum to the walker's propagation. In fact, the walker creates
waves surrounding the particle, and the detailed structure of the
wave configurations influences the walker's path, just as in our approach
the particle, both absorbs heat from and emits heat into its environment,
which can be described in terms of momentum fluctuations.

If we imagine the bouncing of a walker in its fluid environment, the
latter will become excited or heated up wherever the momentum fluctuations
direct the particle to. After some time span \textendash{} which can
be rather short, considering the very rapid oscillations of elementary
particles \textendash{} a whole area of the particle's environment
will be modified by the throughput of energy in this way. Considering
the electron, for example, the fact that it bounces roughly $10^{21}$
times per second, with each bounce eventually providing a slight displacement
from the original path's momentum, one can thus understand the area
filling capacity of any quantum path.

Now, let us assume we have a source of identical particles, which
are prepared in such a way that each one ideally has an initial (classical)
velocity $\VEC v$ moving towards a slit-setup containing at least
one aperture. The latter is assumed to be formed with unsharp edges
to avoid diffraction effects to good approximation. This slit-setup
will be passed by one particle at a time, as usual in quantum mechanical
experiments, thereby generating a probabilistic distribution of particle
locations in the course of time which is the subject of our investigation.
Therefore, our model describes the evolution of said locations from
right after the slit towards a screen (or even beyond) which allows
us to develop and explain the mechanisms of the particle's motion.

At this point we want, however, to point out the difference to Bohmian
theory (see, e.g., Bohm and Hiley~\cite{Bohm.1993undivided}, Holland~\cite{Holland.1993},
or Sanz and Miret-Artéz~\cite{Sanz.2012trajectory,Sanz.2014trajectory}),
which also describes the above-mentioned particle path between a slit
and a screen: The subject of our model is the description of the influence
of an assumed sub-quantum medium on the velocities along the averaged
trajectories and the probability currents\footnote{On the definition of probability currents see chapter~\ref{sec:set-of-current-rules}.}
in the domain between the slit(s) and the screen. In the mean of a
vast number of particles our description converges to Bohm's one.
As will be shown in this chapter, we do not provide a single particle's
path as the outcome of a single experiment because the underlying
(sub-quantum) environment is of statistical nature, similar to a classical
Brownian motion. Accordingly, our model does not predict single-particle
trajectories, instead, the Bohm-type motion emerges from our model
as smoothed out motions of a vast number of single-particle's statistical
hence erratic motions.

Even if we let the particles emerge one at a time only, the local
probability density $P$ right after the slit is assumed to be a Gaussian
one. This comes along with a heat distribution generated by the oscillating
particle, with a maximum at the centre of the aperture $x_{0}$.
\begin{figure}[th]
\centering{}\includegraphics{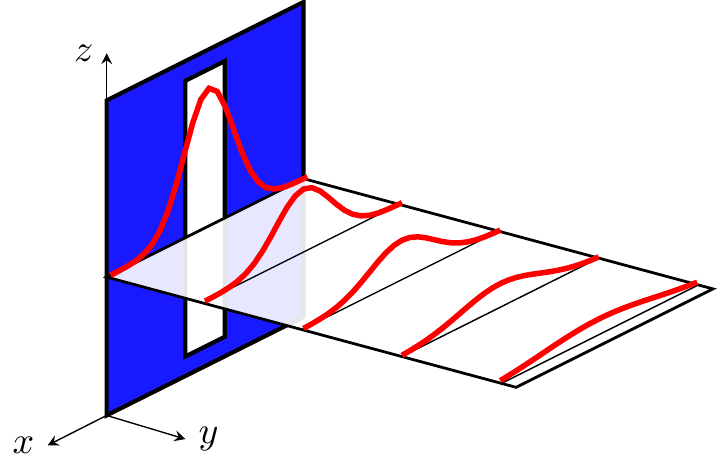}\caption{Setting of a single-slit experiment in three dimensions with sketched
spreading on an exemplary layer\label{fig:single-slit-with-P}}
\end{figure}
 To keep things simple, we describe the Gaussian decay as a function
of its distance $y$ straight ahead from the slit (cf.~Fig\@.~\ref{fig:single-slit-with-P}).
Even more, we connect the $y$-axis with time $t$ by a constant velocity,
\begin{equation}
y\left(t\right)=\frac{\hbar k_{y}t}{m}=v_{y}t\label{eq:y=00003Dv_yt}
\end{equation}
with wave-vector $k_{y}$ in $y$-direction and mass $m$. The idea
behind this constant velocity $v_{y}$ is that the incident sub-quantum
wave before the slit-setup can be considered a plane wave which is
cut by the slits into smaller parts continuing their propagation with
the same, hence constant, velocity $v_{y}$. Any tentative propagation
of the Gaussian shape orthogonal to said straight motion, i.e.\ a
side motion into $x$-direction, will be compensated by an Ehrenfest
motion later on by replacing $x_{0}\to x_{0}+v_{x}t$, i.e.\ an additive
motion of the Gaussian centre along the $x$-axis (cf.~Fig\@.~\ref{fig:td.1}).

\begin{figure}[th]
\centering{}\includegraphics{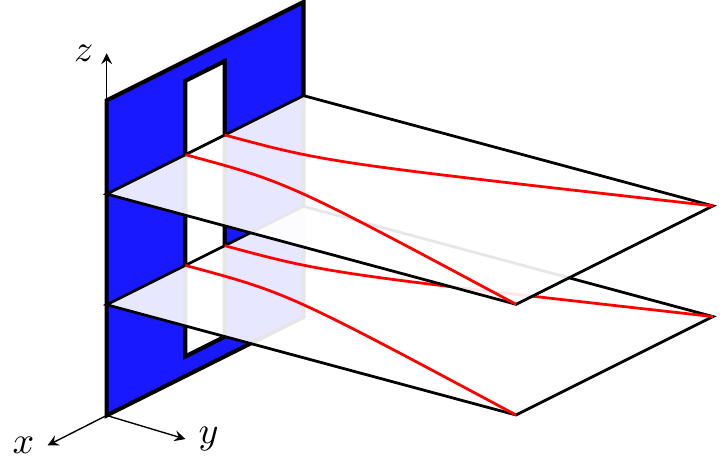}\caption{Setting of a single-slit experiment in three dimensions with Bohm-type
trajectories sketched on different layers\label{fig:diss.2.1}}
\end{figure}

According to the chosen setup, the Gaussian shape broadens only along
the $x$-axis. There is no spreading along the direction of its propagation
because of the assumed steady heat flow from the particle's origin
which is usually an oven in a fixed position far from the slit-setup
in the negative $y$-direction. Further, there is also no spreading
in the $z$-direction which is also the extent of the slit. A thus
assumed spreading of a Gaussian in a plane along the $z$-direction
is compensated by the spreading of a neighboured plane, as sketched
out in Fig.~\ref{fig:diss.2.1}, settled directly above or below
of the current one, respectively, because of equal conditions in neighbouring
plane. For simplicity, we neglect the impact of the slit's edges and
assume for our inquest a sufficiently large distance from the upper
and lower borders, too. 

All problems treated in this thesis contain a Gaussian intensity dispersion
appearing right behind the slit. The dispersion is assumed to be an
ideal Gaussian function not being refracted at the slit's edges. Furthermore,
the Gaussian extends along the whole x-direction, i.e.\ the Gaussian
function is not cut by the slit it runs through, as indicated in Fig.~\ref{fig:single-slit-with-P}
by the left most shape not cut by the slit. Thus one does not need
to consider phase-free spaces along any light-cone-like structures
which would arise otherwise.

In our model, the sub-quantum medium is the mediator between the vacuum
energy and the particle itself. When said sub-quantum medium is excited,
i.e.\ heated up by the oven, it builds immediately a landscape in
the oven's surrounding which includes the setup comprising the slit(s)
and the screen. In terms of an effective theory, the particle, once
sent out by the oven, propagates on average along these trajectories
which are already embedded in said landscape. ``On average'' means
that the particle's propagation is most likely as described, but in
a statistical sense. However, as discussed before, the path of a unique
particle may be completely different.

In other words, when handling the particles' propagations, we make
use of the probabilistic view in that we cannot describe the trajectory
of a single particle but instead have a probability density $P(x,t)$
to find the particle within the interval $[x,x+dx]$ at time $t$.
Even though, quantum mechanics is already a tool to find solutions
of probability density $P(x,t)$ for given setups, i.e.\ in the example
before, the outcome $P(x,t)$ of the measurement at a screen being
at a distance from a slit where the particles passed through, it lacks
a deeper level explanation of this outcome.

\section{Orthogonality relations and fluctuations\label{sec:orthogonality}}

In chapter~\ref{sec:2.fluid-droplet} we have distinguished two velocities:
The osmotic velocity $\VEC u$~\eqref{eq:dp=00003DnabladS=00003Dmu}
and the diffusive velocity $\VEC v$~\eqref{eq:p=00003DnablaS=00003Dmv}.
They have already been provided in textbooks, e.g., Holland~\cite{Holland.1993},
however, herein we will sketch the concise path provided by Grössing~\cite{Groessing.2004hamilton}.
Therefore, within the scope of this single chapter, we extend the
coordinate $x$ to it's three-dimensional equivalent, $\VEC x$, in
order to describe orthogonality relations correctly.

\begin{figure}[!htb]
\centering{}\includegraphics[width=0.75\textwidth]{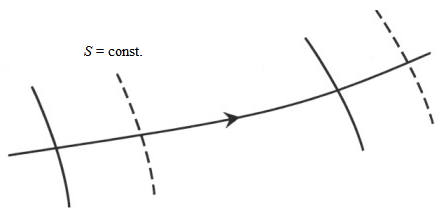}
\caption{Surfaces of constant action function $S(\protect\VEC x,t)$ representing
wave fronts, with orthogonal particle trajectory. Courtesy Gerhard
Grössing~\cite{Groessing.2004hamilton}\label{fig:surface_S}}
\end{figure}

We demand a Gaussian-shaped probability density $P(\VEC x,t)$ to
obey particular requirements, namely the normalization~\eqref{eq:aceq.0.1}
such that the integration over the whole domain $\VEC x$ yields 1,
\begin{equation}
\intop_{t=\mathrm{const.}}P(\VEC x,t)\d^{3}x=1,\label{eq:diss.2.3}
\end{equation}
and the continuity equation
\begin{equation}
\frac{\p P(\VEC x,t)}{\p t}=-\nabla\cdot(\VEC vP(\VEC x,t))\label{eq:continuityEq}
\end{equation}
with the velocity $\VEC v(\VEC x,t)$ along the trajectory derived
from a classical action function $S(\VEC x,t)$ by
\begin{equation}
\VEC v(\VEC x,t)=\frac{\VEC p}{m}=\frac{\nabla S(\VEC x,t)}{m}\thinspace.\label{eq:v=00003Dp_div_m}
\end{equation}
From the assumed uniqueness and differentiability of $S(\VEC x,t)$
follows that the paths don't cross each other. These paths correspond
to particle trajectories orthogonal to surfaces (wave fronts) with
constant action function $S(\VEC x,t)$, as sketched in Fig.~\ref{fig:surface_S}.

\begin{figure}
\begin{centering}
\includegraphics[width=0.75\textwidth]{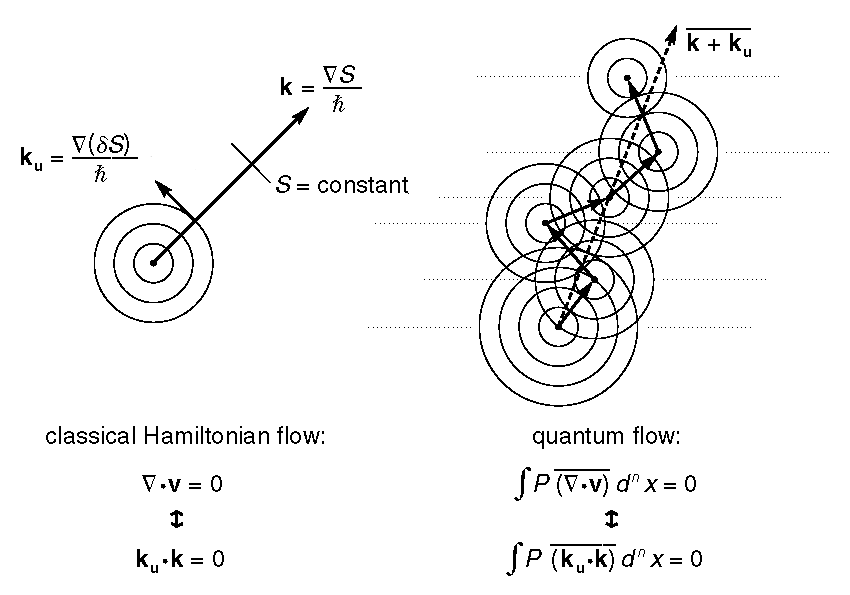}
\par\end{centering}
\caption{Schematic distinction of classical Hamiltonian flow (left) and quantum
flow (right). The dotted lines in the figure on the right indicate
symbolically that the waves pictured represent only the local surroundings
of a generally extending probability field, thus illustrating that
the fluctuations shown are to be seen in the context of the whole
embedding environment. Courtesy Gerhard Grössing~\cite{Groessing.2004hamilton}\label{fig:hamiltonian-flow}}
\end{figure}

The example of Fig.~\ref{fig:surface_S} is but a particular one.
In accordance with Huygens' principle, another wide-spread example
is given by spherical wave surfaces. Here, the surface is initially
concentrated at a point and then expands in a series of closed surfaces,
such that the motion can be compared to that of a shock wave emanating
from a ``disturbing'' point of a surface, i.e., as a travelling
wave front (Fig.~\ref{fig:hamiltonian-flow}).

To emphasise the orthogonality between a particle trajectory and a
wave front, we, firstly, restrict ourselves to  considering the stationary
state of constant flow only, such that the l.h.s.\ of Eq.~\eqref{eq:continuityEq}
is equal to zero. Then dividing by $P$ we get
\begin{equation}
\nabla\cdot\VEC v=-\frac{\nabla P}{P}\cdot\VEC v.\label{eq:diss.2.6}
\end{equation}
In general, however, Eq.~\eqref{eq:diss.2.6} is an expression for
the non-conservation of momentum $\VEC p=m\VEC v$.

Secondly, we observe that the classical, so-called Hamiltonian flow
(i.e.\ of incompressible fluids) given by
\[
\nabla\cdot\VEC v=0
\]
is only obtained if the r.h.s.\ of Eq.~\eqref{eq:diss.2.6} vanishes,
too, i.e.
\begin{equation}
\frac{\nabla P}{P}\cdot\VEC v=0.
\end{equation}
Thus, unless trivially $\nabla P=0$, the Hamiltonian flow can also
be characterized by two orthogonal vectors, the vector $\VEC v=\nabla S/m$
as of Eq.~\eqref{eq:v=00003Dp_div_m}, $\VEC u=\nabla(\delta S)/m$
as of Eq.~\eqref{eq:dp=00003DnabladS=00003Dmu} and the vector
\begin{equation}
\frac{\nabla P}{P}=:\mathrm{const}\cdot\VEC u=\mathrm{const}\cdot\frac{\nabla(\delta S)}{m},\label{eq:nabP_P=00003Dconst.u}
\end{equation}
which can also be set as proportional to a velocity $\VEC u$. In
fact, the totality of all vectors $\VEC u=\nabla(\delta S)/m$ orthogonal
to $\VEC v$ represents the velocity field of the spherical wave fronts
which can be considered to permanently emanate from the particle as
Huygens waves. Using as a result from Grössing~\cite{Groessing.2004hamilton}
that
\begin{equation}
\frac{\Delta Q}{kT}=\frac{2\omega\delta S}{\hbar\omega}\thinspace,\label{eq:diss.2.10}
\end{equation}
where we used relations~\eqref{eq:hw=00003Dkt} and \eqref{eq:diss.1.57}
which fulfil the requirement of equal kinetic energies\footnote{Let us here repeat the note in context with Eq.~\eqref{eq:aceq.3.3a}:
Although we are probably at a length scale where the thermodynamical
laws have not yet proven valid, we use Eq.~\eqref{eq:hw=00003Dkt}
as the sub-quantum analogon to the thermodynamical expression $k_{\mathrm{B}}T/2$.
However, as for today we neither know $T$ nor the constant $k$ \textendash{}
unless it should turn out as identical to $k_{\mathrm{B}}$.} as discussed in chapter~\ref{sec:2.fluid-droplet}, we obtain the
relation between the momentum variation $\delta\VEC p$~\eqref{eq:dp=00003DnabladS=00003Dmu}
and the probability distribution $P$ as
\begin{equation}
\delta\VEC p(\VEC x,t)=m\VEC u(\VEC x,t)=:\hbar\VEC k_{u}(\VEC x,t)=\nabla(\delta S(\VEC x,t))=-\frac{\hbar}{2}\frac{\nabla P(\VEC x,t)}{P(\VEC x,t)}\,,\label{eq:diss.2.7}
\end{equation}
where $\VEC k_{u}$ denotes the wave vector associated to the osmotic
velocity $\VEC u$.\footnote{The r.h.s.\ of Eq.~\eqref{eq:diss.2.7} is readily confirmed by
insertion of the r.h.s.\ of Eq.~\eqref{eq:diss.2.10} into \eqref{eq:diss.2.7}.} Combining with Eq.~\eqref{eq:dp=00003DnabladS=00003Dmu} and using
Eq.~\eqref{nablalnu} we find for the osmotic velocity 
\begin{equation}
\VEC u(\VEC x,t)=-\frac{\hbar}{2m}\frac{\nabla P(\VEC x,t)}{P(\VEC x,t)}=-\frac{\hbar}{2m}\nabla\ln P(\VEC x,t).\label{eq:gb.3.6}
\end{equation}

By setting Eqs.~\eqref{eq:diss.2.6} and \eqref{eq:nabP_P=00003Dconst.u}
we found an orthogonality condition for the velocities $\VEC v$ and
$\VEC u$ which, however, is valid for a classical Hamiltonian flow.
Considering additional fluctuations as discussed by the bouncer\textendash walker
model, we shall demand less stringent requirements, namely the vanishing
of Eq.~\eqref{eq:diss.2.6} \textit{on average},\footnote{The mean value $\meanx{(\cdot)}$ can be understood either as an average
over space or an average over an ensemble at a fixed position, $\meanx{a(x)}=\int_{-\infty}^{\infty}aP(x,t)\d x$.
For a stationary process the mean value is constant because of $P(x,t)=P(t)$
(see e.g.~\cite{Schwabl.2006en,Burgdorfer.2007statistical}).}
\begin{equation}
\meanx{\nabla\cdot\VEC v}=-\meanx{\frac{\nabla P}{P}\cdot\VEC v}=0,\label{eq:mean_ortho_relation}
\end{equation}
as shown in Fig.~\ref{fig:hamiltonian-flow}. The essential difference
is given by a vanishing divergence of the velocity of the probability
current, $\nabla\cdot\VEC v=0$, in the Hamiltonian flow, whereas
in the quantum flow the average over fluctuations and positions of
the average divergence be identical to zero (Eq.~\eqref{eq:mean_ortho_relation}).

The consequences on the averaging process as provided in $\meanx{\VEC u\VEC v}=0$~\eqref{eq:mean_ortho_relation}
are explicated by Grössing~\cite{Groessing.2004hamilton} and later
on by our group~\cite{Groessing.2010emergence} in much deeper detail.
In the latter, we also derived quantum mechanical dispersion as a
consequence of this averaging process.

\section{From classical phase-space distributions to quantum mechanical dispersion\label{sec:3.2.From-classical-phase-space}}

In accordance with the classical model, we shall now relate it more
directly to the walker-bouncer analogy gleaned from Couder and Fort~\cite{Couder.2012probabilities}.
For, as shown, e.g., in Holland~\cite{Holland.1993} or Elze~\cite{Elze.2011general},
one can construct various forms of classical analogies to quantum
mechanical Gaussian dispersion. The two mechanisms may refer to an
early idea of de~Broglie~\cite{DeBroglie.1960book} to model quantum
behaviour by a two-fold process, i.e.\ by the movement of a hypothetical
point-like singularity solution of the Schrödinger equation, and by
the evolution of the usual wave function that would provide the empirically
confirmed statistical predictions. Recently, Couder and Fort~\cite{Couder.2012probabilities}
have used this ansatz to describe the behaviour of their bouncer droplets:
On an individual level, one observes particles surrounded by circular
waves they emit through the phase-coupling with an oscillating bath,
which provides, on a statistical level, the emergent outcome in close
analogy to quantum mechanical behaviour like, e.g., diffraction or
double-slit interference.~\cite{Mesa.2013variable}

In the context of the double solution idea, which is related to correlations
on a statistical level between individual uncorrelated particle positions
$x$ and momenta $p$, respectively, we consider the free Liouville
equation for the probability distribution function $f(x,p,t)$ in
phase-space of a mechanical system
\begin{align}
\frac{\p f}{\p t}+\sum_{i=1}^{3}\frac{p_{i}}{m}\frac{\p f}{\p x_{i}}-\sum_{i=1}^{3}\frac{\p V}{\p x_{i}}\frac{\p f}{\p p_{i}} & =0\label{eq:Liouville}
\end{align}
with potential $V$ and mass $m$. Here, we return to the one-dimensional
description which is sufficient for further investigations. Liouville's
equation~\eqref{eq:Liouville} implies the continuity equation in
phase-space and has the property that precise knowledge of initial
conditions is not lost in the course of time. That is, it provides
a phase-space distribution $f\left(x,p,t\right)$ that shows the emergence
of correlations between $x$ and $p$ from an initially uncorrelated
product function of non-spreading (classical) Gaussian position distributions
as well as momentum distributions,
\begin{equation}
f_{0}\left(x,p\right)=\frac{1}{2\pi\sigma_{0}\pi_{0}}\;\e^{\nicefrac{-x^{2}}{2\sigma_{0}^{2}}}\;\e^{\nicefrac{-p^{2}}{2\pi_{0}^{2}}}\,,\label{eq:td.2}
\end{equation}
where $\sigma_{0}=\sigma(t=0)$, and $\pi_{0}:=mu_{0}$ are the half-widths
in space and momentum, respectively. The general solution of the free
Liouville equation~\eqref{eq:Liouville} for the case where the particles
in the ensemble all have an initial velocity $p/m$ at vanishing potential,
$V=0$, is
\begin{equation}
f\left(x,p,t\right)=f_{0}(x-pt/m,p),\label{eq:diss.2.39}
\end{equation}
inserting Eq.~\eqref{eq:td.2} reads
\begin{equation}
f\left(x,p,t\right)=\frac{1}{2\pi\sigma_{0}mu_{0}}\;\e^{\nicefrac{-\left(x-pt/m\right)^{2}}{2\sigma_{0}^{2}}}\;\e^{\nicefrac{-p^{2}}{2m^{2}u_{0}^{2}}}\,.\label{eq:td.3}
\end{equation}
The probability density in $x$-space is given by
\begin{equation}
P\left(x,t\right):=\int f\d p=\frac{1}{\sqrt{2\pi}\,\sigma(t)}\;\e^{\nicefrac{-x^{2}}{2\sigma^{2}(t)}}\,.\label{eq:p=00003Dgauss}
\end{equation}
whereby the integration has been carried out by completing the square
of $p$ in Eq.~\eqref{eq:td.3}. As a result, we find the variance
at time $t$ given by
\begin{equation}
\sigma^{2}(t)=\sigma_{0}^{2}+u_{0}^{2}\,t^{2}.\label{eq:td.5}
\end{equation}
By superposition of the constant-width Gaussians with a moving centre
we obtain the spreading Gaussian distribution with variance~\eqref{eq:td.5}
which obviously reflects the fact that faster particles move further
in a given time interval.\footnote{We shall use the fact that the Gaussian shape remains a Gaussian in
chapter~\ref{sec:3.4.Spreading-of-the-wave-packet} by replacing
$x\to x-v_{x}t$, $v_{x}=\mathrm{const.}$\label{fn:x->x-vxt}}

The stochastic process described by the Langevin equation~\eqref{eq:Langevin}
involves momentum fluctuations $\delta p=mu$, now described by the
momentum distribution in Eq.~\eqref{eq:td.2}. Therefore, in $u_{0}$
as defined in Eq.~\eqref{eq:diss.u0} we have a connection to our
walker model. This means that $u_{0}$ is related to the sub-quantum
medium and hence to the particle's mass $m$ revealed by the definition
of the walker in Eq.~\eqref{eq:Langevin}. The half-width $\sigma_{0}$
is in turn tightly related to the slit-width as will be discussed
later in chapter~\ref{sec:Calibrating-the-simulation}. Nonetheless,
the half-widths $\sigma_{0}$ and $\pi_{0}$ of the distribution~\eqref{eq:td.2}
are uncorrelated. On the other hand, according to the usual picture
for dispersion~\eqref{eq:td.5} there actually is an initial spread
of velocities $u_{0}=\pi_{0}/m$. According to the minimal uncertainty
principle\footnote{See also Bohm and Hiley~\cite[p.~46]{Bohm.1993undivided} who point
out the fact that eventually the width of the packet corresponds to
the spread of distances covered by the particles which is in turn
determined by the spread of velocities which is equal to $\Delta v$;
velocity $v$ being well-defined in the causal interpretation.} the scale of the fluctuations of $\sigma_{0}$ and $\pi_{0}$ is
given by $\hbar$ via
\begin{equation}
\Delta x\Delta p=\sigma_{0}\pi_{0}=\frac{\hbar}{2}\thinspace.\label{eq:uncertainty}
\end{equation}
Using $\pi_{0}=mu_{0}$, the diffusion constant~\eqref{eq:aceq.4.12},
$D=\hbar/2m$, and Eq.~\eqref{eq:uncertainty} yields
\begin{equation}
u_{0}=\frac{D}{\sigma_{0}}\thinspace.\label{eq:u0=00003DD/sigma0}
\end{equation}
This leads, by substituting Eq.~\eqref{eq:u0=00003DD/sigma0} into
\eqref{eq:td.5}, to
\begin{equation}
\sigma^{2}(t)=\sigma_{0}^{2}\left(1+\frac{D^{2}t^{2}}{\sigma_{0}^{4}}\right)\label{eq:dispersion}
\end{equation}
which explicitly contains $\sigma_{0}$ as an expression for the given
slit which determines $\sigma_{0}$. The properties of the particle
are yet given by the constant $D=\hbar/2m$.

\section{Derivation of the time-dependent diffusion equation\label{sec:3.3.The-derivation-of-Dt}}

In section~\ref{sec:3.2.From-classical-phase-space}, the probability
density $P(x,t)$ is modelled Gaussian shaped. For this class of functions
we can now investigate a generalized diffusion equation with a time-dependent
diffusion coefficient (cf.~\cite{Mesa.2012classical,Mesa.2013variable}).
Therefore, we make an ansatz for a more general relationship of diffusion
equations, 
\begin{equation}
\frac{\partial P}{\partial t}=kt^{\alpha}\frac{\partial^{2}P}{\partial x^{2}}\thinspace,\label{eq:tdde.1}
\end{equation}
with factor alpha, $0\le\alpha\le2$, determining the type of diffusion,
e.g., $\alpha=0$ reduces~\eqref{eq:tdde.1} to the usual heat equation
(cf.~\cite{Bologna.2010density}). Factors $t$ and $k$ denote the
time and a constant factor, respectively. We ask for possible values
of $k$ and $\alpha$.

Inserting $P(x,t)$ of Eq.~\eqref{eq:p=00003Dgauss} as a known solution
into Eq.~\eqref{eq:tdde.1} yields 
\begin{align}
\frac{P\dot{\sigma}}{\sigma}\left(\frac{x^{2}}{\sigma^{2}}-1\right) & =kt^{\alpha}\frac{P}{\sigma^{2}}\left(\frac{x^{2}}{\sigma^{2}}-1\right),\label{eq:diss.2.56}
\end{align}
and by integrating the simplified equation~\eqref{eq:diss.2.56},
$\dot{\sigma}\sigma=kt^{\alpha}$, we find 
\begin{align}
\sigma^{2} & =2k\frac{t^{\alpha+1}}{\alpha+1}+c_{0}.\label{eq:tdde.4}
\end{align}
A comparison of Eq.~\eqref{eq:dispersion} and \eqref{eq:tdde.4}
yields $c_{0}=\sigma_{0}^{2}$, $\alpha=1$, and 
\begin{align}
k & =\frac{D^{2}}{\sigma_{0}^{2}}\thinspace.
\end{align}
Finally, inserting this result into Eq.~\eqref{eq:tdde.1} leads
to
\begin{align}
\frac{\partial P}{\partial t} & =\underbrace{\frac{D^{2}t}{\sigma_{0}^{2}}}_{D_{\mathrm{t}}}\,\frac{\partial^{2}P}{\partial x^{2}}\thinspace,\label{eq:ballisticDE}
\end{align}
where one immediately recognizes the time-dependent diffusion coefficient
\begin{equation}
D_{\mathrm{t}}(t)=\frac{D^{2}}{\sigma_{0}^{2}}\,t=u_{0}^{2}\,t=\frac{\hbar^{2}}{4m^{2}\sigma_{0}^{2}}\,t,\label{eq:td.22}
\end{equation}
which, because of its linearity of time $t$, gives Eq.~\eqref{eq:ballisticDE}
the name \textit{ballistic diffusion equation.} This condition is
only fulfilled by $\alpha=1$, which is the only possible diffusion
equation whose solution has the form~\eqref{eq:p=00003Dgauss}.

If the diffusion depends on space, one has to deal with a diffusion
coefficient $D_{\mathrm{t}}(x,t)$, and thus 
\begin{equation}
\frac{\partial P}{\partial t}=\frac{\p}{\p x}\left(D_{\mathrm{t}}(x,t)\frac{\partial P}{\partial x}\right).\label{eq:td.35}
\end{equation}
However, this is not in the scope of this thesis, though the handling
of space-dependent diffusion equations can be found in, e.g., the
textbook of John C.\ Strikwerda~\cite{Strikwerda.2004finite}.

\section{Spreading of the wave packet\label{sec:3.4.Spreading-of-the-wave-packet}}

\begin{figure}[h]
\centering{}\includegraphics{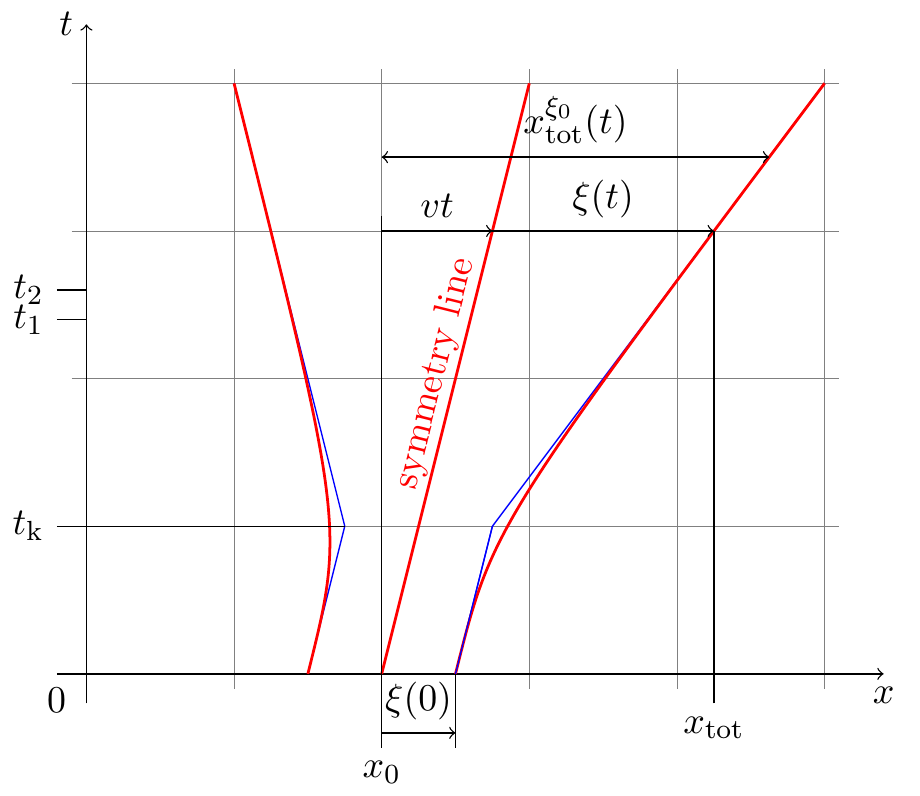}\caption{Bohm-type trajectories for a quantum particle with initial Gaussian
distribution exhibiting the characteristics of ballistic diffusion\label{fig:td.1}}
\end{figure}

Now we generalize the discussion of chapter~\ref{sec:3.2.From-classical-phase-space}
as mentioned in the footnote~\vpageref{fn:x->x-vxt} and add the
displacement\footnote{The particle moves with velocity $v_{y}=\mathrm{const.}$ which is
not relevant to  this one-dimensional examination. The optional additive,
constant motion along the $x$-axis is depicted by $v$ for short.
Accordingly, $v=v_{x}=\mathrm{const.}$, otherwise noted. } $x-vt$ to the Gaussian distributions of Eqs.~\eqref{eq:td.3} and
\eqref{eq:p=00003Dgauss}. The easiest way to follow the decay in
the evolution of time is to observe a point with distance $\xi(t)$
from the centre of the Gaussian shape (see Fig.~\ref{fig:td.1})
defined by
\begin{equation}
\xi(t)=\xi(0)\frac{\sigma(t)}{\sigma_{0}}\label{eq:td.11}
\end{equation}
with
\begin{equation}
\frac{\sigma(t)}{\sigma_{0}}=\sqrt{1+\frac{D^{2}t^{2}}{\sigma_{0}^{4}}}\label{eq:dispersion2}
\end{equation}
being the dispersion~\eqref{eq:dispersion} of the wave packet. Due
to definition~\eqref{eq:td.11} the probability
\begin{equation}
\intop_{vt}^{vt+\xi(t)}P(x,t)\d x\label{eq:int_xi_p=00003Dconst}
\end{equation}
is time-independent.

\noindent In Fig.~\ref{fig:td.1} the spreading according to Eq.~\eqref{eq:td.11}
is sketched. 

The functional relationship~\eqref{eq:int_xi_p=00003Dconst} is clearly
valid for the particular point $\xi(0)=\sigma_{0}$ which, substituted
into~\eqref{eq:td.11}, leads immediately to $\xi(t)=\sigma(t)$,
and hence the evidence that this particular point follows the variance
of the decaying Gaussian. However, the relation $\xi(t)\propto\sigma(t)$
is, for all starting points $\xi(0)$, always true as the Gaussian
remains a Gaussian but broadens during decay for all $t>0$, which
is reflected in Eq.~\eqref{eq:int_xi_p=00003Dconst}.

As the packet spreads according to Eq.~\eqref{eq:dispersion2}, $\xi(t)$
describes the result of the average motion along a trajectory of a
point of this packet that was initially at $\xi(0)$. Depending on
the initial value of $\left|\xi(0)\right|$, i.e.\ the distance from
$x_{0}$ of the initial centre point of the packet, said spreading
happens faster or slower. In our model picture, this is easy to understand:
For a trajectory exactly at the centre of the packet, $x_{\mathrm{tot}}(t)=x_{0}+vt\Leftrightarrow\xi(0)=0$,
the momentum contributions from the heated up environment on average
cancel each other for symmetry reasons. However, the further off a
trajectory is from that centre, the stronger this symmetry will be
broken, i.e.\ leading to a position-dependent net acceleration or
deceleration, respectively, or, in effect, to the decay of the wave
packet. The actual decay of the wave packet starts, roughly spoken,
at a time $t_{\mathrm{k}}$, indicated by a kink in Fig.~\ref{fig:td.1}
which is due to the squared time-behaviour in Eq.~\eqref{eq:dispersion2}.
By dividing the trajectories at $t_{\mathrm{k}}$ into two time domains,
one can see its behaviour for $t\ll t_{\mathrm{k}}$, where $\xi(t)\propto\xi(0)=\mathrm{const.}$,
and $t\gg t_{\mathrm{k}}$, where $\xi(t)\propto t$ \textendash{}
and hence ballistic: The propagations described by $\xi(t)$ are linear
in both domains just kicked off to either side from the symmetry line
(see Fig.~\ref{fig:td.1}).

From Fig.~\ref{fig:td.1} we find $x_{\mathrm{tot}}(t)=x_{0}+vt+\xi(t)$.
Without loss of generality we set $x_{0}=0$ further on. With the
use of Eq.~\eqref{eq:td.11} we obtain 
\begin{equation}
x_{\mathrm{tot}}^{\xi_{0}}(t)=vt+\xi(t)=vt+\xi(0)\frac{\sigma(t)}{\sigma_{0}}=vt+\xi(0)\sqrt{1+\frac{u_{0}^{2}t^{2}}{\sigma_{0}^{2}}}\;.\label{eq:td.14}
\end{equation}
In our model picture, $x_{\mathrm{tot}}^{\xi_{0}}$ maps time $t$
to the position of the \textit{smoothed out trajectories}, i.e.\ those
averaged over a very large number of Brownian motions. 

Moreover, one can now also calculate the \textit{average total velocity
field of a Gaussian wave packet} as 
\begin{equation}
v_{\mathrm{tot}}^{\xi_{0}}(t)=\frac{\d x_{\mathrm{tot}}^{\xi_{0}}(t)}{\d t}=v+\xi(0)\,\frac{u_{0}^{2}t/\sigma_{0}^{2}}{\sqrt{1+u_{0}^{2}t^{2}/\sigma_{0}^{2}}}\;,\label{eq:td.15}
\end{equation}
which describes the velocity $v_{\mathrm{tot}}^{\xi_{0}}$ of a point
along a trajectory at time $t$.

Finally, we derive the \textit{average total acceleration field of
a Gaussian wave packet} is 
\begin{equation}
a_{\mathrm{tot}}^{\xi_{0}}(t)=\frac{\d v_{\mathrm{tot}}^{\xi_{0}}(t)}{\d t}=\xi(0)\,\frac{u_{0}^{2}/\sigma_{0}^{2}}{\sqrt{\left(1+u_{0}^{2}t^{2}/\sigma_{0}^{2}\right)^{3}}}\;,\label{eq:td.16}
\end{equation}
describing the acceleration of a point along the trajectory at time
$t$. Eqs.~\eqref{eq:td.14} to \eqref{eq:td.16} allow us to calculate
the quantities along a trajectory only out of a given starting point,
indicated by $\xi(0)$.

Actually we are interested in the dynamics at any given position $(x,t)$
directly. Using 
\begin{equation}
\xi(t)=x-vt\label{eq:td.17}
\end{equation}
 and Eq.~\eqref{eq:td.11} we rewrite
\begin{equation}
\xi(0)=\frac{x-vt}{\sqrt{1+u_{0}^{2}t^{2}/\sigma_{0}^{2}}}
\end{equation}
which leads to the generalized fields,
\begin{align}
x_{\mathrm{tot}}(x,t) & =x,\vphantom{\intop_{0}^{0}}\label{eq:td.19}\\
v_{\mathrm{tot}}(x,t) & =v+\xi(t)\,\frac{u_{0}^{2}t/\sigma_{0}^{2}}{1+u_{0}^{2}t^{2}/\sigma_{0}^{2}}=v+(x-vt)\,\frac{u_{0}^{2}t}{\sigma^{2}(t)}\,,\vphantom{\intop_{0}^{0}}\label{eq:td.20}\\
a_{\mathrm{tot}}(x,t) & =\xi(t)\,\frac{u_{0}^{2}/\sigma_{0}^{2}}{\left(1+u_{0}^{2}t^{2}/\sigma_{0}^{2}\right)^{2}}=(x-vt)\,\frac{u_{0}^{2}\sigma_{0}^{2}}{\sigma^{4}(t)}\;,\vphantom{\intop_{0}^{0}}\label{eq:td.21}
\end{align}
which will be used in the simulations.
\begin{sidewaysfigure}
\centering{}\subfloat[Initial half width $\sigma_{0}$\label{fig:1a}]{\includegraphics[width=0.48\textwidth]{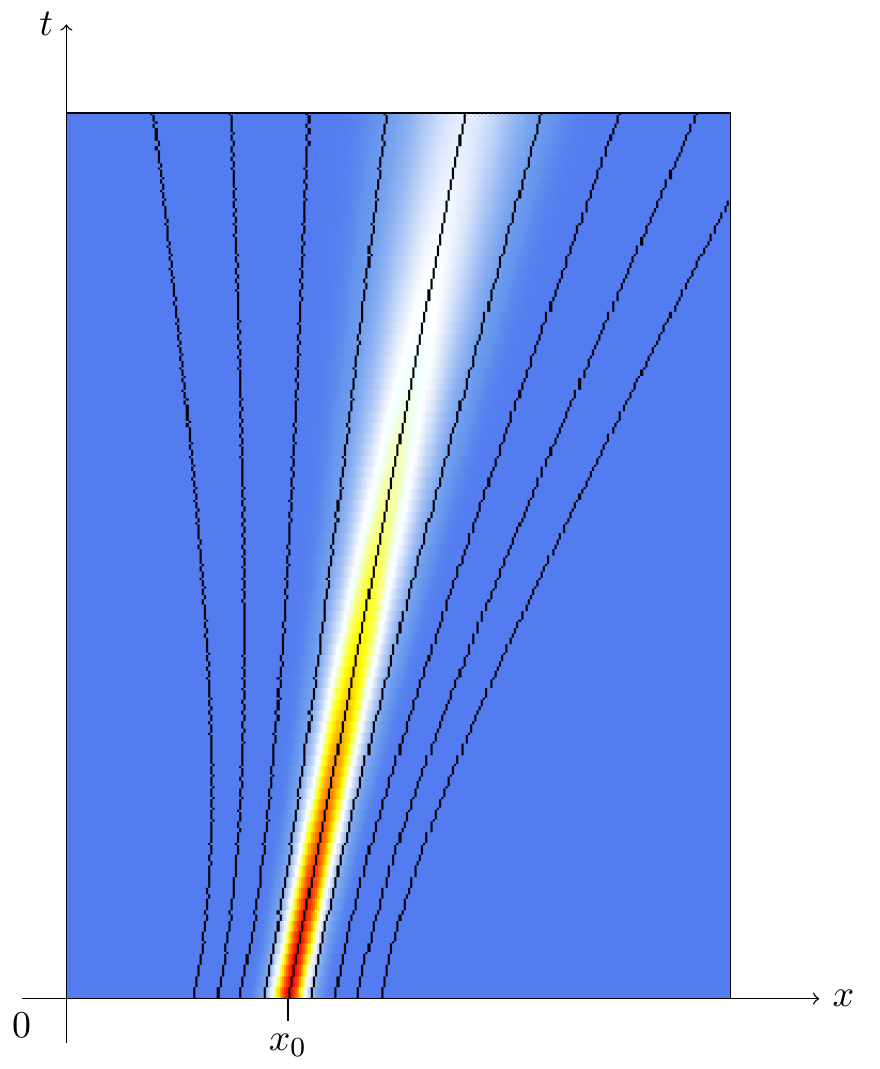} }\hfill{}\subfloat[Initial half width $2\sigma_{0}$\label{fig:1b}]{ \includegraphics[width=0.48\textwidth]{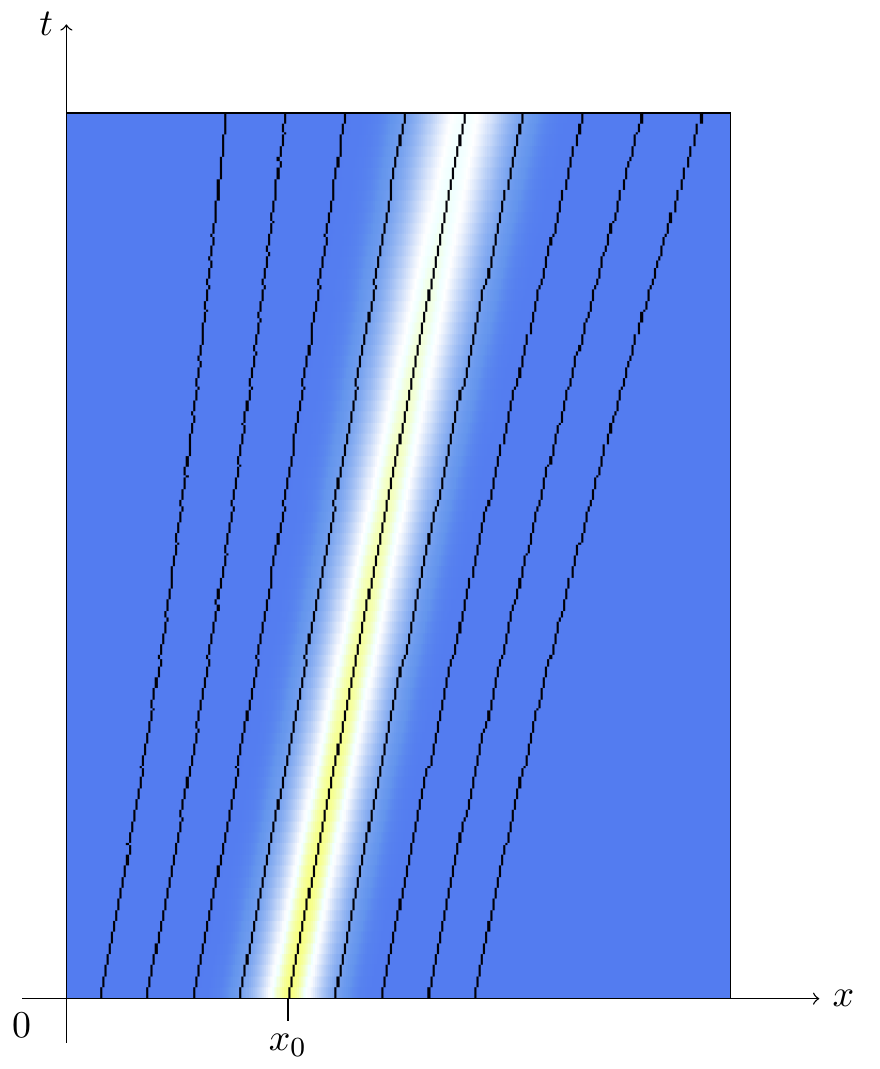} }\caption{Dispersion of a free Gaussian wave packet with trajectories representing
the averaged behaviour of a statistical ensemble for two different
initial spreadings\label{fig:1}}
\end{sidewaysfigure}

Eqs.~\eqref{eq:td.14} to \eqref{eq:td.21} provide the trajectory
distributions and the velocity field of a Gaussian wave packet as
derived solely from classical physics. The trajectories here only
represent the averaged behaviour of a statistical ensemble, i.e.\ averaged
over many single trajectories of ballistic diffusion assuming Eq.~\eqref{eq:uncertainty},
i.e.\ a relation between the initial spatial and momentum distributions.
The results are in full accordance with quantum theory, and in particular
with Bohmian trajectories (see, for example, Holland~\cite{Holland.1993}
or Sanz~\cite{Sanz.2008trajectory}, or the figures for the Gaussian
wave packet example of von~Bloh~\cite{vonBloh.2010}, which are
in excellent agreement with our Fig.~\ref{fig:1}). This is so despite
the fact that neither a quantum mechanical wave function, nor the
Schrödinger equation, nor a guiding wave equation, nor a quantum potential
has been used yet.

Fig.~\ref{fig:1} provides a graphic representation of Eq.~\eqref{eq:td.14}
for an exemplary set of trajectories. Considering the particles of
a source as oscillating bouncers, they can be shown to heat up their
\textendash{} generally nonlocal \textendash{} environment in such
a way that the particles leaving the source are guided through the
thus created thermal landscape. In the Fig.~\ref{fig:1}, the classically
simulated evolution of exemplary \textit{averaged} trajectories is
shown.

The figures show results of simulations with coupled map lattices
(cf.\ section~\ref{subsec:Coupled-map-lattices}) of classical diffusion
and a time-dependent diffusivity as given by Eq.~\eqref{eq:td.22}.
Two examples are shown, with different half-widths of the initial
Gaussian distribution, respectively: space-time diagrams, providing
the intensity field with time development from bottom to top and averaged
trajectories in agreement with Eq.~\eqref{eq:td.14}. In Fig.~\ref{fig:1a},
the initial $\sigma_{0}$ is half the value in Fig.~\ref{fig:1b}.
Note that the narrower the Gaussian distribution is concentrated initially
around the central position, the more the thus stored heat energy
tends to push trajectories apart.

\section{Conclusions and perspectives\label{sec:conclusions-gaussborn}}

As a follow-up of chapter~\ref{sec:2.fluid-droplet}, the constituting
single-slit setup has been introduced in this chapter. A distant oven
has been supposed to be the particle source for the later experiment.
Before a particle ever drops out of the oven and would be taken into
account, a continuously emitted energy wave, borne by the sub-quantum
medium, has been assumed to be produced by the oven. The wave itself,
when reaching the setup, has been approximated by a plane wave being
cut-out and sliced when passing the slit. Immediately after the slit,
the remaining, cut wave has been assumed taking shape of a Gaussian.

Two velocities, the osmotic velocity $\VEC u$ and the diffusive velocity
$\VEC v$, have been assumed to be orthogonal on average. This kind
of orthogonality \textendash{} not valid for a single event but for
a vast number of events \textendash{} has also been stated to be the
main difference to the Bohmian philosophy. As on average our results
converge to the Bohmian ones, the characteristic of our ansatz may
be called a phenomenological one.

From classical phase-space distribution comprising non-spreading Gaussian
position and momentum distributions, the quantum mechanical dispersion
has been derived. This then has led to the time-dependent diffusion
equation, or more precisely, the ballistic diffusion equation. With
these tools available, the spreading of a wave packet could be established,
founded on the ballistic diffusion equation only, which in turn allowed
for quickly performed simulations of said spreading fields.

Yet no phase relations have been required because the setup has comprised
of a single slit only. Consequently, the next step shall be expanding
the setup by at least one further slit and studying the then importantly
needed phase relations.

\chapter{Current-based theory on interference effects\label{sec:Interference-effects}}
\begin{quote}
In this chapter we investigate the phase relations due to adding one
further slit and eventually an arbitrary number of slits. By considering
both the classical and the emergent distributions' relations as well
as the orthogonality relations between the convective and osmotic
currents discussed in chapter~\ref{sec:3.probability-distributions},
we shall derive a set of current-based rules providing calculation
recipes for both, the total intensity $P_{\mathrm{tot}}$ and the
total current $\VEC J_{\mathrm{tot}}$ in a systematic way.

As an application of these current-based rules, we shall provide simulation
results of double-slit setups and discuss the sub-quantum behaviour
according to our phenomenological approach.

In a final step, we shall extend the current-based rules to multi-slit
scenarios and discuss the Talbot effect by means of simulations based
on these rules.\global\long\def\Rw#1{R_{\VEC w_{#1}}}
\global\long\def\RRw#1{R(\VEC w_{#1})}
\global\long\def\Pw#1{P_{\VEC w_{#1}}}
\global\long\def\PPw#1{P(\VEC w_{#1})}
\global\long\def\Pcw#1#2{P(\VEC w_{#1}|\VEC w_{#2})}
\global\long\def\Jw#1{\VEC J_{\VEC w_{#1}}}
\global\long\def\JJw#1{\VEC J(\VEC w_{#1})}
\end{quote}

\section{Interference and emergence at a Gaussian double-slit\label{sec:path-1}}

\begin{figure}[!htb]
\centering{}\includegraphics[width=0.98\textwidth]{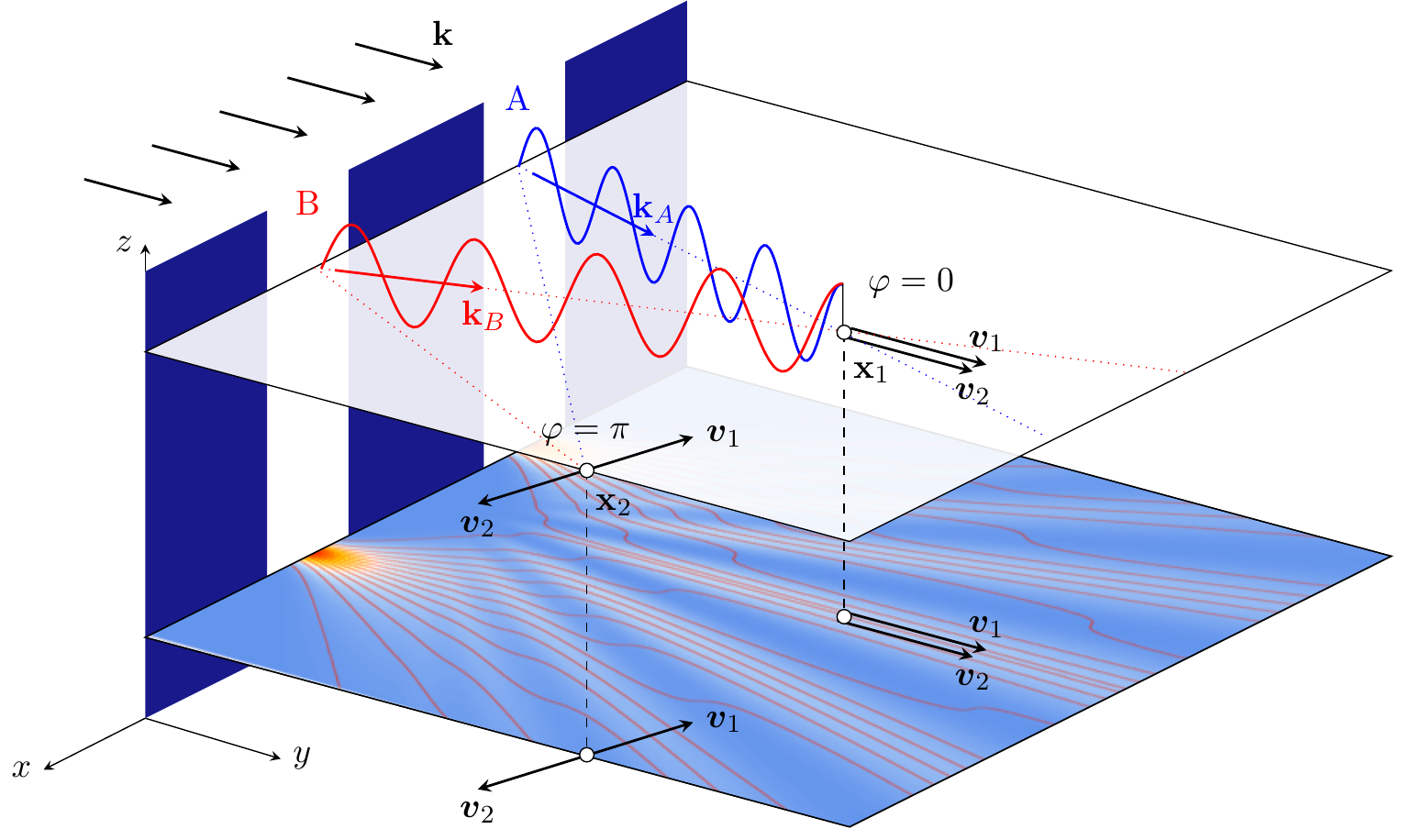}\caption{Geometry of interference at a double-slit at exemplary points $\protect\VEC x_{1}$
and $\protect\VEC x_{2}$\label{fig:interf}}
\end{figure}

In Fig.~\ref{fig:interf} 
the underlying geometry for the wave vectors in a double-slit setup
is sketched, both for the classical interference and the emergent\footnote{Emergence is a process whereby larger entities arise through interactions
among smaller or simpler entities such that the larger entities exhibit
properties the smaller/simpler entities do not exhibit (cf.\ \cite{Wikipedia.2016emergence}).
The interference pattern in Fig.~\ref{fig:interf} is considered
to be emergent in this sense.} case (cf.~\cite{Fussy.2014multislit,Groessing.2016emqm}). For illustration,
we show the three-dimensional setup with two exemplary planes emphasised.
The upper one contains a sketch of the classical picture according
to wave optics, the lower one contains a simulated resulting image
and trajectories to illustrate the emergent picture. The incident
wave\footnote{To get a picture of what it is that is oscillating, we stress the
walker-bouncer picture again and consider the wave to comprise the
oscillating sub-quantum medium having the properties known from wave
optics.} is indicated by parallel wave vectors $\VEC k$ of a plane wave in
the $xz$-plane propagating in $y$-direction as is used in our simplified
model to keep things clearly arranged. All vectors are assumed to
be located in the $xy$-plain, i.e.\ they are independent of $z$,
whereby $y\propto t$ as defined by Eq.~\eqref{eq:y=00003Dv_yt}.

Let us start with the upper plain. In the classical picture the incoming
wave vector 
\begin{equation}
\VEC k=\frac{\mathrm{2\pi}}{\lambda}\VEC{\hat{k}},
\end{equation}
with $\VEC{\hat{k}}=\VEC k/|\VEC k|$ being the unit vector and $\lambda$
the wavelength, splits up at the Gaussian slits\footnote{The distribution after the slit is assumed to be an ideal Gaussian
not being refracted at the slit's edges as explained in chapter~\ref{sec:The-constituting-setup}.} $\mathit{A}$ and $\mathit{B}$ into $\VEC k_{A}$ and $\VEC k_{B}$,
both are orthogonal to the particular propagating wave fronts. As
the slits $\mathit{A}$ and $\mathit{B}$ act like coherent sources
the resulting interference pattern is time-independent. The respective
phases for each of the beams are usually denoted as\footnote{We use this notation for short,with $A(B)$ meaning that either the
left character is to be used for the whole equation, or the character
inside the parentheses. However, they must not be mixed up, i.e.~$\varphi_{A(B)}=\VEC k_{A(B)}\cdot\VEC r_{A(B)}$
means $\varphi_{A}=\VEC k_{A}\cdot\VEC r_{A}$ and $\varphi_{B}=\VEC k_{B}\cdot\VEC r_{B}$,
but $\varphi_{A}\neq\VEC k_{B}\cdot\VEC r_{A}$ \textit{etc}. } 
\begin{equation}
\varphi_{A(B)}=\VEC k_{A(B)}\cdot\VEC r_{A(B)},\label{eq:diss.3.2}
\end{equation}
with $\VEC r_{A(B)}$ being a position vector from source $A(B)$
to point $\VEC x$, marked as dotted lines in Fig.~\ref{fig:interf}.

With Eq.~\eqref{eq:diss.3.2} together with plane wave amplitudes
at an arbitrary point $\VEC x$ of the spatio-temporal plane we aim
at describing relations known from Bohmian theory like Eqs.~\eqref{eq:Jfinal}
and \eqref{eq:vtot}. The amplitudes 
\begin{equation}
R_{A(B)}(\VEC x)=\sqrt{P_{A(B)}^{S}(\VEC x)},\label{eq:R=00003Dsqrt_P}
\end{equation}
with $P^{S}(\VEC x)$ being an intensity distribution function of
a single slit as defined in Eq.~\eqref{eq:p=00003Dgauss}, allow
for describing the beams coming from slits $A$ and $B$ as
\begin{equation}
\tilde{R}(\VEC x,t)=R(\VEC x)\,\mathrm{Re}\left\{ \e^{\i(\VEC k\cdot\VEC r-\omega t)}\right\} =R(\VEC x)\cos(\VEC k\cdot\VEC r-\omega t)
\end{equation}
wherein describing $R(\VEC x)$ thereby omitting the frequency $\omega$
is sufficient. Combining the beams of, say, two slits by simply adding
the two components leads to 
\begin{equation}
R(\VEC x)=R_{A}(\VEC x)\cos\varphi_{A}+R_{B}(\VEC x)\cos\varphi_{B}.\label{eq:11a-1}
\end{equation}
Even though Eq.~\eqref{eq:11a-1} is a usual method to describe the
distribution correctly, we want to introduce in this chapter the results
from the last chapters instead, namely the ballistic diffusion and
the associated velocities derived for the single slit system.

Therefore, we turn towards the lower plane, the ``emergent'' scenario.
We have to treat the two slits, or respective beam paths, as the sources
of a flow of probability densities which we want to express by the
involved wave vectors, or equivalently\footnote{For the relation between wave vectors and velocities is about equation
$\VEC p=m\VEC v=\hbar\VEC k$ and the quantities used therein. See
Eqs.~\eqref{eq:y=00003Dv_yt} and \eqref{eq:diss.2.7}.}, by the involved velocities. For this picture, we have in the foregoing
chapters already introduced the (emergent) convective velocity $\VEC v_{i}(x)$
and the (emergent) osmotic velocity $\VEC u_{i}(\VEC x)$, both of
which have its source originated in the slits $A$ and $B$. However,
the impacting velocities shall be denoted with numbers $1$ and $2$,
instead of the letters $A$ and $B,$ respectively, in order to distinguish
them from the classical picture. The osmotic velocities have to fulfil
the condition of being unbiased w.r.t.\ the convective velocities,
i.e.\ the orthogonality relation~\eqref{eq:mean_ortho_relation}
for the \textit{averaged} velocities, $\VEC{\overline{vu}}=0$, since
any fluctuations $\VEC u=\nabla(\delta S)/m$ are shifts along the
surfaces of action $\mathit{S=\mathrm{\mathrm{const}}.}$, as shown
in Fig.~\ref{fig:surface_S}.

\begin{figure}[!htb]
\centering{}\includegraphics{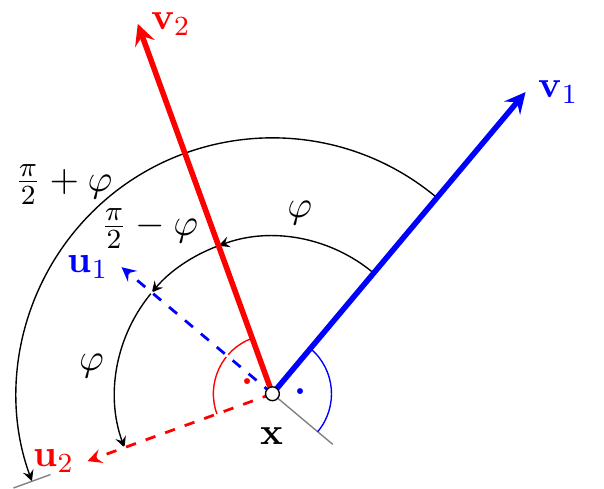}\caption{Geometry of emergent velocities and relative phases for a two-beam
setup.\label{fig:vektor}}
\end{figure}

Each point of the probability (or amplitude) landscape evolves on
the spatial plane according to the convective velocities $\VEC v_{i}(\VEC x)$,
$i=1,2$ (exemplarily shown at $\VEC x_{1}$ and $\VEC x_{2}$ in
Fig.~\ref{fig:interf}). In addition, the osmotic velocity $\VEC u(\VEC x)$
describes the dispersion~\ref{eq:dispersion} of the Gaussian and
split up into $\VEC u_{\mathrm{1}}(\VEC x)$ and $\VEC u_{\mathrm{2}}(\VEC x)$
dependent of the slit which causes the respective osmotic velocity
(Fig.~\ref{fig:vektor}). 
Since $\VEC u_{i}$ are orthogonal to $\VEC v_{i}$, $\sphericalangle(\VEC v_{i},\VEC u_{i})=\frac{\pi}{2}$,
all enclosed angles can be expressed in terms of $\varphi=\sphericalangle(\VEC v_{1},\VEC v_{2})$.
As can be seen in Fig.~\ref{fig:vektor} we get
\begin{equation}
\begin{aligned}\sphericalangle(\VEC v_{1},\VEC u_{2}) & =\frac{\pi}{2}+\varphi,\\
\sphericalangle(\VEC v_{2},\VEC u_{1}) & =\frac{\pi}{2}-\varphi.
\end{aligned}
\begin{gathered}\end{gathered}
\end{equation}

\section{A set of current rules\label{sec:set-of-current-rules}}

In the following\footnote{We will omit the variable $\VEC x$ in the argument of any vector,
amplitude, probability density, and probability current to improve
readability.} we shall show how the trajectories representing the paths of the
averaged velocities can be calculated with the help of a set of current
rules leading to the expressions for the total current $\VEC{J_{\mathrm{tot}}}$
and the total probability density $P_{\mathrm{tot}}$ at point $\VEC x$.

As we have to deal with two velocities caused by the same slit, we
introduce the term \textit{channel} here, i.e.\ we have two channels
per slit. To account for the different velocity channels $i=1,\ldots,2N$,
$N$ being the number of slits, we now introduce for general cases
generalized velocity vectors $\VEC w_{i}$, with 
\begin{equation}
\begin{array}{cc}
\VEC w_{1}:=\VEC v_{\mathrm{1}},\quad & \VEC w_{2}:=\VEC u_{\mathrm{1}},\\
\VEC w_{3}:=\VEC v_{\mathrm{2}},\quad & \VEC w_{4}:=\VEC u_{\mathrm{2}},
\end{array}\begin{gathered}\end{gathered}
\label{eq:w_i}
\end{equation}
for the first (upper line) and second (lower line) channel in the
case of $N=2$. This renumbering procedure will turn out as an important
practical bookkeeping tool.

For the weighting procedure to be introduced next, each amplitude
$R_{i}$ according to Eq.~\eqref{eq:R=00003Dsqrt_P} is assumed to
have its corresponding $P_{i}^{S}$ of the interference-free single-slit,
as if none of the probability distributions has interfered with any
other hitherto. For the bookkeeping we apply the same nomenclature
as before, i.e.
\begin{equation}
\begin{array}{cc}
\Rw 1=\Rw 2=R_{1},\\
\Rw 3=\Rw 4=R_{2},
\end{array}\begin{gathered}\end{gathered}
\label{eq:R_wi}
\end{equation}
again, for the case of $N=2$. It should be noted that any $\Rw i$
is the amplitude of the sub-quantum medium at point $\VEC x$ moving
with velocity $\VEC w_{i}$.

Now, we apply the usual definition of a probability current, which
reads
\begin{equation}
\Jw i=\VEC w_{i}\Pw i,\quad i=1,\ldots,4,\label{eq:J=00003DwP}
\end{equation}
wherein the index runs from $1,\ldots,2N$ with $N$ being the numbers
of slits. Here the number of slits is $N=2$. The general velocity
vectors $\VEC w_{i}$ are defined in Eq.~\eqref{eq:w_i}, such that
a probability current $\Jw i$ at point $\VEC x$ is caused by the
sub-quantum medium moving with velocity $\VEC w_{i}$ at that point.
The total probability current is the sum over all partial currents~\eqref{eq:J=00003DwP}
which reads
\begin{equation}
\VEC J_{\mathrm{tot}}=\sum_{i=1}^{4}\Jw i={\displaystyle \sum_{i=1}^{4}}\VEC w_{i}\Pw i.\label{eq:Jtot4}
\end{equation}
The local intensity of a partial current is dependent on all other
currents, thus the total current composes of all partial components.
This mutual dependence of a current's totality and its parts constitutes
the essential part that leads to a convenient set of current rules.\ \cite{Walleczek.2000self-organized,Groessing.2014relational,Walleczek.2016is,Jizba.2012emergence,Jizba.2012quantum}
Notable, this concept uses the peculiarity of using currents as \textit{basic}
ingredient and not as derivation of some elementary entity like, e.g.,
an elementary particle.

However, in Eqs.~\eqref{eq:J=00003DwP} and \eqref{eq:Jtot4} we
shall define $\Pw i$ different from the previously used $P_{i}^{S}$
in that we want to incorporate interference processes between the
channels. To account for that, we assume the probability density to
be caused by $\VEC w_{i}$ under the influence of $\VEC w_{j}$. 
\begin{figure}[!tbh]
\begin{centering}
\includegraphics{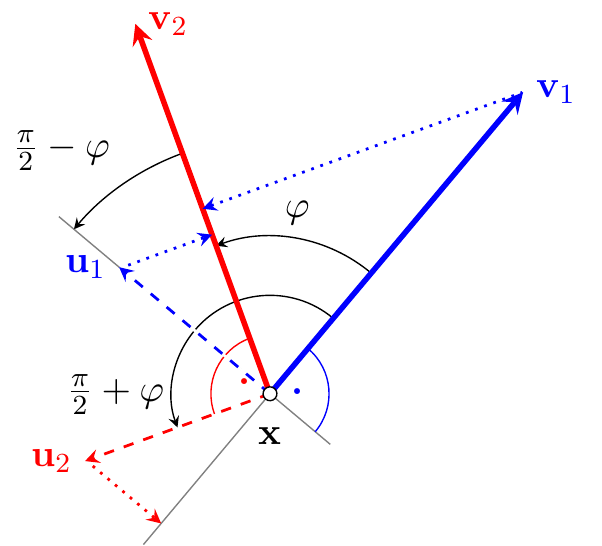}
\par\end{centering}
\caption{Scheme for the construction of the the projections.\label{fig:Scheme}}
\end{figure}
We stick herein to the theory proposed by Fussy~\cite{Fussy.2014multislit}
but adapt his procedure\footnote{In Fussy~\cite{Fussy.2014multislit} the procedure works by splitting
up the velocities $\VEC u_{i}$ in two parts, right and left, $\VEC u_{i\mathrm{R}}$
and $\VEC u_{iL}$, respectively, and hence associated unit vectors
$\VEC{\hat{u}}_{i\mathrm{R}}$ and $\VEC{\hat{u}}_{iL}$ that cancel
each other during the projection. This is equivalent to the procedure
shown herein, however, the argumentation is different.} to a rather straightforward scheme that works as follows: The influencing,
convective currents $\VEC v_{1}$ and $\VEC v_{2}$ determine the
causing currents $\VEC w_{i}$ in such a way that only their projection
\begin{equation}
\cos\varphi_{i,j}:=\VEC{\hat{w}}_{i}\cdot\VEC{\hat{w}}_{j}\label{eq:w_i.w_j}
\end{equation}
takes effect. The principle of the projection scheme is  sketched
in Fig.~\ref{fig:Scheme}. Furthermore, the causing and influencing
amplitudes, $\Rw i$ and $\Rw j$, respectively, both contribute to
the resulting probability density. In this spirit, we define
\begin{equation}
\begin{gathered}\begin{aligned}\Pw i & =\Rw i\VEC{\hat{w}}_{i}\cdot\left(\VEC{\hat{v}}_{1}R_{1}+\VEC{\hat{v}}_{2}R_{2}\right)\end{aligned}
\end{gathered}
\label{eq:tot_prob_law}
\end{equation}
and the total intensity as
\begin{equation}
\begin{aligned}P_{\mathrm{tot}} & ={\displaystyle \sum_{i=1}^{4}}\Pw i={\displaystyle \sum_{i=1}^{4}}\VEC{\hat{w}}_{i}\Rw i\cdot\left(\VEC{\hat{v}}_{1}R_{1}+\VEC{\hat{v}}_{2}R_{2}\right)\\
\vphantom{\intop^{0}} & =\left(\VEC{\hat{v}}_{1}R_{1}+\VEC{\hat{v}}_{2}R_{2}\right)^{2}=P_{\VEC v_{1}}+P_{\VEC v_{2}}
\end{aligned}
\begin{gathered}\end{gathered}
\label{eq:Ptot4}
\end{equation}
and obtain
\begin{equation}
P_{\mathrm{tot}}=R_{1}^{2}+R_{2}^{2}+2R_{1}R_{2}\cos\varphi.\label{eq:Ptot2slit}
\end{equation}
From $\VEC J=\VEC wP$ we get the \textit{emergent total velocity}
\begin{equation}
\VEC v_{\mathrm{tot}}=\frac{\VEC J_{\mathrm{tot}}}{P_{\mathrm{tot}}}=\frac{{\displaystyle \sum_{i=1}^{4}}\VEC w_{i}\Pw i}{{\displaystyle \sum_{i=1}^{4}}\Pw i}\,.\label{eq:vtot_fin}
\end{equation}

Thus we obtain amplitude contributions of the total system's wave
field projected on each channel's amplitude at point $\VEC x$ via
$\Pw i$. Then, the usual symmetry, even in the classical interference
case above, between $\Pw i$ and $\Rw i$ is broken: 
\begin{equation}
\Pw i\neq\Rw i^{2},
\end{equation}
i.e.\ although each velocity component $\VEC w_{i}$ has an associated
amplitude $\Rw i$, the \textit{partial} term $\Pw i$ is not the
mere squared amplitude any more. That is why $\Pw i$ should rather
be referred to as \textit{relational intensity} since the intensities
$\Pw i$ of Eq.~\eqref{eq:tot_prob_law} may assume negative values
which works well for contributions to the overall probability density
$P_{\mathrm{tot}}$~\eqref{eq:Ptot4} but lacks an interpretation
as a probability itself.

Returning now to our previous notation of the four velocity components,
$\VEC v_{i}$ and $\VEC u_{i}$, $i=1,2$, the partial current associated
with $\VEC v_{\mathrm{1}}$ is generated by constructing the scalar
product of $\VEC{\hat{v}}_{1}$ with all other unit vector components
which reads (see Fig.\ref{fig:Scheme})
\begin{equation}
\VEC J_{\VEC v_{\mathrm{1}}}=\VEC v_{\mathrm{1}}P_{\VEC v_{\mathrm{1}}}=\VEC v_{\mathrm{1}}R_{1}\VEC{\hat{v}}_{1}\cdot(\VEC{\hat{v}}_{1}R_{1}+\VEC{\hat{v}}_{2}R_{2})=\VEC v_{\mathrm{1}}\left(R_{1}^{2}+R_{1}R_{2}\cos\varphi\right)\label{eq:Jv1}
\end{equation}
and analogously
\begin{equation}
\VEC J_{\VEC v_{\mathit{\mathrm{2}}}}=\VEC v_{\mathrm{2}}P_{\VEC v_{\mathrm{2}}}=\VEC v_{2}\left(R_{2}^{2}+R_{1}R_{2}\cos\varphi\right).\label{eq:Jv2}
\end{equation}
The same applied to currents $\VEC u_{i}$ leads to (see Fig.\ref{fig:Scheme})
\begin{equation}
\begin{aligned}\VEC J_{\VEC u_{1}} & =\VEC u_{1}P_{\VEC u_{1}}=\VEC u_{\mathrm{1}}R_{1}\VEC{\hat{u}}_{1}\cdot(\VEC{\hat{v}}_{1}R_{1}+\VEC{\hat{v}}_{2}R_{2})\\
\vphantom{\intop_{0}^{0}} & =\VEC u_{1}R_{1}R_{2}\cos\left(\frac{\pi}{2}-\varphi\right)=\VEC u_{1}R_{1}R_{2}\sin\varphi
\end{aligned}
\begin{gathered}\end{gathered}
\end{equation}
and 
\begin{equation}
\begin{aligned}\VEC J_{\VEC u_{2}} & =\VEC u_{2}P_{\VEC u_{2}}=\VEC u_{\mathrm{2}}R_{2}\VEC{\hat{u}}_{2}\cdot(\VEC{\hat{v}}_{1}R_{1}+\VEC{\hat{v}}_{2}R_{2})\\
\vphantom{\intop_{0}^{0}} & =\VEC u_{2}R_{1}R_{2}\cos\left(\frac{\pi}{2}+\varphi\right)=-\VEC u_{2}R_{1}R_{2}\sin\varphi
\end{aligned}
\begin{gathered}\end{gathered}
\end{equation}
with an asymmetry in the last line which is obvious from the geometry
sketched in Fig.~\ref{fig:vektor}. 

By summing up all current contributions according to Eq.~\eqref{eq:Jtot4}
we obtain the final expression for the total density current built
from the remaining $2N=4$ velocity components
\begin{equation}
\VEC J_{\mathrm{tot}}=R_{1}^{2}\VEC v_{\mathrm{1}}+R_{2}^{2}\VEC v_{\mathrm{2}}+R_{1}R_{2}\left(\VEC v_{\mathrm{1}}+\VEC v_{2}\right)\cos\varphi+R_{1}R_{2}\left(\VEC u_{1}-\VEC u_{2}\right)\sin\varphi.\label{eq:Jfinal}
\end{equation}
The total velocity $\VEC{v_{\mathrm{tot}}}$ according to Eq.~\eqref{eq:vtot_fin}
now reads as
\begin{equation}
\VEC v_{\mathrm{tot}}=\frac{R_{1}^{2}\VEC v_{\mathrm{1}}+R_{2}^{2}\VEC v_{\mathrm{2}}+R_{1}R_{2}\left(\VEC v_{\mathrm{1}}+\VEC v_{2}\right)\cos\varphi+R_{1}R_{2}\left(\VEC u_{1}-\VEC u_{2}\right)\sin\varphi}{R_{1}^{2}+R_{2}^{2}+2R_{1}R_{2}\cos\varphi}\,.\label{eq:vtot}
\end{equation}

The obtained total probability current field $\VEC J_{\mathrm{tot}}$
spanned by the various velocity components $\VEC v_{i}$ and $\VEC u_{i}$
we have denoted as the path excitation field (cf.~chapter~\ref{sec:3.4.Spreading-of-the-wave-packet},
and \cite{Groessing.2012doubleslit}). It is built by the sum of its
partial currents, which themselves are built by an amplitude weighted
projection of the total current. Furthermore, we observe that the
superposition principle is violated for $\VEC J$, and, analogously
for $P,$ in the following sense: In quantum mechanics the amplitudes
of the wave function components have to be summed up coherently, i.e.\ superpositioned,
in the case of undisturbed paths, and for calculation of the probability
density this sum has to be taken as absolute value squared. In other
words, the Schrödinger equation is linear, and observation of a state
is regularized by Born's rule. In our case, all the relevant variables,
i.e.\ $\Pw i$ and $\Jw i$, are nonlinear. Consequently, to obtain
the correct total probability density $P_{\mathrm{tot}}$ or total
current $\VEC J_{\mathrm{tot}}$, respectively, one has to take into
account \textit{all} elementary, i.e.\ partial, contributions to
the corresponding variable.

Summarizing, the shift to a new projection rule of Eq.~\eqref{eq:tot_prob_law}
build the kernel for a set of relations of current rules. It is characterized
by summing up the nonlinear partial currents, where each of the latter
contains information about the total field via the  projection rule.
This property is characterized in that any change in a local field
affects the total field, and \textit{vice versa}. 

The trajectories or streamlines, respectively, are obtained according
to $\VEC{\dot{x}}=\VEC v_{\mathrm{tot}}$ in the usual way by integration.
By re-inserting the expressions for convective velocities from Eq.~\eqref{eq:v=00003Dp_div_m},
\begin{equation}
\VEC v_{i}=\frac{\nabla S_{i}}{m}\thinspace,\label{eq:diss.4.40}
\end{equation}
and diffusive velocities from Eq.~\eqref{eq:gb.3.6} together with
\eqref{eq:nablap2p=00003D1_2nablar2r},
\begin{equation}
\VEC u_{i}=-\frac{\hbar}{m}\frac{\nabla R_{i}}{R_{i}}\thinspace,\label{eq:diss.4.41}
\end{equation}
one immediately identifies Eq.~\eqref{eq:vtot} with the Bohmian
guiding equation and Eq.~\eqref{eq:Jfinal} with the quantum mechanical
pendant for the probability current~\cite{Bohm.1993undivided,Sanz.2008trajectory}. 

\section{Double-slit interference\label{sec:Double-slit-interference}}

It is straightforward to now also describe and explain quantum interference
with our approach (cf.~\cite{Groessing.2012doubleslit,Groessing.2012vaxjo,Groessing.2013dice}).
We choose a textbook scenario in the form of the calculation of the
intensity distribution and the particle trajectories in an electron
interferometer. As we are also interested in the trajectories, we
refer to, and compare our results with, the well-known work by Philippidis~\emph{et\,al}.~\cite{Philippidis.1979quantum},
albeit in the form as presented by Holland~\cite{Holland.1993}.

We choose similar initial situations as Holland, i.e.\ electrons,
represented by plane waves in the forward $y$-direction, from a source
passing through soft-edged slits~$1$ and $2$ in a barrier, located
along the $x$-axis, and recorded at a screen. In our model, we therefore
note two Gaussians representing the totality of the effectively heated-up
path excitation field, one for slit~1 and one for slit~$2$, whose
centres have the distances $+X$ and $-X$ from the plane spanned
by the source and the centre of the barrier along the $y$-axis, respectively. 

\begin{figure}[!t]
\centering{}\includegraphics[width=1\textwidth,height=1\textwidth]{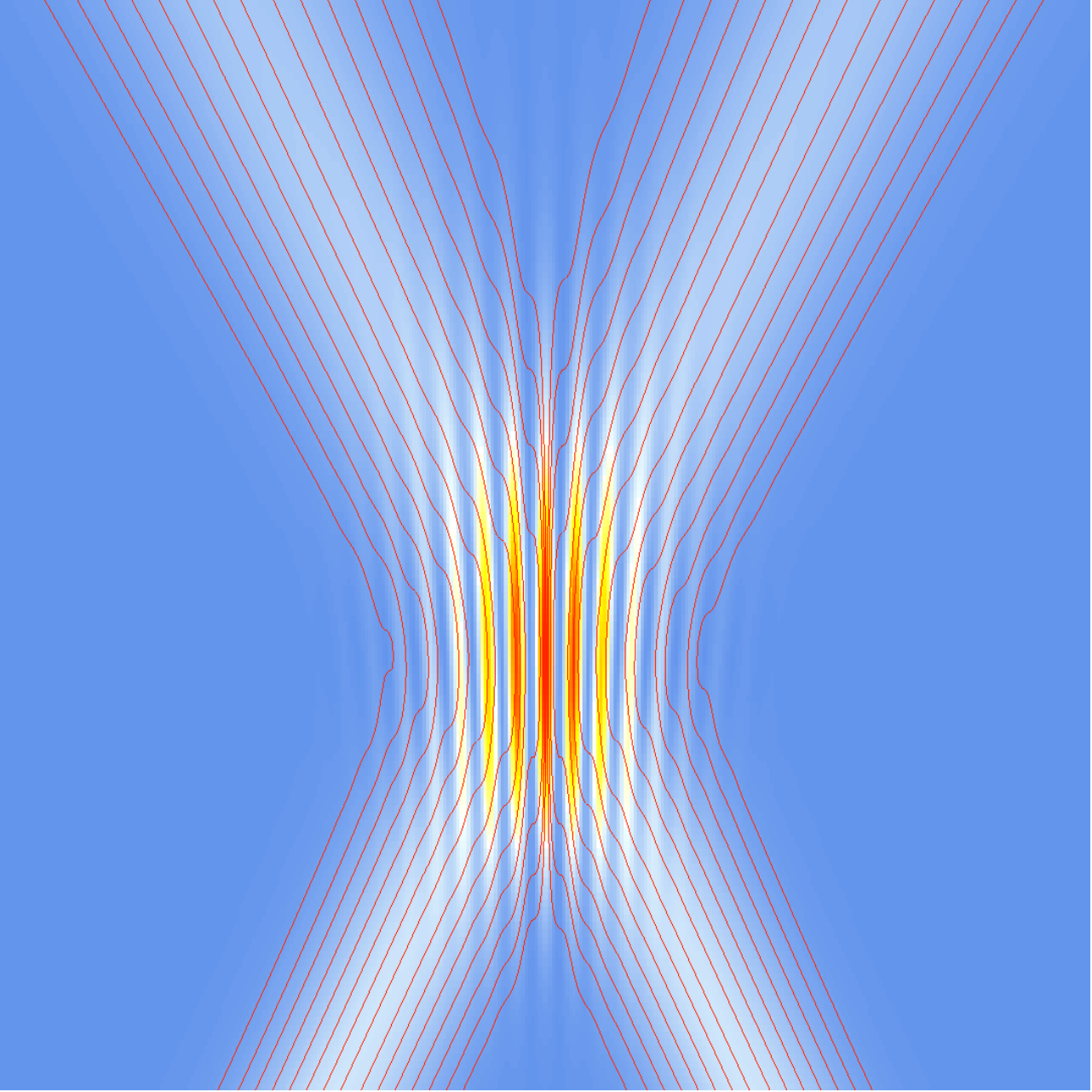}
\caption{Classical computer simulation of the interference pattern: Intensity
distribution with increasing intensity from white through yellow and
orange, with trajectories (red) for two Gaussian slits, and with small
dispersion (evolution from bottom to top; $v_{x,1}=-v_{x,2}$).\label{fig:2}}
\end{figure}

The results according to Eq.~\eqref{eq:Ptot2slit} is shown in Fig.~\ref{fig:2}
which depicts the interference of two beams emerging from Gaussian
slits\footnote{For details on how the simulations have been carried out see chapter~\ref{sec:8.2.Simulation-procedure},
on the construction of the trajectories see chapter~\ref{sec:Trajectories}.
Initial values for all simulations are $P_{1}=P_{2}$, $\sigma_{1}=\sigma_{2}$,
$v_{x,1}=v_{x,2}=0$, otherwise noted.}. The trajectories are the flux lines obtained by choosing a set of
appropriate initial points at $y=0$. The trajectories follow a no-crossing
rule\footnote{From the assumed uniqueness and differentiability of $S(\VEC x,t)$
follows that the paths don't cross each other. See section~\ref{sec:orthogonality}
for further explanations. However, at this stage we are discussing
an ontological point of view on how the no-crossing phenomenon can
be explained.}: Particles from the left slit stay on the left side and \textit{vice
versa} for the right slit. This feature is explained here by a sub-quantum
build-up of kinetic (heat) energy acting as an emergent repellent
along the symmetry line.

In Fig.~\ref{fig:2} one can observe a basic characteristic of the
averaged particle trajectories, which, just because of the averaging,
are identical with the Bohmian trajectories. To fully appreciate this
surprising characteristic, we remind the reader of the severe criticism
of Bohmian trajectories as put forward by Scully and others~\cite[and references therein]{Scully.1998bohm}.
The critics claimed that Bohmian trajectories would have to be described
as ``surreal'' ones because of their apparent violation of momentum
conservation. In fact, due to the no-crossing rule for Bohmian trajectories
in Young's double-slit experiment, for example, the particles coming
from, say, the right slit~\textendash{} and expected at the left
part of the screen if momentum conservation should hold on the corresponding
macro-level \textendash{} actually arrive at the right part of the
screen \textendash{} and \textit{vice versa} for the other slit. In
Bohmian theory, this no-crossing rule is due to the action of the
non-classical quantum potential, such that, once the existence of
a quantum potential is accepted, no contradiction arises and the trajectories
may be considered ``real'' instead of ``surreal''.

Here we can note that in our sub-quantum approach an explanation of
the no-crossing rule is even more straightforward and actually a consequence
of a detailed \textit{microscopic momentum conservation} as discussed
in section~\ref{sec:orthogonality} and in~\cite{Groessing.2012doubleslit}.
As can be seen in Fig.~\ref{fig:2}, the trajectories are repelled
from the central symmetry line. However, in our case this is only
implicitly due to a quantum potential, but actually due to the identification
of the latter with a kinetic rather than a potential energy: As has
already been stressed in~\cite{Groessing.2009origin}, it is the
\textit{heat of the compressed vacuum} that accumulates along said
symmetry line, i.e.\ as reservoir of outward oriented kinetic energy,
and therefore repels the trajectories. Fig.~\ref{fig:2} is in full
concordance with the Bohmian interpretation (see, for example,~\cite{Sanz.2008trajectory}
for comparison).

This can be shown even in greater detail. Whereas in the example of
Fig.~\ref{fig:2} the small amount of dispersion is practically negligible,
we now discuss in more detail an interference scenario with significant
dispersion of the two Gaussians, i.e.\ where initially the two Gaussians
spread independently from each other due to the action of the diffusive
velocities $u_{1}$ and $u_{2}$, respectively. 
\begin{figure}[!t]
\centering{}\includegraphics[width=1\textwidth,height=1\textwidth]{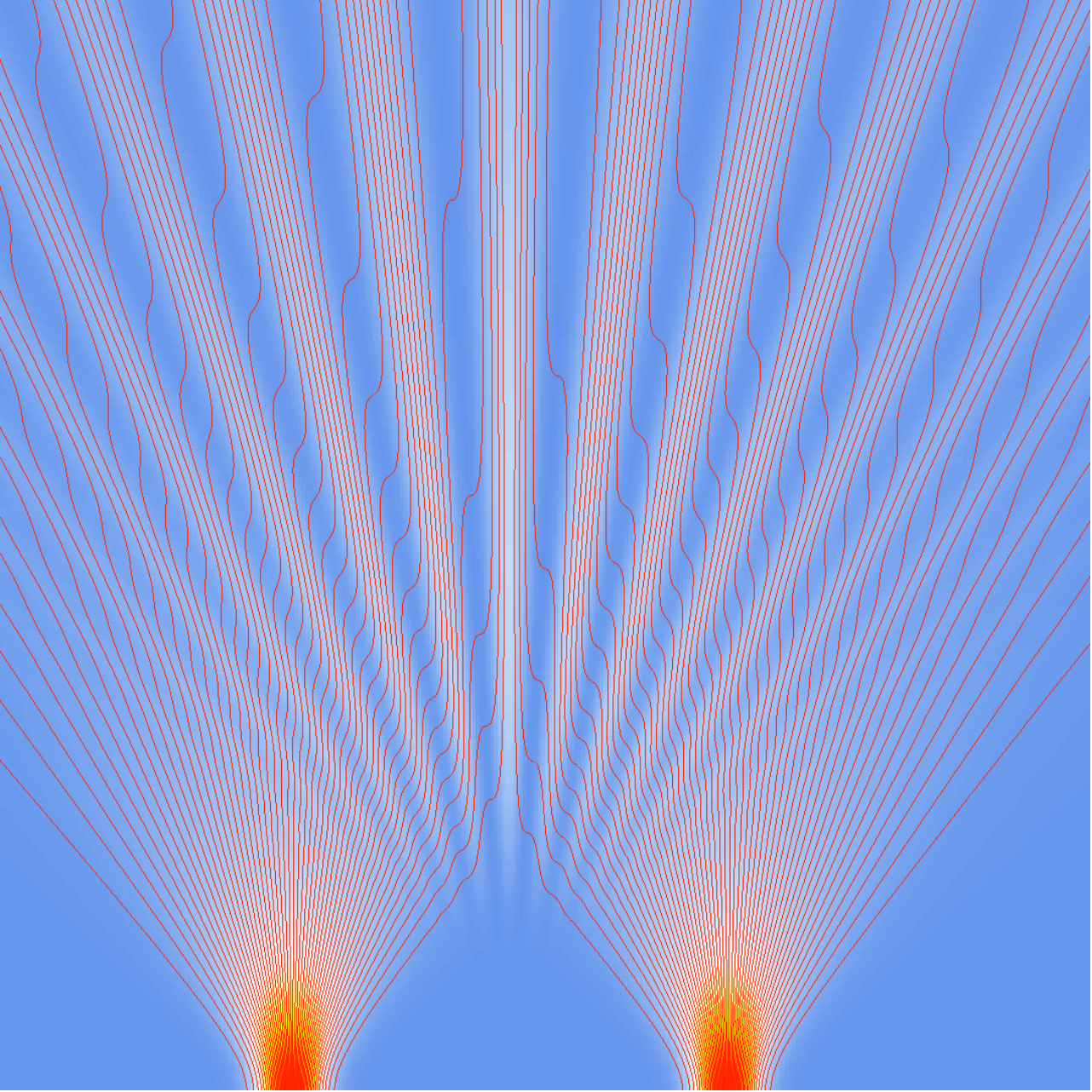}
\caption{Classical computer simulation of the interference pattern: Intensity
distribution with increasing intensity from white through yellow and
orange, with trajectories (red) for two Gaussian slits, and with large
dispersion (evolution from bottom to top; $v_{x,1}=v_{x,2}=0$)\label{fig:3}}
\end{figure}

In Fig.~\ref{fig:3}, trajectories according to Eq.~\eqref{eq:vtot}
for the two Gaussian slits are shown. The interference hyperbolas
for the maxima characterize the regions where the phase difference
$\varphi=2n\pi$, and those with the minima lie at $\varphi=(2n+1)\pi$,
$n=0,\pm1,\pm2,\ldots$ Note in particular the kinks of trajectories
moving from the centre-oriented side of one relative maximum to cross
over to join more central relative maxima. The trajectories are in
full accordance with those obtained from the Bohmian approach, as
can be seen by comparison with references~\cite{Holland.1993,Bohm.1993undivided,Sanz.2009context},
for example. In our classical explanation of double-slit interference,
a detailed micro-causal account of the corresponding kinematics can
be given: Firstly, we note that the last term in Eq.~\eqref{eq:Jfinal},
which is responsible for the genuinely quantum behaviour, is characterized
by the vector $\VEC u_{1}-\VEC u_{2}$ which in the interference region
is always oriented into the forward direction away from the slits
(Fig.~\ref{fig:3}). This means that said last term is modulated
by $\sin\varphi$, with the result that the term alternates between
forward directions where $\sin\varphi>0$ and backward directions
where $\sin\varphi<0$.

Thus, in the backward cases, the trajectories coming from the relative
maxima (bright fringes) loose velocity/momentum in the forward direction
and cross over into the area of the relative minimum (dark fringes).
Alternatively, in the forward cases, the trajectories coming from
the relative minima (dark fringes) gain velocity/momentum in the forward
direction and thus align with the other trajectories of the bright
fringes. In other words, for the areas where $\sin\varphi<0$, part
of the current (along a relative maximum) is being removed (depletion),
whereas for $\sin\varphi>0$, parts of currents flow together to produce
a newly formed bright fringe (accumulation). This is in accordance
with an earlier description of quantum interference, where the effects
of diffusion wave fields (cf.~\cite{Mandelis.2000diffusion,Mandelis.2001structure})
were explicitly described by alternating zones of heat accumulation
or depletion, respectively~\cite{Groessing.2009origin}. Towards
the central symmetry line, then, one observes heat accumulation from
both sides, and due to big momentum kicks from the central accumulation
of heat energy, the forward particle velocities' directions align
parallel to the symmetry axis. With the crossing-over of particle
trajectories being governed by the last, diffusion-related, term on
the right hand side of Eq.~\eqref{eq:Jfinal}, one finds that for
$\varphi=0$ the resulting diffusive current is zero and thus, as
total result of the overall kinematics, no-crossing is possible. Further,
we note that our results are also in agreement with the experimental
results by Kocsis~\emph{et\,al}.~\cite{Kocsis.2011observing}.

Finally, to illustrate the straightforward applicability of our model
to more general situations, i.e.\ as compared to the simple symmetrical
scenarios of 
\begin{figure}[!th]
\centering{}\includegraphics[width=0.5\textwidth]{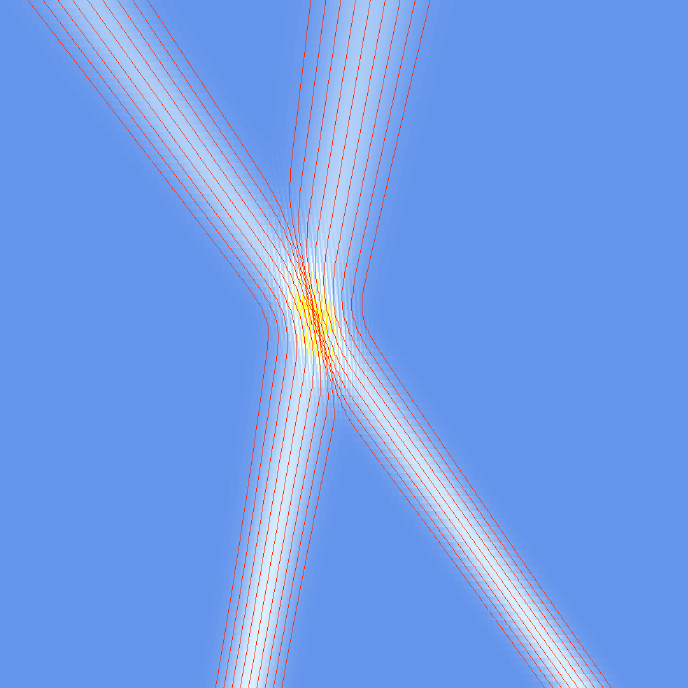}
\caption{Same as Fig.~\ref{fig:2}, but with different initial average velocities:
$v_{x,2}=-4v_{x,1}$. Note again the no-crossing behaviour, with the
two trajectory bundles repelling each other due to the kinetic (heat)
energy along the slanted central line.\label{fig:2b}}
\end{figure}
\begin{figure}[!th]
\centering{}\includegraphics[width=0.5\textwidth]{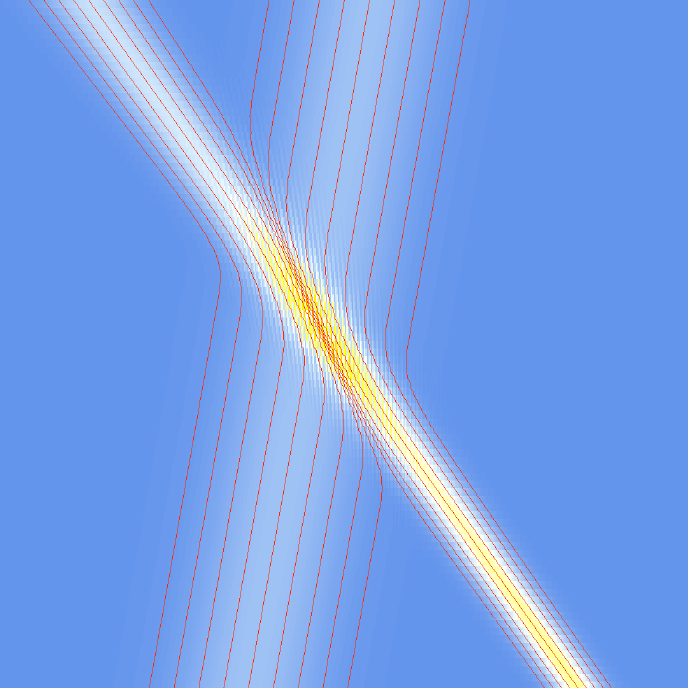}
\caption{Same as Fig.~\ref{fig:2b}, but with different initial spreading:
$\sigma_{1}=3\sigma_{2}$. Although the two partial beams altogether
reflect off each other, one can clearly observe the effect of \textit{microscopic
momentum conservation}: The path excitation field of the right beam
is propagated over to the micro-kinematics of the left beam, and \textit{vice
versa}.\label{fig:2c}}
\end{figure}
\begin{figure}[!th]
\centering{}\includegraphics[width=0.5\textwidth]{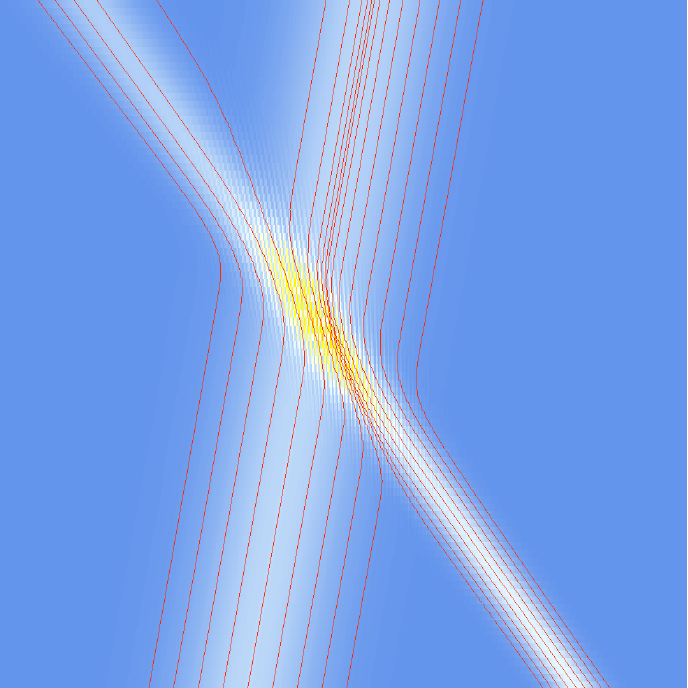}
\caption{Same as Fig.~\ref{fig:2c}, but with different probability densities:
$P_{1}=2P_{2}$. Note that the emerging beam behaviour compares more
with inelastic scattering than with the elastic-type scattering of
Fig.~\ref{fig:2c}, as part of the left beam merges with the right
one.\label{fig:2d}}
\end{figure}
Figs.~\ref{fig:2} and \ref{fig:3}, we now employ our simulation
schema to cases where neither the Gaussians are identical nor their
weights. We thus study asymmetric coherent superpositions as discussed
in ref.~\cite{Sanz.2008trajectory}, and in our Figs.~\ref{fig:2b}
to \ref{fig:2d} we show results in accordance with the figures~4\textendash 6
of ref.~\cite{Sanz.2008trajectory}. The analysis of ref.~\cite{Sanz.2008trajectory}
holds identically in our approach, so that we here restrict ourselves
to pointing out that our figures display the following cases of varied
properties for the beams emerging from the two slits: 
\begin{itemize}
\item different modulus of the initial velocity/momentum, 
\item different initial spreading, 
\item different weights for the probability densities.
\end{itemize}

\section{Entangling currents in the double-slit experiment\label{sec:4.6.Entangling-currents}}

Because of the mixing of diffusion currents from both channels, we
call the following decisive term in $\mathbf{J_{\mathrm{\mathit{\mathrm{tot}}}}}$~\eqref{eq:Jfinal}
the \textit{entangling current~}\cite{Mesa.2013variable,Groessing.2012vaxjo,Groessing.2013dice}
\begin{equation}
\mathbf{J_{\mathit{\mathit{\mathrm{e}}}}}=R_{1}R_{2}\left(\VEC u_{1}-\VEC u_{2}\right)\sin\varphi=\frac{\hbar}{m}\left(R_{1}\nabla R_{2}-R_{2}\nabla R_{1}\right)\sin\varphi\label{eq:Je}
\end{equation}
where Eq.~\eqref{eq:diss.4.41} has been substituted.

\begin{sidewaysfigure}
\subfloat[Intensity distribution $P$ with increasing intensity from blue and
white through yellow and orange, with averaged trajectories (red)\label{fig:dice.1a}]{\centering{}\includegraphics[width=0.48\textwidth]{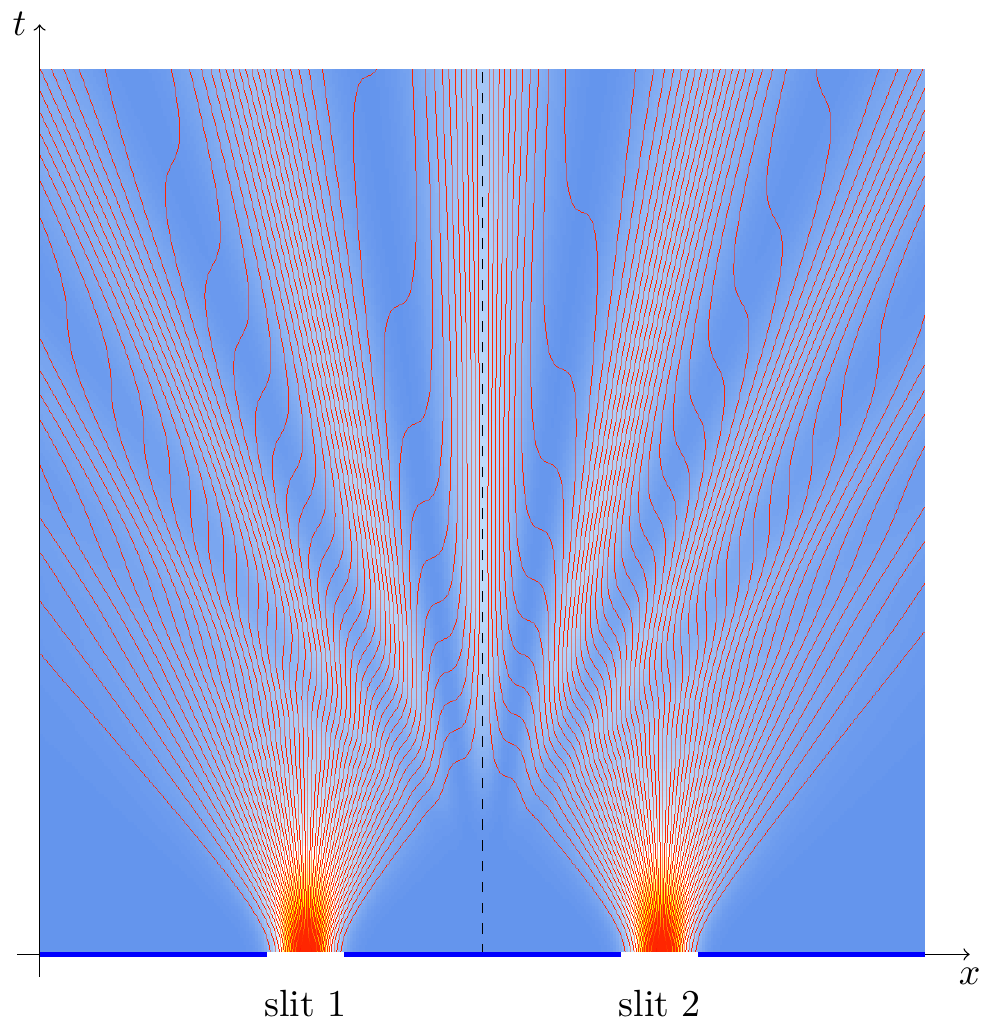}}\hfill{}\subfloat[Corresponding entangling current density $\mathbf{J}_{\mathrm{e}}$\label{fig:dice.1c}]{\centering{}\includegraphics[width=0.48\textwidth]{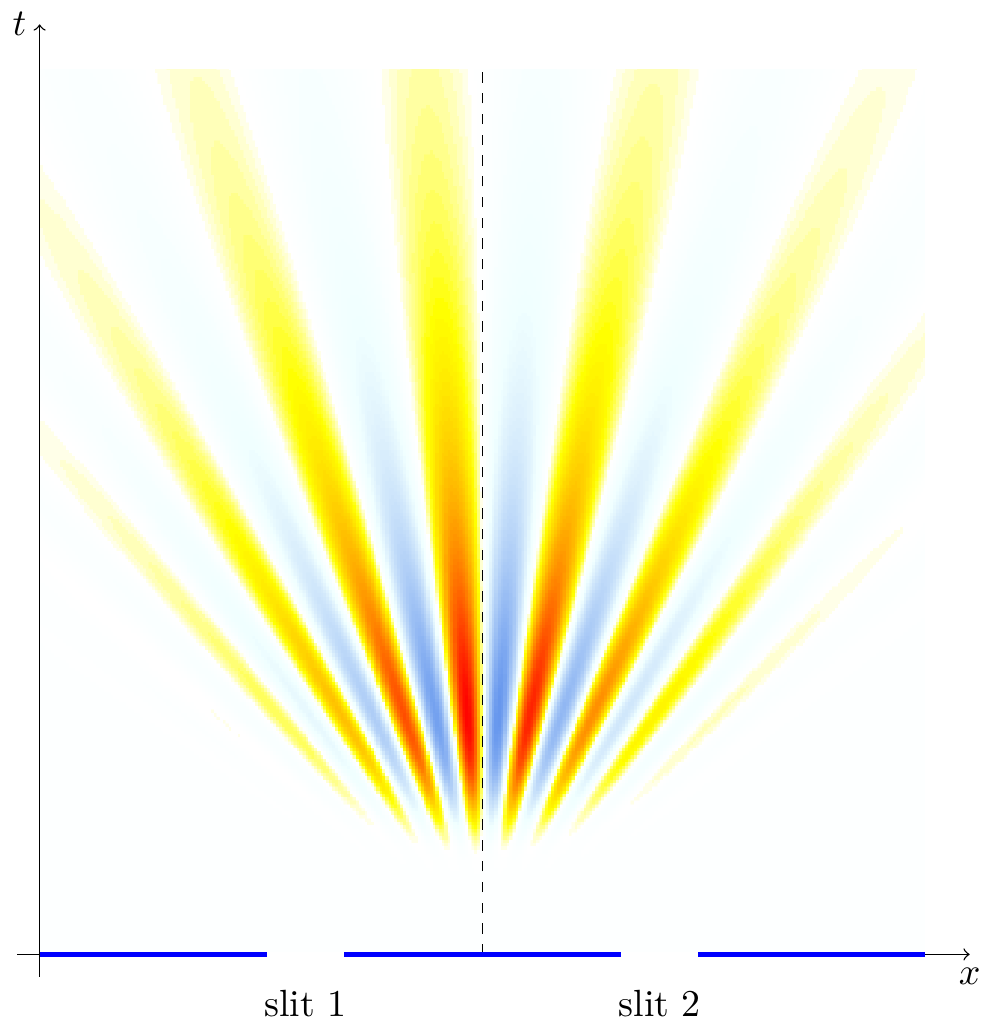}}\caption{Classical computer simulation of the interference pattern with equal
slit widths for two Gaussian slits ($v_{x,1}=v_{x,2}=0$)\label{fig:diss-dice.1}}
\end{sidewaysfigure}
\begin{sidewaysfigure}
\subfloat[Intensity distribution $P$\label{fig:dice.2a}]{\centering{}\includegraphics[width=0.48\textwidth]{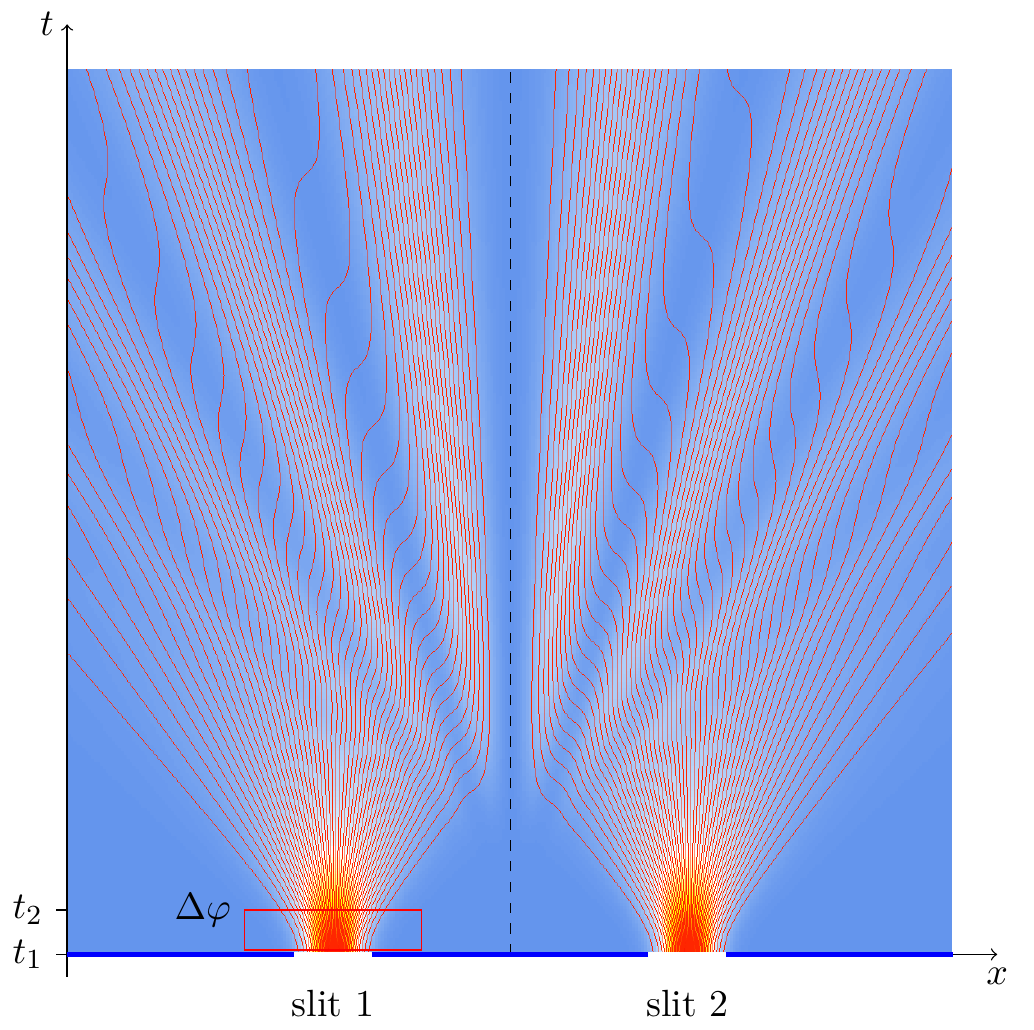}}\hfill{}\subfloat[Corresponding entangling current $\mathbf{J}_{\mathrm{e}}$. Note
the shift of maxima and minima in the emerging pattern, as compared
to Fig.~\ref{fig:dice.1c}.\label{fig:dice.2c}]{\centering{}\includegraphics[width=0.48\textwidth]{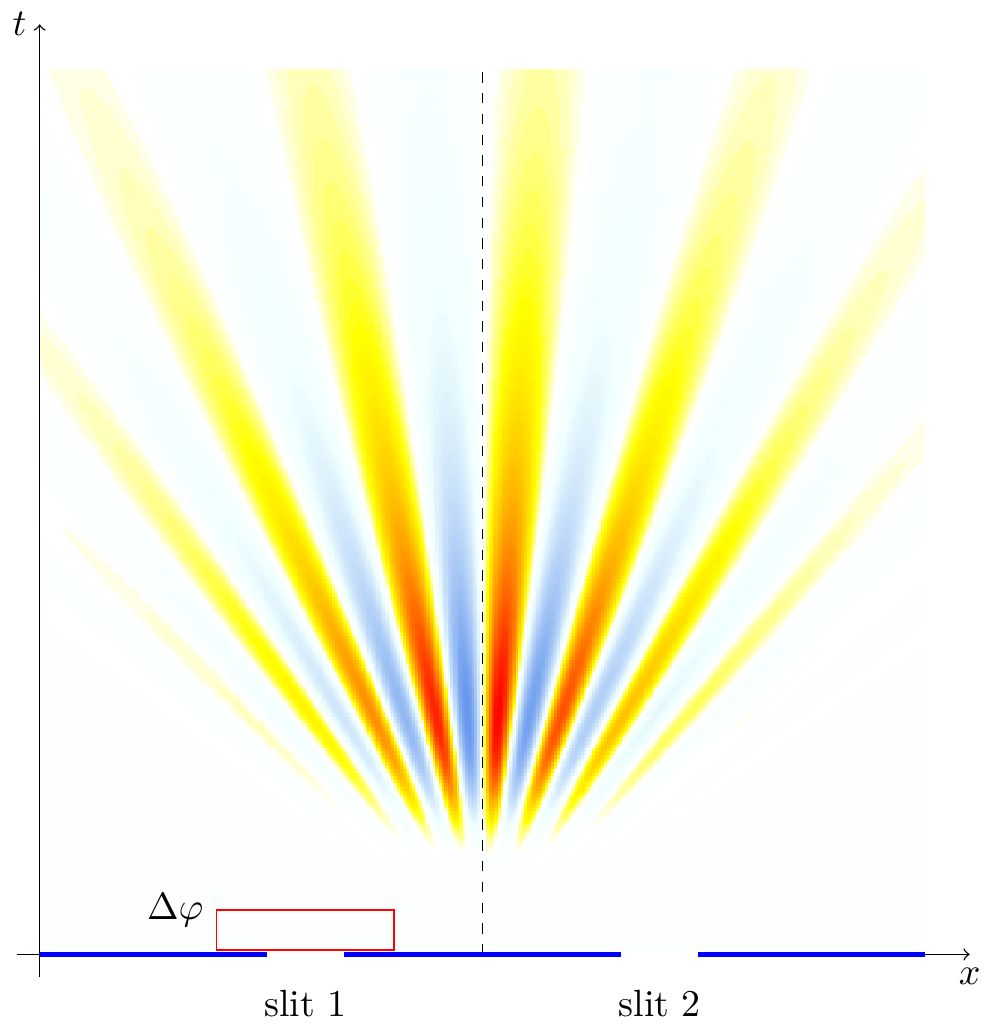}}\caption{Classical computer simulation as in Fig.~\ref{fig:diss-dice.1},
but with additional phase shift $\Delta\varphi=3\pi$ gradually accumulated
during the time interval between $t_{1}$ and $t_{2}$ at slit~1\label{fig:diss-dice.2}}
\end{sidewaysfigure}
\begin{sidewaysfigure}
\subfloat[Intensity distribution $P$\label{fig:dice.3a}]{\centering{}\includegraphics[width=0.48\textwidth]{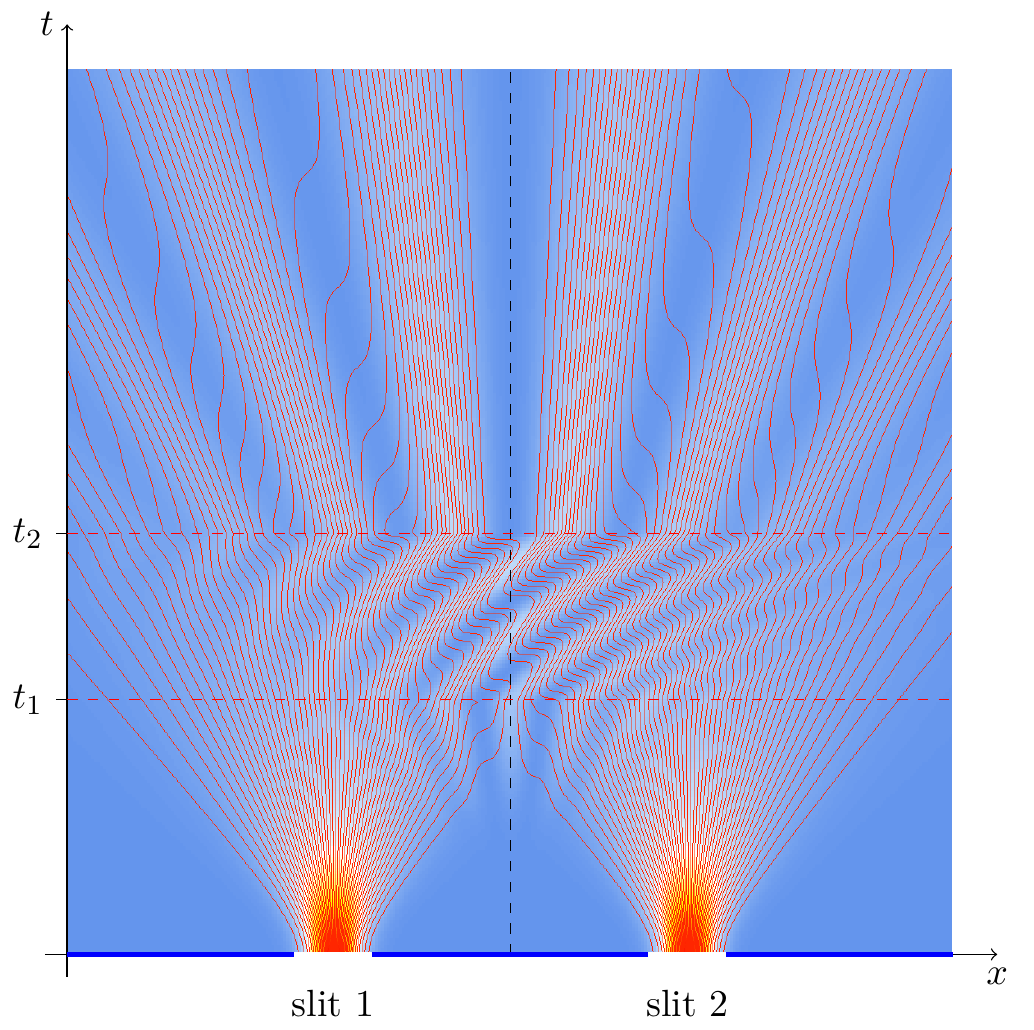}}\hfill{}\subfloat[Corresponding entangling current $\mathbf{J}_{\mathrm{e}}$\label{fig:dice.3c}]{\centering{}\includegraphics[width=0.48\textwidth]{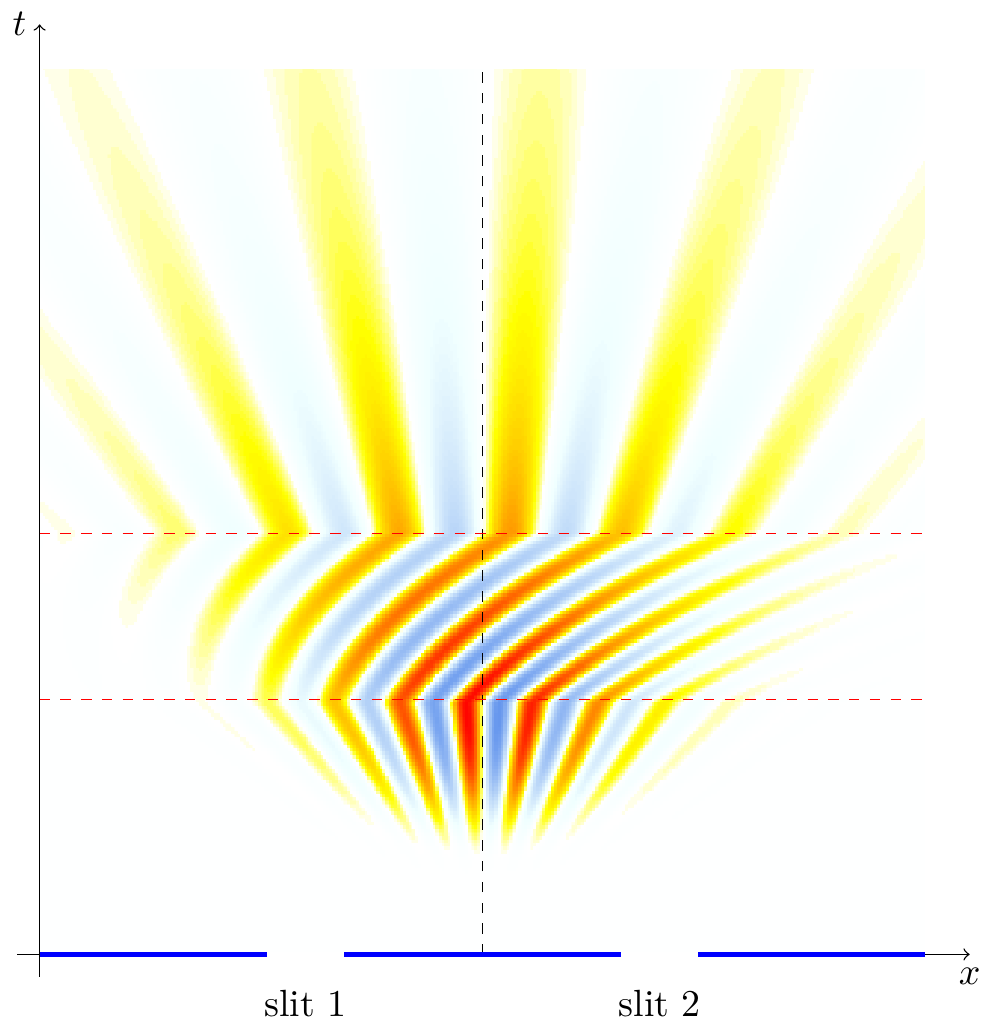}}\caption{Classical computer simulation as in Fig.~\ref{fig:diss-dice.2},
but with different times $t_{i}$ and with gradually accumulated additional
phase $\Delta\varphi=5\pi$. Values for both, $P$ and $\protect\VEC J_{\mathrm{e}}$
being equal to that of Fig.~\ref{fig:diss-dice.1} for $t<t_{1}$
and Fig.~\ref{fig:diss-dice.2} for $t>t_{2}$, respectively\label{fig:diss-dice.3}}
\end{sidewaysfigure}
\begin{figure}[!tbh]
\begin{centering}
\subfloat[$\Delta\varphi$ in Fig.~\ref{fig:diss-dice.2}\label{fig:dice.2-phase-shift}]{\centering{}\qquad{}\includegraphics{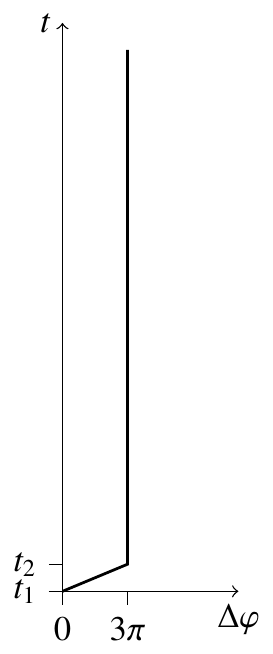}\qquad{}}\subfloat[$\Delta\varphi$ in Fig.~\ref{fig:diss-dice.3}\label{fig:dice.3-phase-shift}]{\centering{}\qquad{}\includegraphics{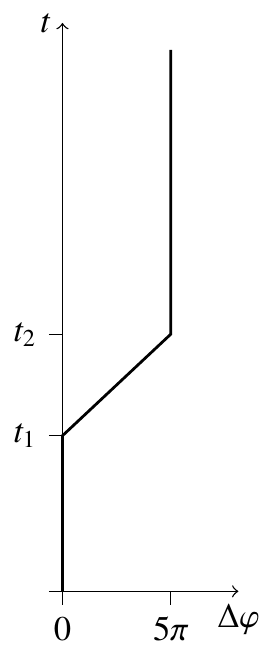}\qquad{}}
\par\end{centering}
\caption{Additional phase shift $\Delta\varphi$ accumulated during the time
interval between $t_{1}$ and $t_{2}$ at slit~1. Different phase
shifts of $\Delta\varphi=3\pi$ and $\Delta\varphi=5\pi$, respectively,
lead to identical distributions of $P$ and $J_{\mathrm{tot}}$ at
last.\label{fig:dice-phase-shift}}
\end{figure}

For illustration, Figs.~\ref{fig:diss-dice.1}\textendash \ref{fig:diss-dice.3}
show our classical computer simulations of interference and the role
of the entangling current $\mathbf{J_{\mathit{\mathit{\mathrm{e}}}}}$
in different situations. The entangling current $\mathbf{J}_{\mathrm{e}}$~\eqref{eq:Je}
is characterized by the difference of the diffusive velocities $\VEC u_{i}$,
and is hence responsible for the nature of the process forming the
interference pattern. Fig.~\ref{fig:diss-dice.1} shows the emerging
interference pattern and the average trajectories without, and Fig.~\ref{fig:diss-dice.2}
with an applied extra phase shift (according to Fig.~\ref{fig:dice.2-phase-shift})
at one slit. To bring out the shifting of the interference pattern
more clearly, in Fig.~\ref{fig:diss-dice.3} we apply \textendash{}
mainly for didactic reasons, as it is not clear what applying the
phase to a slit in the distance means \textendash{} the phase shift
at much later times (according to Fig.~\ref{fig:dice.3-phase-shift})
than in Fig.~\ref{fig:diss-dice.2}. Thereby, also a decoupling of
wave and particle behaviours becomes visible.

\begin{figure}
\begin{centering}
\subfloat[No additional phase\label{fig:diss.4.12a}]{\centering{}\includegraphics[height=0.44\textheight]{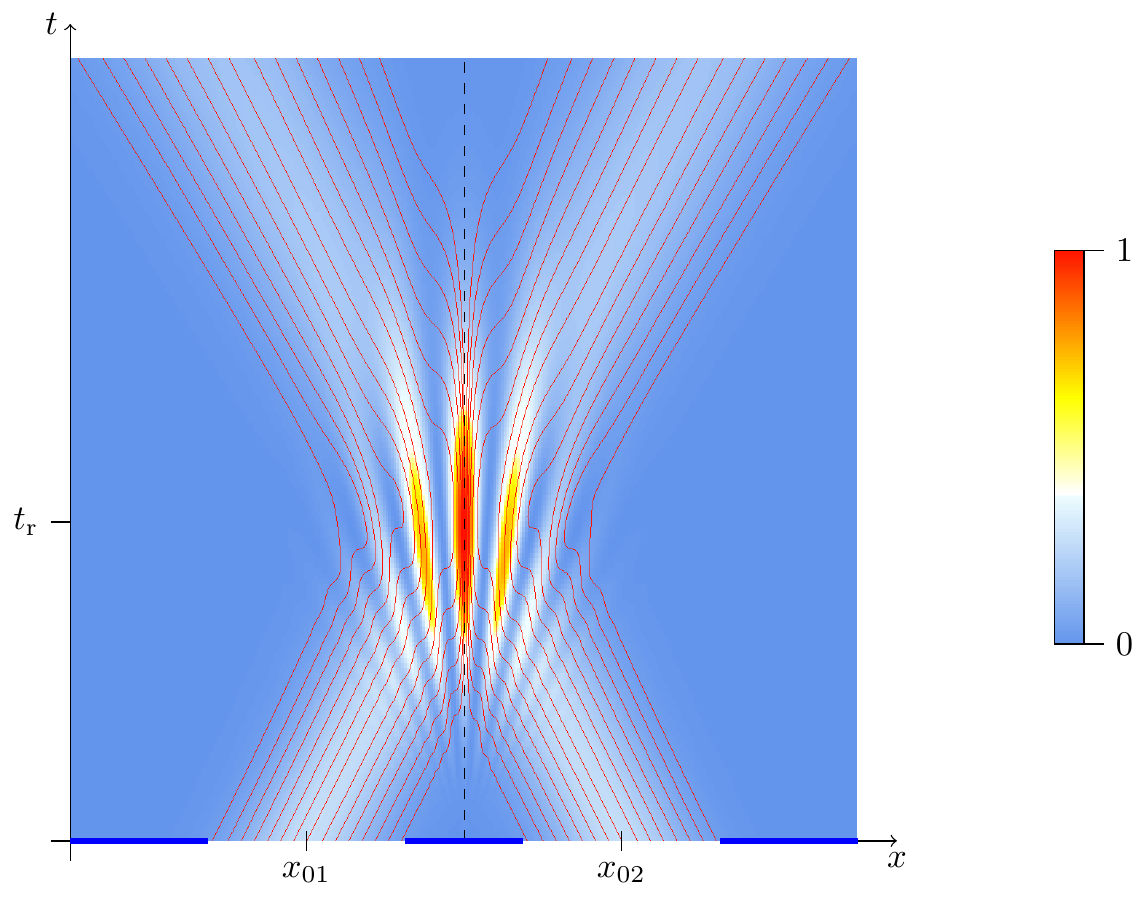}}
\par\end{centering}
\begin{centering}
\subfloat[Same as Fig.~\ref{fig:diss.4.12a}, with an additional phase $\Delta\varphi=\pi$
at slit~1\label{fig:diss.4.12b}]{\centering{}\includegraphics[height=0.44\textheight]{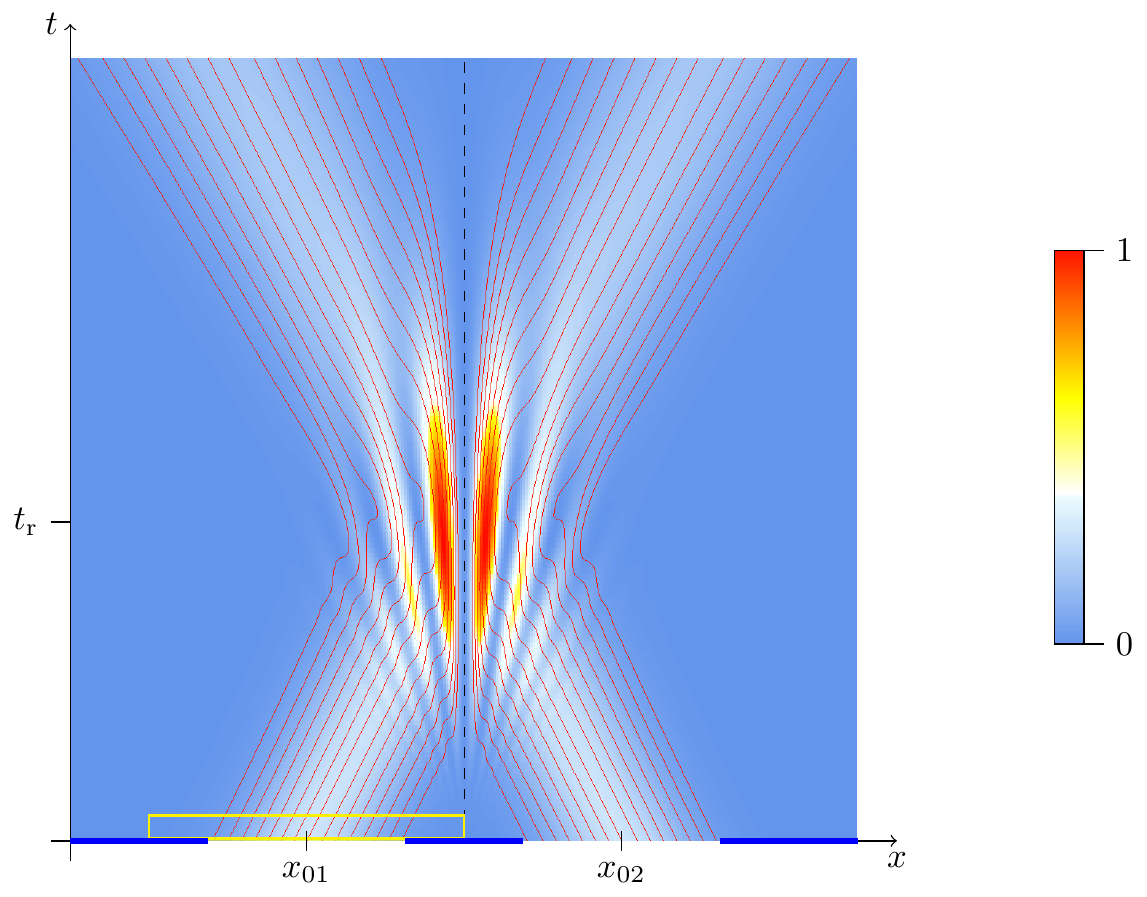}}
\par\end{centering}
\caption{Classical computer simulation of the interference pattern in a double-slit
experiment; with $v_{x,1}=-v_{x,2}$\label{fig:diss.4.12}}
\end{figure}
\begin{figure}
\begin{centering}
\subfloat[Same setup as in Fig.~\ref{fig:diss.4.12a}, with arbitrary normalization
and $v_{x,1}=-v_{x,2}$\label{fig:traj.4}]{\centering{}\includegraphics[height=0.44\textheight]{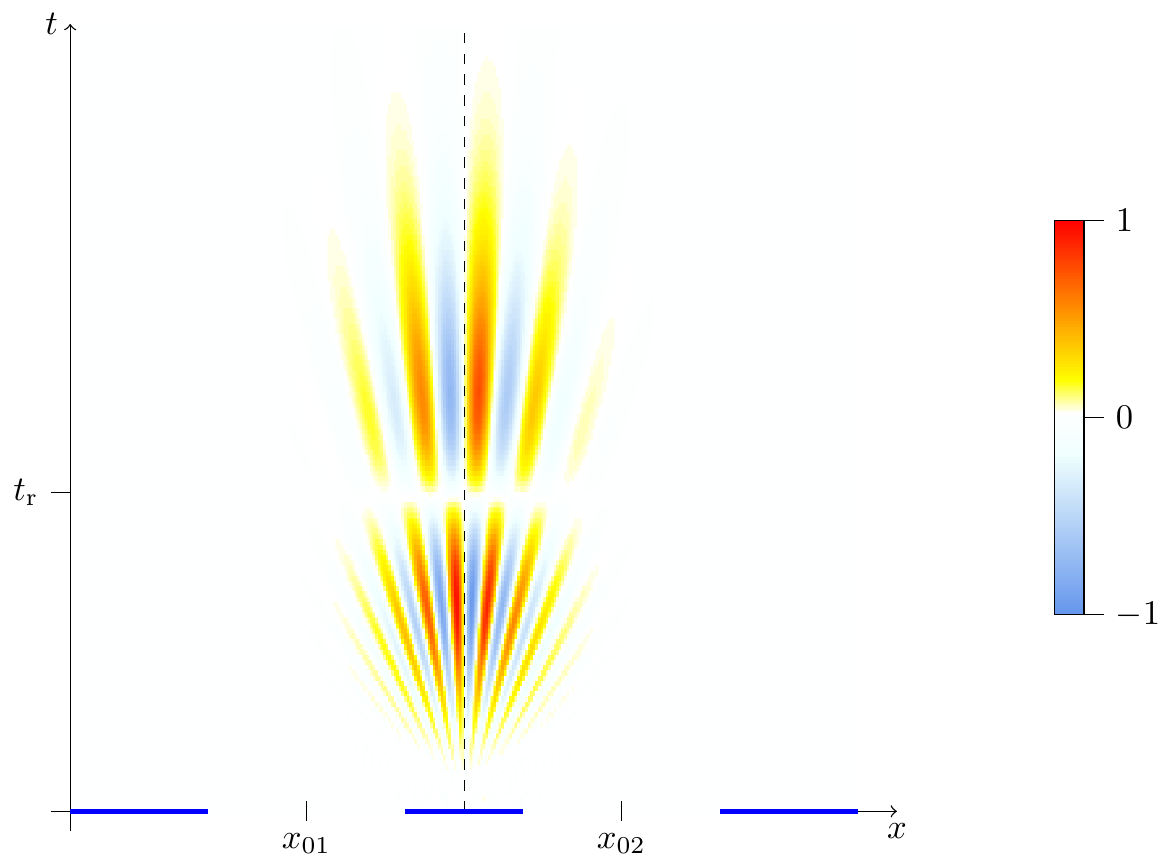}}
\par\end{centering}
\begin{centering}
\subfloat[Same as Fig.~\ref{fig:traj.4}, with an additional phase shift of
$\Delta\varphi=\pi$ at slit~1\label{fig:traj.5}]{\centering{}\includegraphics[height=0.44\textheight]{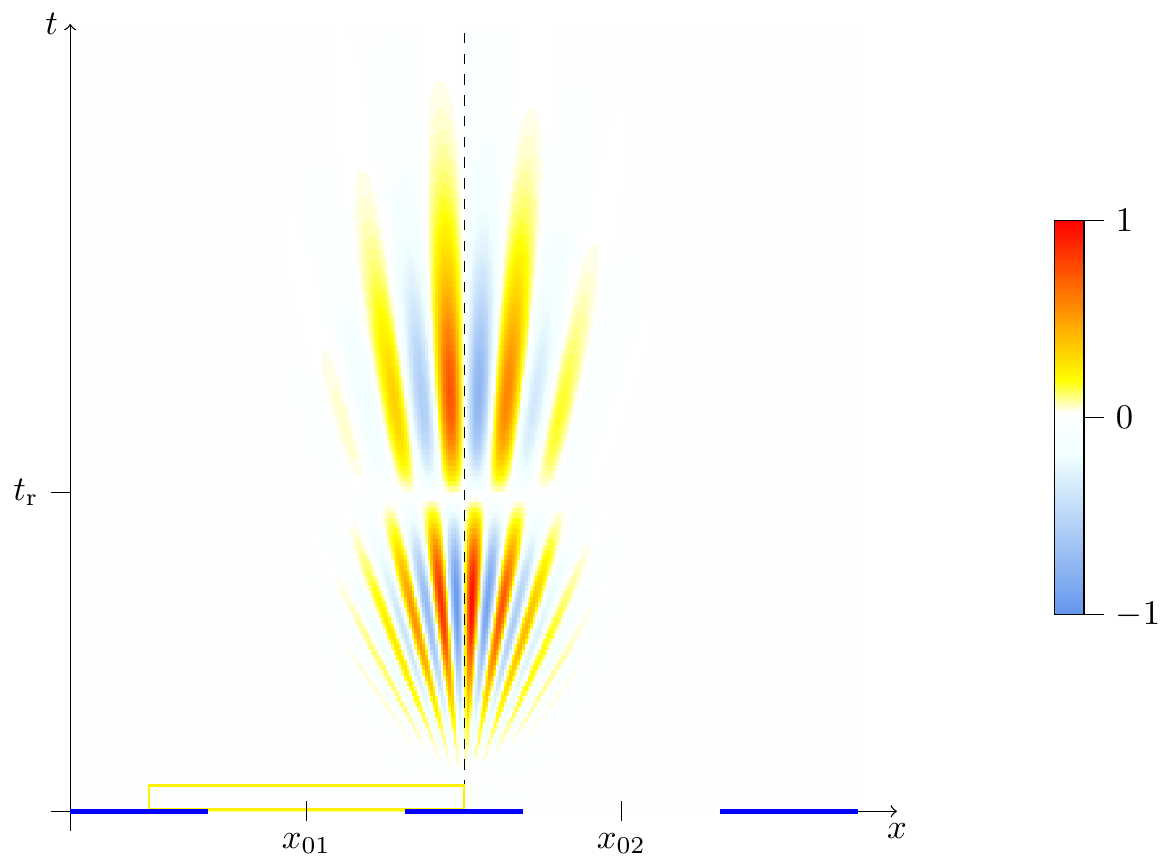}}
\par\end{centering}
\caption{Classical computer simulation of the total average entangling current
density $\mathbf{J}_{\mathrm{e}}$ in a double-slit experiment\label{fig:diss.4.13}}
\end{figure}

The distributions of $P$ and $\mathbf{J}_{\mathrm{e}}$ in Fig.~\ref{fig:diss-dice.3}
are the same as in Fig.~\ref{fig:diss-dice.1} for times $t<t_{1}$
and as in Fig.~\ref{fig:diss-dice.2} for times $t>t_{2}$, respectively,
and show the effect of the shifting of the interference fringes more
clearly than Fig.~\ref{fig:diss-dice.2}. Note the radically different
behaviours of the probability density related to wave-like interference
on the one hand, and that of the average particle trajectories on
the other hand. Although the currents $\mathbf{J}_{\mathrm{e}}$ dramatically
cross the central symmetry line separating the areas of the two slits,
the average particle trajectories (Fig.~\ref{fig:dice.3a}) strictly
obey the no-crossing rule familiar from, but not restricted to, the
de~Broglie\textendash Bohm interpretation. This is a clear demonstration
of the partial decoupling of wave and particle behaviour as envisaged
in our model.

As a further example, we use a similar setup as in Fig.~\ref{fig:2}.
The graphical result of a classical computer simulation of the interference
pattern in a double-slit experiment, including the average trajectories,
with evolution from bottom to top, is shown in Fig.~\ref{fig:diss.4.12a}.
The Gaussian wave packets characterized by moderate spreading at the
same standard deviations $\sigma$ move towards each other with constant
velocities $v_{x,1}=-v_{x,2}$. In Fig.~\ref{fig:diss.4.12b}, we
use the same double-slit arrangement as in Fig.~\ref{fig:diss.4.12a},
but now include a phase shifter affecting slit~1, as sketched by
the yellow rectangle on the left hand side. The exemplary choice of
$\Delta\varphi=\pi$ results in a shift of the interference fringes.
Comparing with Fig.~\ref{fig:diss.4.12a}, we recognize now a minimum
of the resulting distribution along the central symmetry line.

Comparing Figs.~\ref{fig:traj.4} with \ref{fig:traj.5}, one notes
that the dramatic shift from maxima to minima, and \emph{vice versa},
as observed in the interference patterns of Fig.~\ref{fig:diss.4.12a}
and Fig.~\ref{fig:diss.4.12b}, respectively, is essentially caused
by the changes in these entangling currents. This corresponds to a
sub-quantum account of the processes underlying quantum interference.

The result of our computer simulation of the probability current~\eqref{eq:Jfinal}
is shown in Fig.~\ref{fig:diss.4.13} corresponding to the intensity
distributions of Fig.~\ref{fig:diss.4.12}. One recognizes the change
of the maximum values of the probability current along the central
symmetry line in Fig.~\ref{fig:traj.5} in comparison with those
of Fig.~\ref{fig:traj.4}. Since the figures display the one-dimensional
case, the current flow is along the $x$-axis only. Interestingly,
at the time $t_{\mathrm{r}}$ of the reversal of the trajectories,
the current flow comes to a hold, and starts again for times $t>t_{\mathrm{r}}$
with reversed signs. This can be understood as a reversal of the relative
flow of heat $Q_{2}-Q_{1}$ between the two channels, since we have
from~\eqref{eq:aceq.4.8a} that 
\begin{equation}
\VEC u_{i}=-\frac{1}{2\omega m}\nabla Q_{i},
\end{equation}
such that the last term of $\VEC J_{\mathrm{e}}$~\eqref{eq:Je}
reads as 
\begin{equation}
\frac{1}{2\omega m}\sqrt{P_{1}P_{2}}\nabla(Q_{2}-Q_{1})\sin\varphi_{12}.
\end{equation}

The probability current $\VEC J_{\mathrm{tot}}$ in both figures essentially
only consists of its last terms, i.e.\ $\VEC J_{\mathrm{e}}$~\eqref{eq:Je},
as the convective velocities $\VEC v_{i}$ and the osmotic velocities
$\VEC u_{i}$ typically differ by many orders of magnitude. In other
words, the probability current $\VEC J_{\mathrm{tot}}$ is \textit{always}
dominated by the quantum mechanical \textit{entangling current} $\VEC J_{\mathrm{e}}$~\eqref{eq:Je}
which is connected to the osmotic velocities, $\VEC u_{1}$ and $\VEC u_{2}$,
and implies the existence of strong correlations. As we have just
seen, this entangling current can also be understood as describing
the heat flow between the two channels. As opposed to the average
total probability current $\VEC J_{\mathrm{tot}}$, in the distribution
of the probability density $P_{\mathrm{tot}}$~\eqref{eq:Ptot2slit}
alone the entangling part is not explicitly visible.

The phenomenon of entanglement is thus possibly rooted in the existence
of the path excitation field. In other words, one can say that the
entanglement characteristic for two-particle interferometry is a natural
consequence of the fact demonstrated here, i.e.\ that already in
single-particle interferometry one deals with entangling currents,
which generally are of a nonlocal nature.

\section{Multi-slit interference and the quantum Talbot effect\label{sec:quantum-talbot-effect}}

We can already infer from the three-slit device that due to the pairwise
selection of the velocity field components $\VEC v_{i}$ and $\VEC u_{i}$,
$i=1,\ldots,N$, $N$ being the number of slits, the interference
effect of every higher order grating can be reduced to successive
double-slit algorithms (cf.~\cite{Fussy.2014multislit}). For a compact
description of the $\mathit{N}$-slit case we return to the notation~\eqref{eq:w_i}
of general velocity vectors $\VEC w_{i}$ with
\begin{equation}
\begin{array}{rclrcl}
\VEC w_{1} & := & \VEC v_{\mathrm{1}},\quad & \VEC w_{2} & := & \VEC u_{\mathrm{1}},\\
\VEC w_{3} & := & \VEC v_{\mathrm{2}}, & \VEC w_{4} & := & \VEC u_{\mathrm{2}},\\
 & \vdots &  &  & \vdots\\
\VEC w_{2N-1} & := & \VEC v_{N}, & \VEC w_{2N} & := & \VEC u_{N},
\end{array}\begin{gathered}\end{gathered}
\end{equation}
with $\VEC w_{2i-1}:=\VEC v_{i}$ and $\VEC w_{2i}:=\VEC u_{i}$,
$i=1,\ldots,N$, denoting the convective and osmotic velocities, respectively,
for each slit~$i$. Analogously, we define the amplitudes
\begin{equation}
\begin{array}{rclcl}
\Rw 1 & = & \Rw 2 & = & R_{1},\\
\Rw 3 & = & \Rw 4 & = & R_{2},\\
 & \vdots &  & \vdots\\
R_{\VEC w_{2N-1}} & = & R_{\VEC w_{2N}} & = & R_{N}.
\end{array}\begin{gathered}\end{gathered}
\end{equation}
 According to the Eqs.~\eqref{eq:tot_prob_law} to \eqref{eq:Jtot4},
now with a general number $\mathit{N}$ of slits, the calculation
for the total probability density is straightforward, as only contributions
of the convective velocities are involved. We get
\begin{equation}
\begin{aligned}P_{\mathrm{tot}}^{N} & ={\displaystyle \sum_{i=1}^{2N}}\Pw i={\displaystyle \sum_{i=1}^{2N}}\VEC{\hat{w}}_{i}\Rw i\cdot\sum_{j=1}^{N}\VEC{\hat{v}}_{j}R_{j}=\left(\overset{{\scriptstyle N}}{\underset{{\scriptstyle i=1}}{\sum}}\VEC{\hat{v}}_{i}R_{\VEC v_{i}}\right)^{2}\\
 & ={\displaystyle \sum_{i=1}^{N}}P_{\VEC v_{i}}=\sum_{i=1}^{N}\left(R_{i}^{2}+\sum_{j=i+1}^{N}2R_{i}R_{j}\cos\varphi_{i,j}\right).
\end{aligned}
\begin{gathered}\end{gathered}
\end{equation}
For the generalized current density we obtain 
\begin{equation}
\VEC J_{\mathrm{tot}}^{N}=\sum_{i=1}^{2N}\Jw i=\sum_{i=1}^{2N}\left(\Rw i\VEC w_{i}{\displaystyle \cdot\sum_{j=1}^{N}}\VEC{\hat{v}}_{j}R_{\VEC v_{j}}\right),
\end{equation}
which leads after a short calculation to
\begin{equation}
\VEC J_{\mathrm{tot}}^{N}=\sum_{i=1}^{N}\left(R_{i}^{2}\VEC v_{i}+\overset{}{\sum_{j=i+1}^{N}R_{i}R_{j}\left\{ \vphantom{\sum_{i=1}^{N}}\left(\VEC v_{i}+\VEC v_{j}\right)\cos\varphi_{i,j}+\left(\VEC u_{i}-\VEC u_{j}\right)\sin\varphi_{i,j}\right\} }\right)
\end{equation}
with $\varphi_{i,j}=\sphericalangle(\VEC v_{i},\VEC v_{j})=\sphericalangle(\VEC u_{i},\VEC u_{j})$
as sketched in Fig.~\ref{fig:vektor}.

From these results we can clearly see that the addition of an arbitrary
number of slits represents a simple inductive extension from the double-slit
case as we had stated in the previous
\begin{sidewaysfigure}
\centering{}\includegraphics{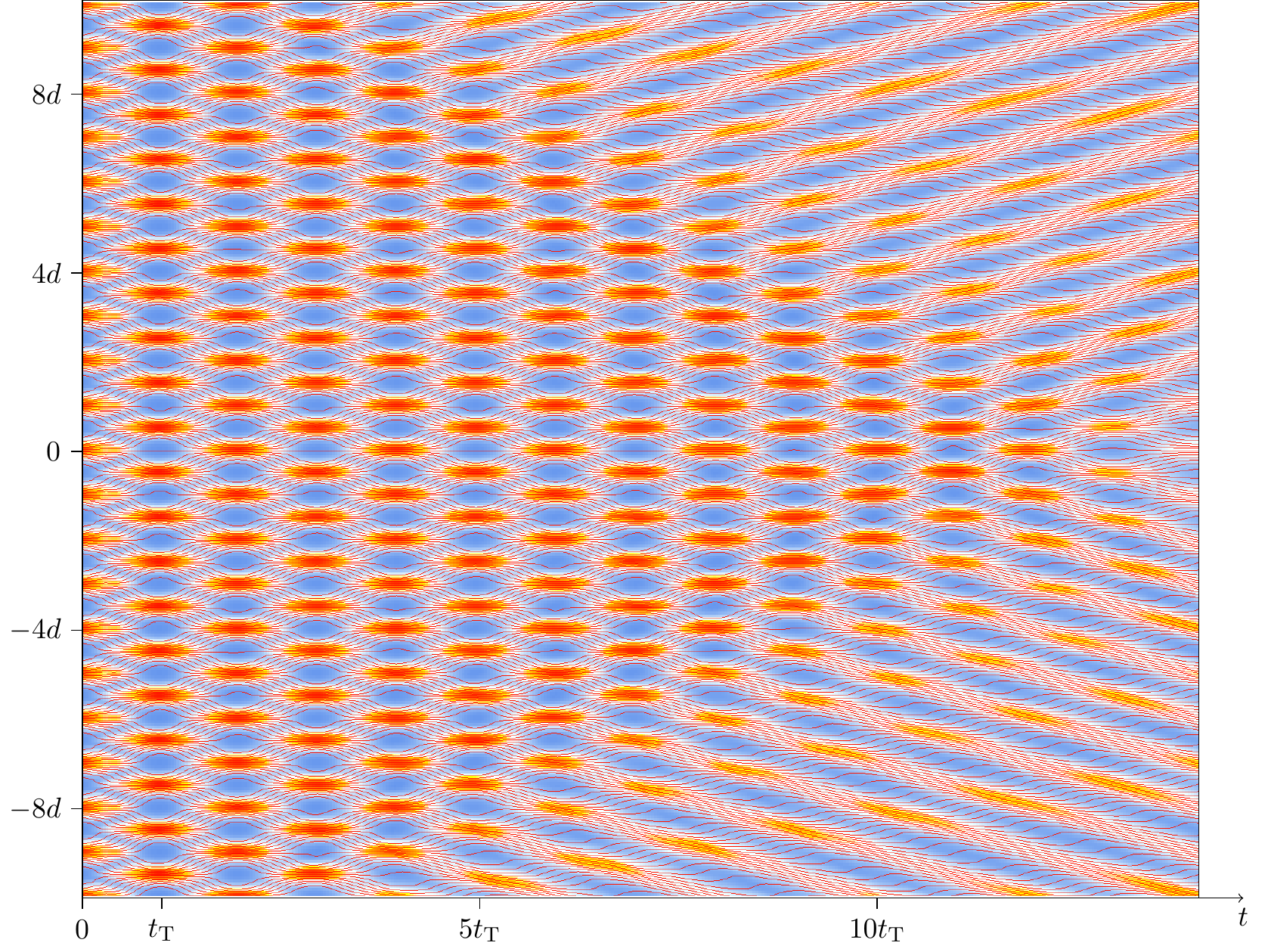}\caption{Intensity distribution via classical computer simulation of the Talbot
carpet for a 27-slit ($d=0.53\,\mathrm{nm}$) setup of Table~\ref{tab:TalbotCarpetParameters},
respectively. Averaged particle trajectories are displayed in red.\label{fig:Born.3.15}}
\end{sidewaysfigure}
 section.
\begin{sidewaysfigure}
\centering{}\includegraphics{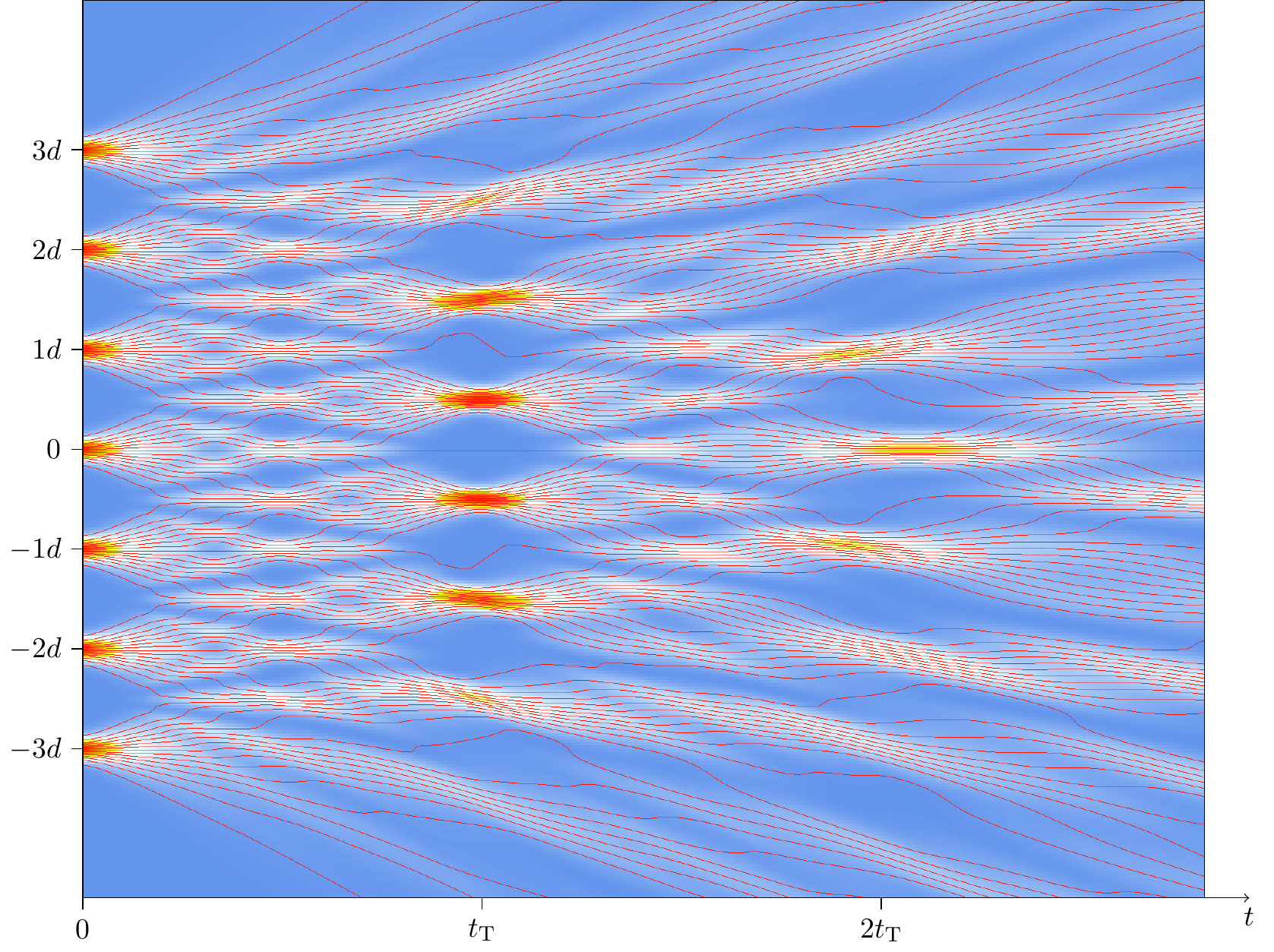}\caption{Intensity distribution via classical computer simulation of the Talbot
carpet for a 7-slit ($d=1.06\,\mathrm{nm}$) setup of Table~\ref{tab:TalbotCarpetParameters},
respectively. Averaged particle trajectories are displayed in red.\label{fig:Born.3.16}}
\end{sidewaysfigure}
\begin{sidewaysfigure}
\centering{}\includegraphics{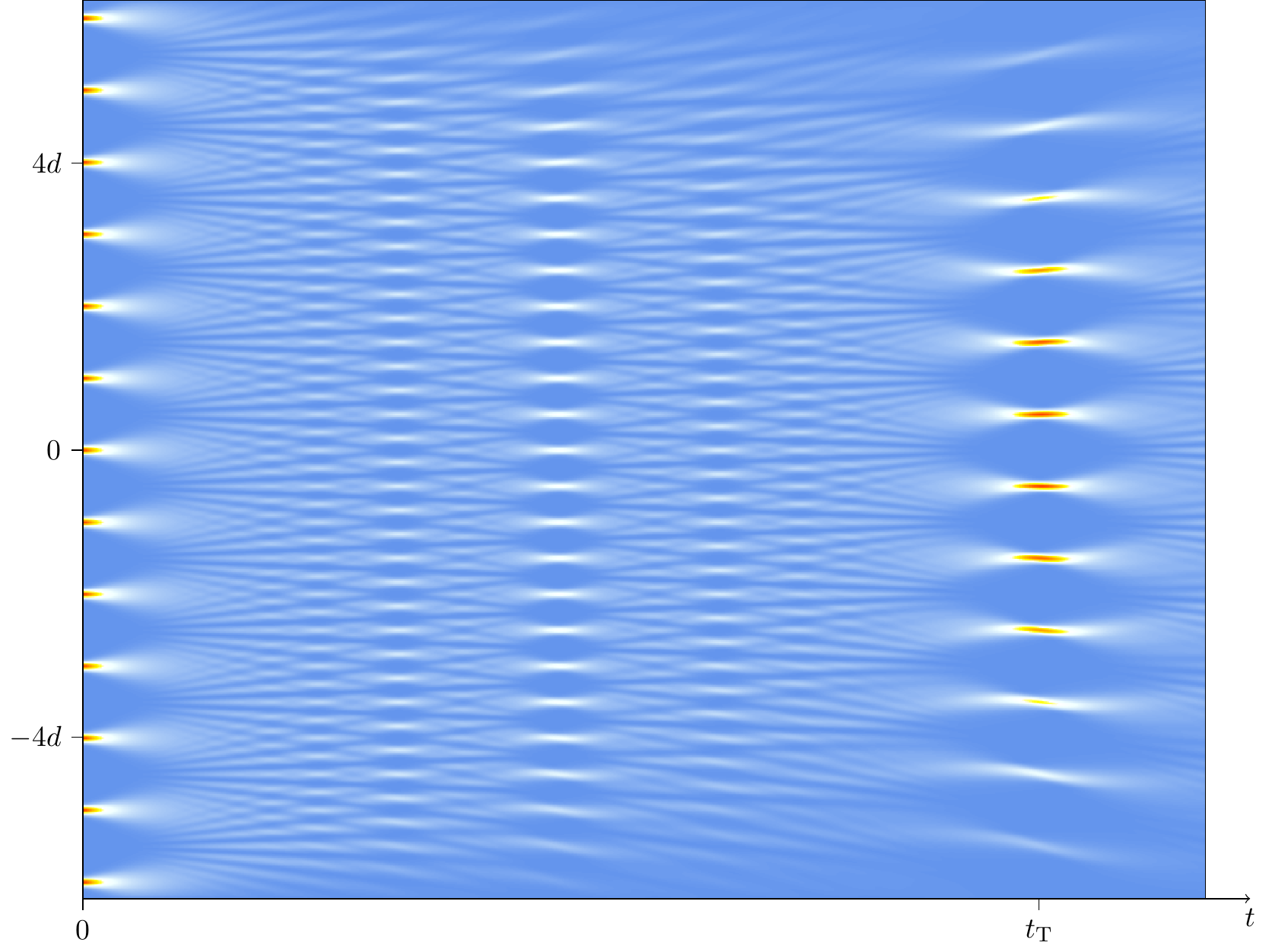}\caption{Intensity distribution via classical computer simulation of the Talbot
carpet for a 13-slit ($d=1.59\,\mathrm{nm}$) setup of Table~\ref{tab:TalbotCarpetParameters},
respectively.\label{fig:Born.3.17}}
\end{sidewaysfigure}
\begin{sidewaysfigure}
\centering{}\includegraphics{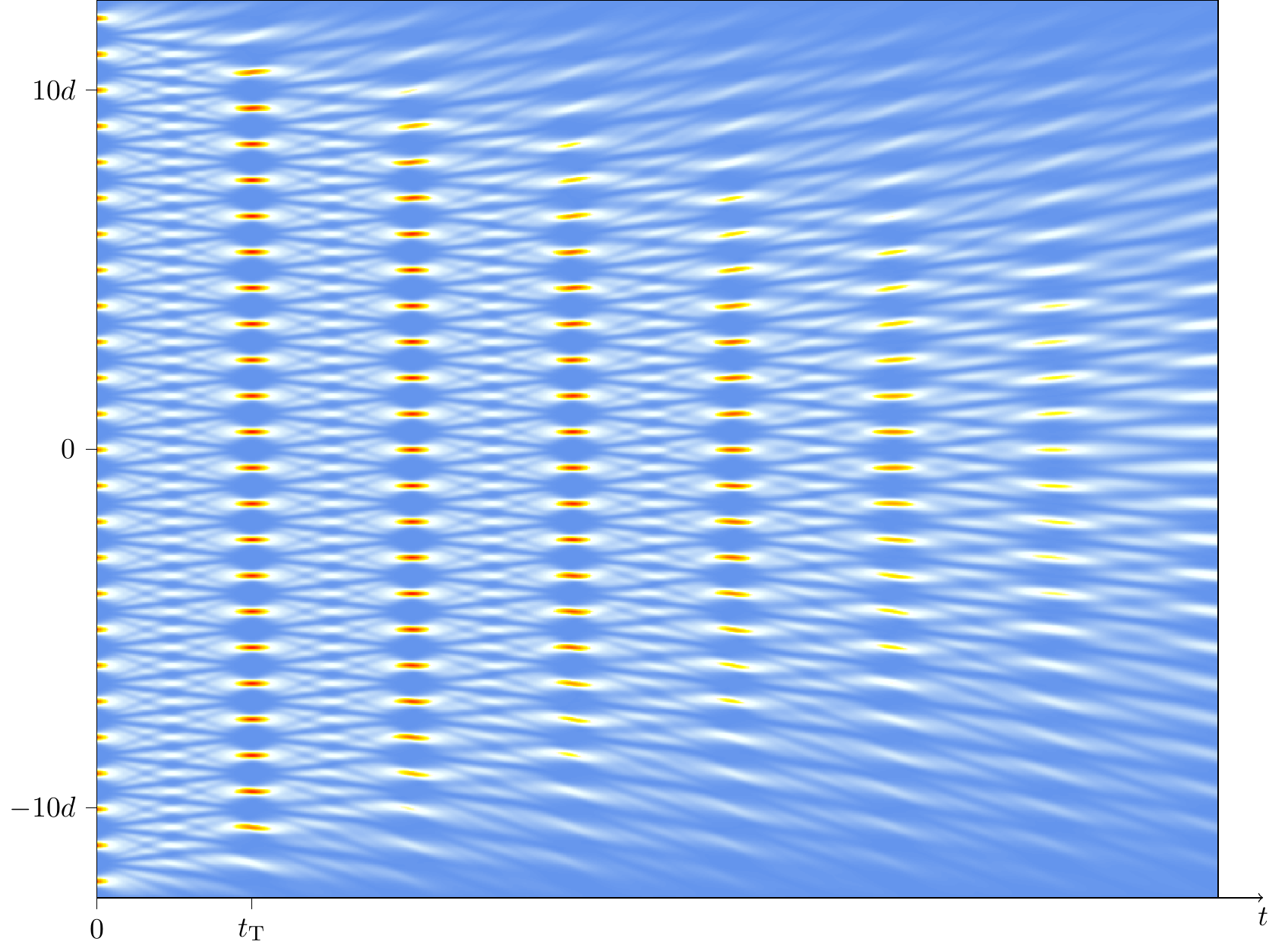}\caption{Intensity distribution via classical computer simulation of the Talbot
carpet for a 25-slit ($d=2.12\,\mathrm{nm}$) setup of Table~\ref{tab:TalbotCarpetParameters},
respectively.\label{fig:Born.3.18}}
\end{sidewaysfigure}

In well-known manner one obtains the trajectories from $\VEC{\dot{x}}_{\mathrm{tot}}=\VEC v_{\mathrm{tot}}=\frac{\VEC J_{\mathrm{tot}}}{P_{\mathrm{tot}}}$
\cite{Sanz.2008trajectory}. As opposed to this analytical procedure,
we use our simulation tools, which are displayed in the computer simulations
of Figs.~\ref{fig:Born.3.15} to \ref{fig:Born.3.18} for 7-, 13-,
25-, and 27-slit setup, respectively. Already for the 7-slit case
one can observe the emergence of a repetitive short range pattern
until the Fraunhofer regime\footnote{The patterns arise in the short range or Fresnel region, gradually
disappear in the transition region and end up in the far-field or
Fraunhofer region, cf.~\cite{Sanz.2012trajectory}.} is reached. At the so-called Talbot distance 
\begin{equation}
z_{\mathrm{T}}=d^{2}/\lambda,\label{eq:talbot}
\end{equation}
where $\mathit{d}$ denotes the grating period and $\lambda$ the
wavelength of the incident plane wave, the initial patterns of the
7 vertically arranged slit openings reappear with a shift of $d/2$.
Table~\ref{tab:TalbotCarpetParameters} shows the results for different
values of $\lambda$ and $d$, compares them with the observed values
$y_{\mathrm{T}}$ of the Talbot distance for various $N$-slit cases. 

To explain these results, we use the parameters for neutrons according
to Table~\ref{tab:TalbotCarpetParameters}: 
\begin{table}
\begin{centering}
\renewcommand*\arraystretch{1.5}%
\begin{tabular}{l|llll}
Setup & Fig.~\ref{fig:Born.3.15} & Fig.~\ref{fig:Born.3.16} & Fig.~\ref{fig:Born.3.17} & Fig.~\ref{fig:Born.3.18}\tabularnewline
\hline 
$\lambda$ & 1 nm & 1 nm & 1 nm & 1 nm\tabularnewline
$d$ & 0.53 nm & 1.06 nm & 1.59 nm & 2.12 nm\tabularnewline
$z_{\mathrm{T}}$ & 0.28 nm & 1.13 nm & 2.53 nm & 4.5 nm\tabularnewline
$y_{\mathrm{T,7-slit}}$ & 0.28 nm & 1.14 nm & 2.53 nm & 4.52 nm\tabularnewline
$y_{\mathrm{T,}N\mathrm{-slit}}$ & 0.29 nm ($N=27$) & 1.13 nm ($27)$ & 2.53 nm ($25$) & 4.49 nm ($13)$\tabularnewline
\end{tabular}
\par\end{centering}
\vspace{5mm}

\caption{Parameters for the Talbot carpet simulations\label{tab:TalbotCarpetParameters}}
\end{table}
$d=1.06\,\mathrm{nm}$, $\lambda=1\,\mathrm{nm}$, with mass $m_{\mathrm{n}}=1.675\cdot10^{-27}\,\mathrm{kg}$.
The spatial step width is chosen as $\Delta x=0.0378\,\mathrm{nm}$,
the time resolution is set to $\Delta t=1.92\cdot10^{-14}\,\mathrm{s}$.
Then, said shifted reappearance of the pattern occurs for the first
time at time step 150, i.e.\ at $t_{\mathrm{T}}=150\cdot\Delta t=2.88\cdot10^{-12}\,\mathrm{s}$.
The standard transformation into the two-dimensional case by re-parametrizing
the $t$-axis according to Eq.~\eqref{eq:y=00003Dv_yt}, $y=\hbar k_{\mathrm{n}}\Delta t/m_{\mathrm{n}}=h\Delta t/(\lambda m_{\mathrm{n}})$,
leads to the observed distance $y_{\mathrm{T}}=ht_{\mathrm{T}}/(\lambda m_{\mathrm{n}})=1.14\,\mathrm{nm}$,
which matches nicely with the formula of the Talbot distance $z_{\mathrm{T}}$~\eqref{eq:talbot}.
The observed values for the Talbot distance $y_{\mathrm{T}}$ in our
discretised model agree for any $N$-slit setup as expected in accordance
with Eq.~\eqref{eq:talbot}, which only depends on $d$ and $\lambda$.
Moreover, we also obtain the correct results for any other choice
of $m$ or $\lambda$.

For multiples of $2z_{\mathrm{T}}$ the recurrence of the original
state is observed, as it is particularly obvious in the case of 27
slits. Due to the non-crossing of all trajectories, as has been discussed
in section~\ref{sec:Double-slit-interference}, the caverns in the
middle stay confined until they are broken up by the influence of
the boundary area via the light-like cone. In the limit of an indefinitely
extended grating the pattern clearly would be maintained \textit{ad
infinitum}. 

Since the averaged trajectories obtained with our derived current
set are identified with the Bohmian trajectories of Sanz \textit{et\,al.}~\cite{Sanz.2007causal},
we have thus shown that the emerging quantum carpet for $\mathit{N}$
slits constituted by characteristic repetitive patterns can be reproduced
without any quantum mechanical state function.

\section{Conclusions and outlook\label{sec:conclusion-current-based}}

It has also been shown how our model entails the existence of a path
excitation field, i.e.\ a field spanned by the average velocity fields
$\VEC v$ and $\VEC u$, respectively, where the latter refers to
diffusion processes reflecting also the stochastic parts of the zero-point
field. Then, on the basis of classical physics, the exact intensity
distribution on a screen behind a double-slit has been derived, as
well as the details of the more complicated particle current, or of
the Bohmian particle trajectories, respectively.

Furthermore, general formulas for the $N$-slit current densities
have been derived, thus enabling us to give a micro-causal account
for the kinematics of the quantum Talbot effect. The Talbot distance
could be reproduced also quantitatively in this model.

\chapter{Beam attenuation in double-slit experiments\label{sec:6.extreme-beam-attenuation}}
\begin{quote}
In this chapter we shall employ a double-slit setup with one slit's
probability density being attenuated by a huge factor. Therefore,
we start with a survey on different absorber types used in interference
experiments and discuss the resulting consequences of using these.

In a phenomenological approach we shall study the probability distribution
of said double-slit and show the emergence of a lateral drift of the
interference zone due to increasing attenuation factors applied to
one of the slits. This drift phenomenon, the \textit{quantum sweeper
effect,} will be compared to both coherent and incoherent beams and
shown to be existing in either case.

As a result of our investigations we shall propose an advanced measurement
method comprising a side-screen which is oriented along the spreading
direction, i.e.\ the side-screen turned by an angle of $90^{\circ}$
compared to its usual position.
\end{quote}

\section{Outline}

In the search of new, and perhaps surprising, features of quantum
systems, one option is to steadily decrease the intensity of a slit
into one spatially constrained area, as compared to a reference intensity
in another, equally constrained area. For example, one can employ
the usual double-slit experiments and modify one of the two slits'
channels in such a way that the corresponding outgoing probability
density is very low compared to that of the other slit. We call a
combination of such distributions of high and low probability densities,
or intensities, respectively,\emph{ intensity hybrids} (cf.~\cite{Groessing.2014attenuation,Mesa.2016emqm,Groessing.2016emqm15-book,Groessing.2015dice}).

Since the 1980ies, one possibility to experimentally establish and
probe such hybrids has been through the introduction of beam attenuation
techniques, as demonstrated in the well-known papers by Rauch's group
in neutron interferometry~\cite{Rauch.1984static,Rauch.1990low-contrast}.
Here, we re-visit these experiments and results from a new perspective,
and we also discuss new, previously unexpected effects. Our main result
is that in employing ever weaker channel intensities, nonlinear effects
become ever more important, which are a crucial characteristic of
sub-quantum models such as the one developed by our group. Whereas
the intensity distributions are predicted to be the same for the standard
quantum mechanical as well as our  approach, respectively, more detailed
information is available when the behaviour of average trajectories
is studied.

\section{Deterministic and stochastic beam attenuation\label{sec:BeamAttenuation}}

\subsection{Beam attenuation in neutron interferometry\label{subsec:2.1 neutron interferometry}}

Deterministic and stochastic beam attenuation have been studied extensively
in neutron interferometry, beginning with the work by Rauch and Summhammer
in 1984~\cite{Rauch.1984static}. More recently, an interesting model
of these results has been presented by De\,Raedt\emph{ et\,al}.~\cite{DeRaedt.2012discrete-event}
with the aid of event-by-event simulations, thus confirming the possibility
to describe the known results even without the use of quantum mechanics. 

\begin{figure}
\begin{centering}
\includegraphics[width=1\textwidth]{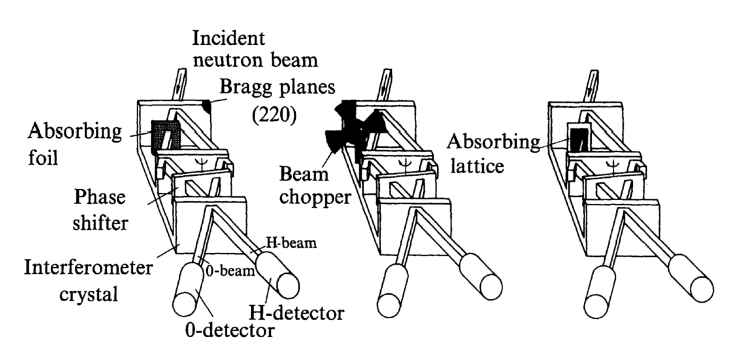}
\par\end{centering}
\caption{Stochastic type absorber (left and right) and deterministic type absorber
(middle). From~\cite{Summhammer.1987stochastic}\label{fig:interferometers}}
\end{figure}

In~\cite{Rauch.1984static,Rauch.1990low-contrast}, a beam chopper
(Fig.~\ref{fig:interferometers})
was used as a deterministic absorber in one arm of a two-armed interferometer,
whereas for stochastic absorption semitransparent foils of various
materials were used. Despite the net effect of the same percentage
of neutrons being attenuated, the quantum mechanical formalism predicts
the following different behaviours for the two cases. Introducing
the \emph{transmission factor} $a$ as the beam's transmission probability,
in the case of a (deterministic) chopper wheel it is given by the
temporal open-to-closed ratio, 
\begin{equation}
a=\frac{t_{\mathrm{open}}}{t_{\mathrm{open}}+t_{\mathrm{closed}}}\thinspace,\label{eq:ratio-deterministic}
\end{equation}
whereas for a (stochastic) semitransparent material defined by its
absorption cross section, it is simply the relation of the intensity
$I$ with absorption compared to the intensity $I_{0}$ without, i.e.\ 
\begin{equation}
a=\frac{I}{I_{0}}.\label{eq:ratio-stochastic}
\end{equation}
In a quantum mechanical description the beam modulation behind the
interferometer is obtained in the following two forms. For the deterministic
chopper system the intensity~$I_{\mathrm{det}}$ is, with $\varphi$
denoting the phase difference, given by\footnote{The quantum mechanical wave function $\psi_{j}$, for slits $j=1$
or $2$, is connected with the probability density $P_{j}$ and the
amplitude $R_{j}$ by
\[
P_{j}=R_{j}^{2}=|\psi_{j}|^{2}=\psi_{j}^{*}\psi_{j}
\]
}~\cite{Rauch.1984static}
\begin{align}
I_{\mathrm{det}} & \propto\left(1-a\right)\left|\psi_{1}\right|^{2}+a\left|\psi_{1}+\psi_{2}\right|^{2}\propto1-a+a\left|1+\e^{\i\varphi}\right|^{2}=1+a+2a\cos\varphi,\label{eq:sw.1}
\end{align}
whereas for stochastic beam attenuation with the semitransparent material
the intensity~$I_{\mathrm{sto}}$ is
\begin{equation}
I_{\mathrm{sto}}\propto\left|\psi_{1}+\psi_{2}\right|^{2}\propto\propto1+a+2\sqrt{a}\cos\varphi.\label{eq:sw.2}
\end{equation}
Although the same number of neutrons is observed in both cases, in
the first one the contrast of the interference pattern is proportional
to $a$, whereas in the second case it is proportional to $\sqrt{a}$.

In our accounting for the just described attenuation effects, we choose
the usual double-slit scenario, primarily because this will be very
useful later on when discussing more extreme intensity hybrids.

\subsection{Application to deterministic and stochastic beam attenuation experiments}

Let us now display some typical results from our double-slit approach,
as presented in chapter~\ref{sec:Interference-effects}, to beam
attenuation. We can simulate the propagation of a Gaussian whose variance
increases due to the ballistic diffusion process (see chapter~\ref{sec:3.3.The-derivation-of-Dt}).
To begin with, we consider deterministic attenuation first. Therefore,
we use the ratio $a$~\eqref{eq:ratio-deterministic} and simulate
indirectly as a combination of 
\begin{enumerate}
\item a single-slit experiment resulting in distribution $P_{\mathrm{single}}=P_{1}=R_{1}^{2}$,
according to Eq.~\eqref{eq:R=00003Dsqrt_P}, as slit~2 is closed
during time $t_{\mathrm{closed}}$, and 
\item a double-slit experiment resulting in $P_{\mathrm{double}}=P_{\mathrm{tot}}$~\eqref{eq:Ptot2slit}
with both slits are opened, both beams having equal intensities, during
time $t_{\mathrm{open}}$. 
\end{enumerate}
As the ratio of the two intensities is set to $P_{1}=P_{2}$ the resulting
distribution after incoherent summing up reads
\begin{align}
P_{\mathrm{det}} & =(1-a)P_{\mathrm{single}}+aP_{\mathrm{double}}\nonumber \\
 & =(1-a)P_{1}+a(P_{1}+P_{1}+2\sqrt{P_{1}P_{1}}\cos\varphi)\nonumber \\
 & =P_{1}+aP_{1}+2aP_{1}\cos\varphi\nonumber \\
 & =P_{1}(1+a+2a\cos\varphi).\label{eq:P_DET}
\end{align}
Accordingly, we have in complete agreement with Eq.~\eqref{eq:sw.1}
that
\begin{equation}
I\propto1+a+2a\cos\varphi.
\end{equation}

For stochastic attenuation we find with the intensity ratio $a$~\eqref{eq:ratio-stochastic},
i.e.\ $P_{2}=aP_{1}$, thus with the amplitude of the attenuated
slit~2, and according to Eq.~\eqref{eq:Ptot2slit} that
\begin{align}
P_{\mathrm{sto}} & =P_{1}+P_{2}+2\sqrt{P_{1}P_{2}}\cos\varphi\nonumber \\
 & =P_{1}+aP_{1}+2\sqrt{aP_{1}P_{1}}\cos\varphi\nonumber \\
 & =P_{1}(1+a+2\sqrt{a}\cos\varphi).\label{eq:P_STO}
\end{align}
Again, we have complete agreement with Eq.~\eqref{eq:sw.2}, i.e.\ 
\begin{equation}
I\propto1+a+2\sqrt{a}\cos\varphi.\label{eq:sw.14}
\end{equation}

\begin{figure}[!htb]
\subfloat[$a=0.25$\label{fig:sw.1a}]{\centering{}\includegraphics[width=0.5\textwidth]{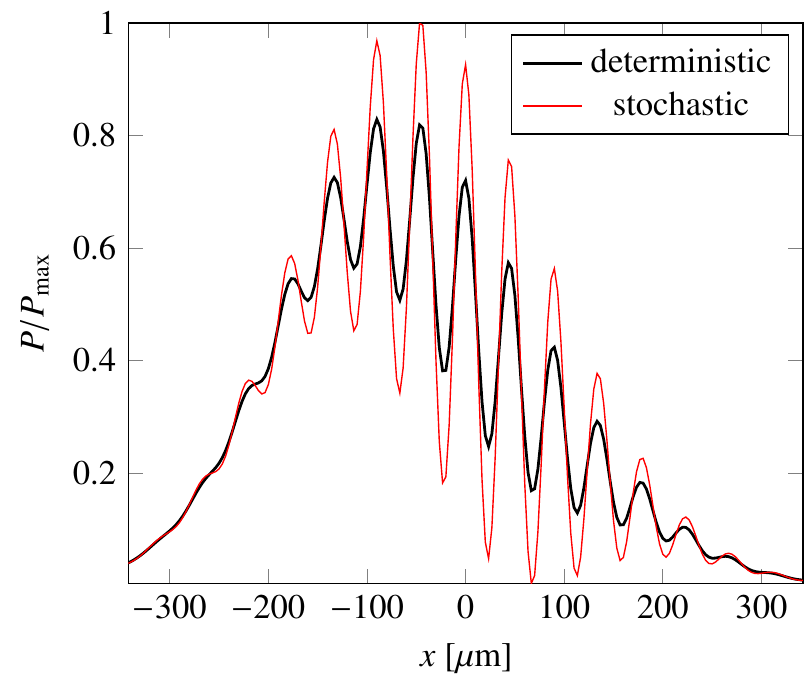}}\subfloat[$a=0.025$\label{fig:sw.1b}]{\centering{}\includegraphics[width=0.5\textwidth]{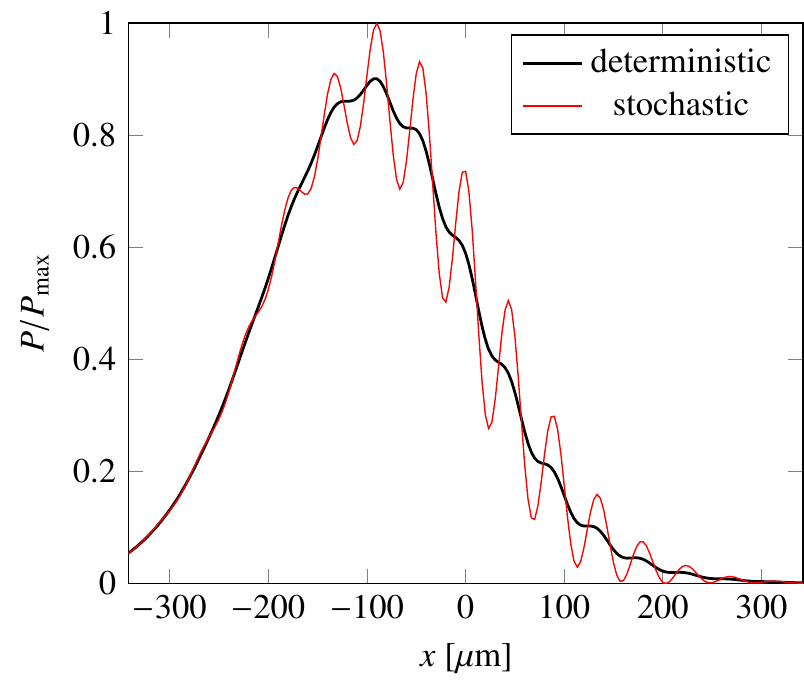}}
\begin{centering}
\subfloat[$a=0.0025$\label{fig:sw.1c}]{\centering{}\includegraphics[width=0.5\textwidth]{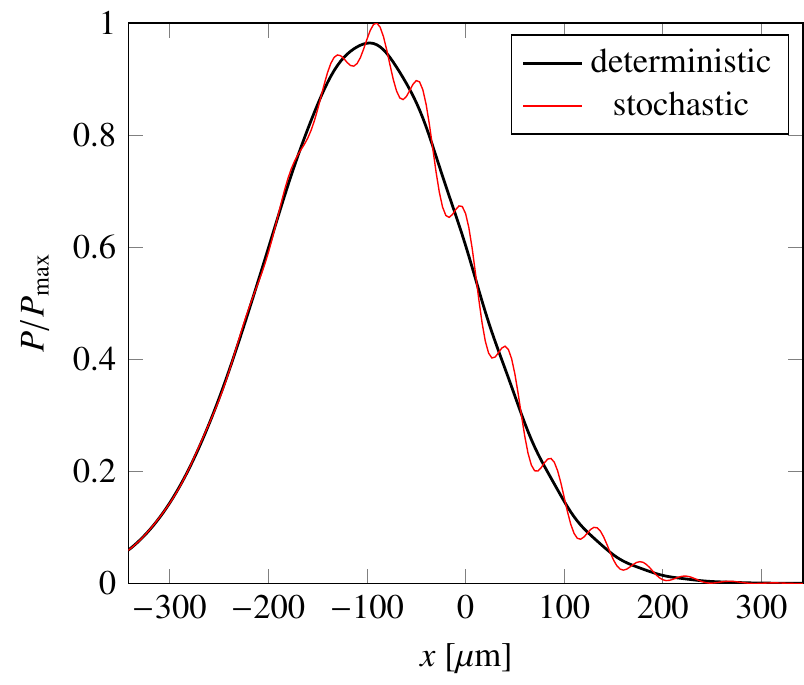}}
\par\end{centering}
\caption{Simulation of probability density distributions with beam attenuation
$a$ at slit~2, in complete accordance with standard quantum mechanics.\label{fig:sw.1}}
\end{figure}

In Fig.~\ref{fig:sw.1} we show the results of our computer simulations
following Eqs.~\eqref{eq:P_DET} and \eqref{eq:P_STO}, respectively,
for the probability density distributions of a neutron beam for three
different values of the beam transmission factor $a$. The typical
wavelength used is $\lambda=1.8\thinspace\mathrm{nm}$ (cf.~\cite{Rauch.2000neutron}).
The Gaussian slits each are $22\thinspace\mu\mathrm{m}$ wide, with
their centres being $200\thinspace\mu\mbox{m}$ apart, and the intensity
distributions are recorded on a screen located in the forward direction
at a distance of $5\thinspace\mbox{m}$ from the double-slit. Corresponding
to the different behaviours of the contrast in deterministic and stochastic
attenuation, respectively, one can see the different contributions
to the overall probability density distribution, with the differences
becoming smaller and smaller with ever decreasing transmission factor
$a$. For consistency, we have also checked and confirmed that the
total areas below the two curves are identical, as they must be in
order to represent the same overall throughput of the number of neutrons.

\section{Phenomenology of the quantum sweeper for coherent and incoherent
beams\label{sec:quantum sweeper}}

We assume a coherent beam in a double-slit experiment, with the intensity
distribution being recorded on a screen, and we are going to discuss
a particular effect of the stochastic attenuation of one of the two
emerging Gaussians at very small transmission factors. With the appropriate
filtering of the particles going through one of the two slits, the
recorded probability density in the surroundings of the experiment
will appear differently compared to what one would normally expect.
That is, if one had a low beam intensity coming from one slit, one
would expect that the contributions from the fully open slit would
become dominant until such a low counting rate from the attenuated
slit is arrived at that essentially one would have a one-slit distribution
of recorded particles on the screen. This tendency is at least clearly
visible in Fig.~\ref{fig:sw.1}. One would thus expect for ever smaller
values of $a$ that the oscillatory behaviour of the stochastic case
would more and more disappear to finally merge with the smoothed-out
pattern of an essentially one-slit distribution pattern, and that
no other effects would be observed.

\begin{figure*}[!t]
\begin{centering}
\subfloat[$a=10^{-1}$\label{fig:sw.2a}]{\begin{centering}
\includegraphics[width=0.49\textwidth]{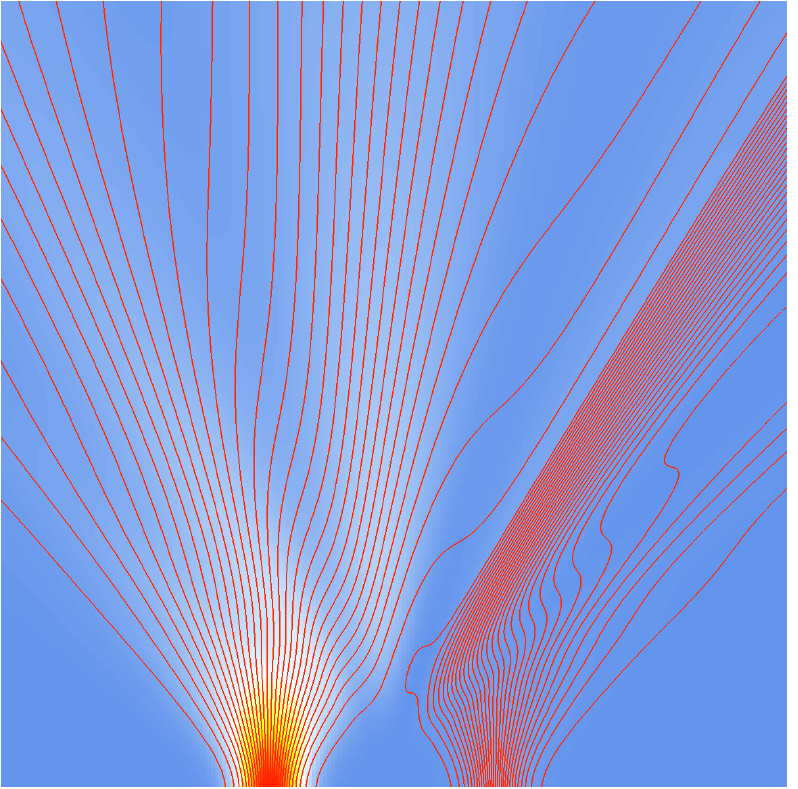} 
\par\end{centering}
}\subfloat[$a=10^{-2}$\label{fig:sw.2b}]{\begin{centering}
\includegraphics[width=0.49\textwidth]{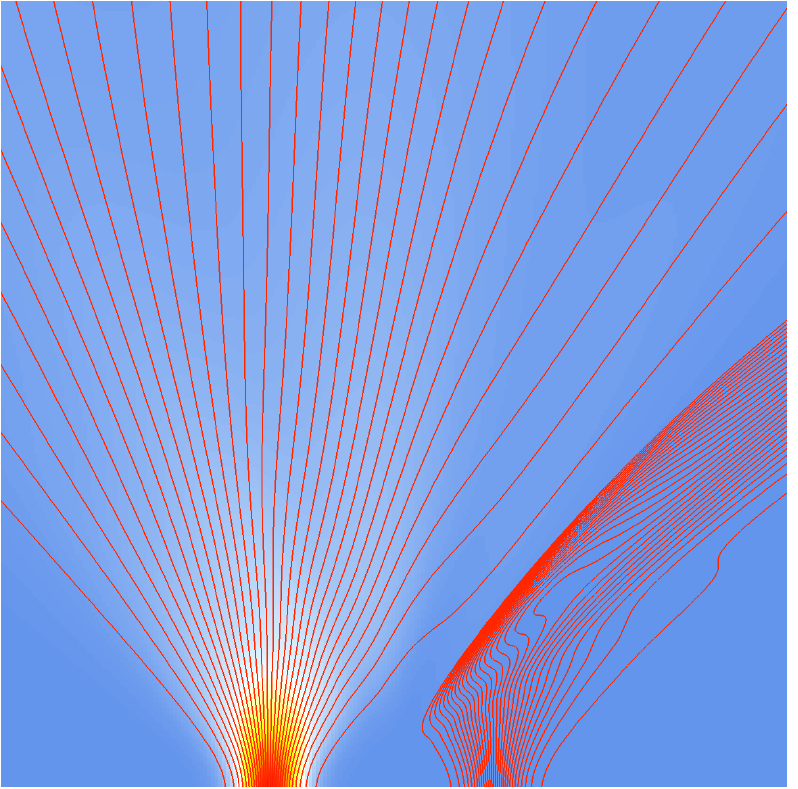}
\par\end{centering}
}
\par\end{centering}
\begin{centering}
\subfloat[$a=10^{-4}$\label{fig:sw.2c}]{\begin{centering}
\includegraphics[width=0.49\textwidth]{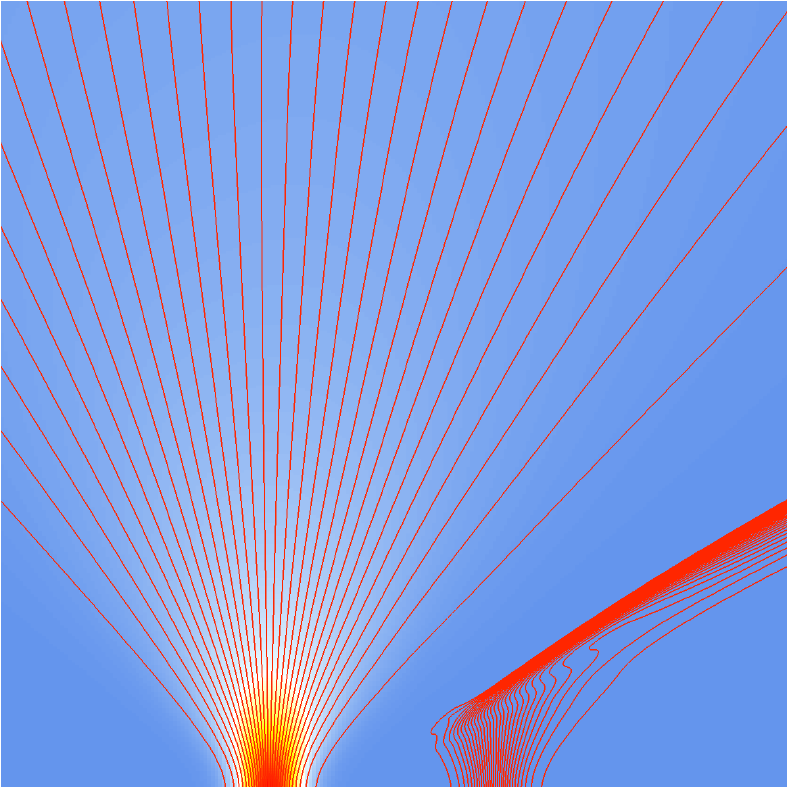}
\par\end{centering}
}\subfloat[$a=10^{-10}$\label{fig:sw.2d}]{\begin{centering}
\includegraphics[width=0.49\textwidth]{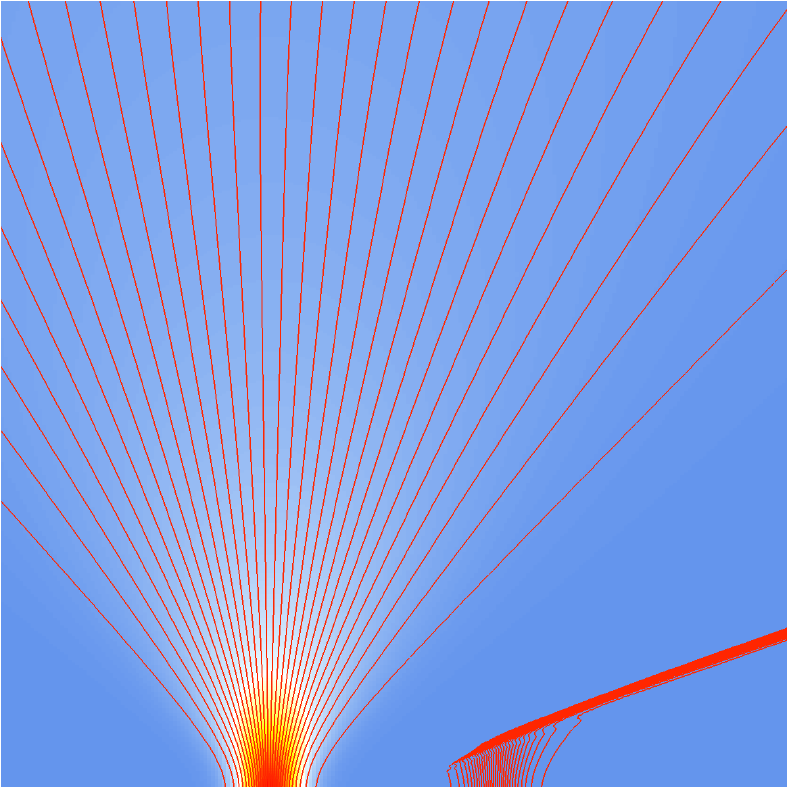}
\par\end{centering}
}
\par\end{centering}
\caption{The quantum sweeper effect for different transmission factors $a$.
To demonstrate the effect more clearly, we use the same number of
trajectories for each slit.\label{fig:sw.2}}
\end{figure*}

Interestingly, this is not exactly what one obtains at least for very
low values of $a$ when going through the calculations and computer
simulations\footnote{See chapter~\ref{chap:Numerical-methods} on how simulations have
been practically realized.} with our bouncer model. The latter encompasses, among other features,
an explicit form of the velocity field $\VEC v_{\mathrm{tot}}$~\eqref{eq:vtot}
emerging from the double slit, as well as of the probability current
$\VEC J_{\mathrm{tot}}$~\eqref{eq:Jfinal} associated with it. Whereas
full agreement exists with the standard quantum mechanical prediction
of the probability density $P_{\mathrm{tot}}$, viz.\ Eqs.~\eqref{eq:P_DET}
and \eqref{eq:P_STO}, respectively, the probability current $\VEC J_{\mathrm{tot}}$
exhibits an unexpected behaviour, which we are going to discuss now.

Fig.~\ref{fig:sw.2} shows the quantum sweeper effect: A series of
probability density distributions plus averaged trajectories for the
case that the intensity in slit~2 is gradually diminished. We use
the same model as described in section~\ref{sec:3.4.Spreading-of-the-wave-packet}:
Wave packets, represented by plane waves in the forward $y$-direction,
from a coherent source passing through soft-edged slits in a barrier,
located along the $x$-axis, and recorded at a screen in the forward
direction, i.e.\ parallel to the barrier. This situation is described
by two Gaussians representing the totality of the effectively heated-up
path excitation field, one for slit~1 and one for slit~2, whose
centres have the same distances from the plane spanned by the source
and the centre of the barrier along the $y$-axis, respectively (see
Fig.~\ref{fig:vektor}). 

Now, with ever lower values of the transmission factor $a$ during
beam attenuation, one can see a steadily growing tendency for the
low counting rate particles of the attenuated beam to become swept
aside. In our model, this is straightforward to understand, because
we have the analytical tools to differentiate between the forward
convective $\VEC v_{i}$~\eqref{eq:diss.4.40} and the osmotic influences
of velocities $\VEC u_{i}$~\eqref{eq:diss.4.41}, as distinguishable
contributions from the different slits~$i$. Thus, it is processes
of diffusion which are seen in operation here, due to the presence
of accumulated heat, i.e.\ kinetic energy, primarily in the strong
beam. So, in effect, we understand Fig.~\ref{fig:sw.2} as the result
of the vacuum heat sweeping aside the very low intensity beam, with
a no-crossing line\footnote{From the assumed uniqueness and differentiability of $S(\VEC x,t)$
follows that the paths don't cross each other. See section~\ref{sec:orthogonality}
for further explanations. However, at this stage we are discussing
an ontological point of view on how the no-crossing phenomenon can
be explained.} defined by the balancing out of the diffusive momenta, $m\left(\VEC u_{1}+\VEC u_{2}\right)=0$. 

Importantly, for certain slit configurations and sizes of the transmission
factor, the sweeper effect leads to a bunching of trajectories which
may become deflected into a direction almost orthogonal to the original
forward direction. In other words, one would need much wider screens
in the forward direction to register them, albeit then weakened due
to a long travelling distance. On the other hand, if one installed
a screen orthogonal to the forward screen, i.e.\ one that is parallel
to the original forward motion \textendash{} and thus to the $y$-axis
\textendash{} one could significantly improve the contrast and thus
register the effect more clearly (see also Fig.~\ref{fig:sw.4} further
below). Further, we note that changing the distance between the two
slits does not alter the effect, but demonstrates the bunching of
the low counting rate arrivals in essentially the same narrow spatial
area even more drastically. So, again, if one places a screen orthogonally
to the forward direction, one registers an increased local density
of particle arrivals in a narrow spatial area under an angle that
is independent of the slit distance.

\begin{figure*}
\subfloat[$a=1$\label{fig:sw.3a}]{\centering{}\includegraphics[width=0.49\textwidth]{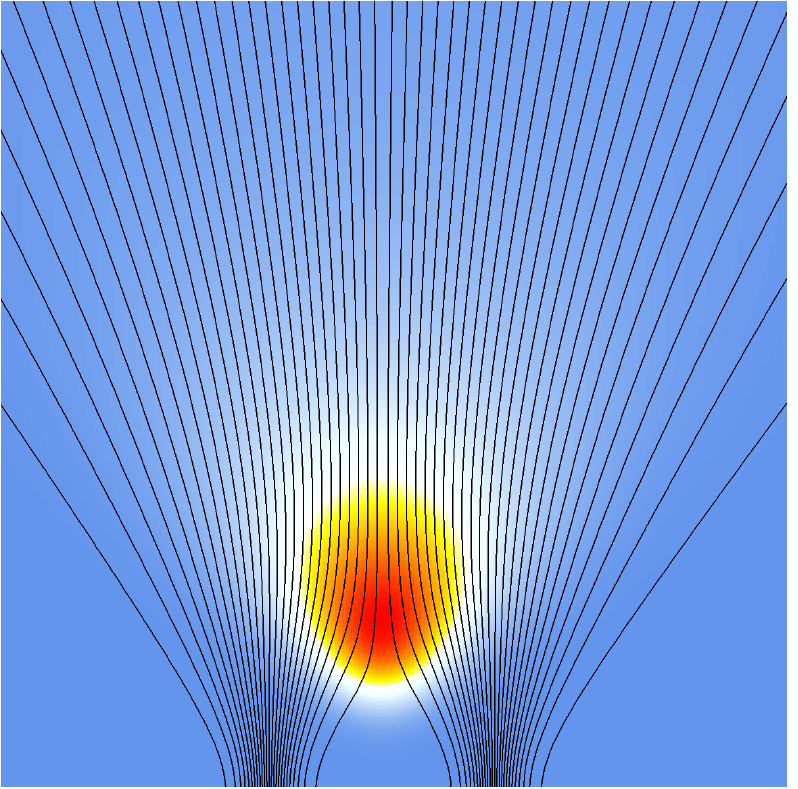}}\hfill{}\subfloat[$a=10^{-8}$\label{fig:sw.3b}]{\centering{}\includegraphics[width=0.49\textwidth]{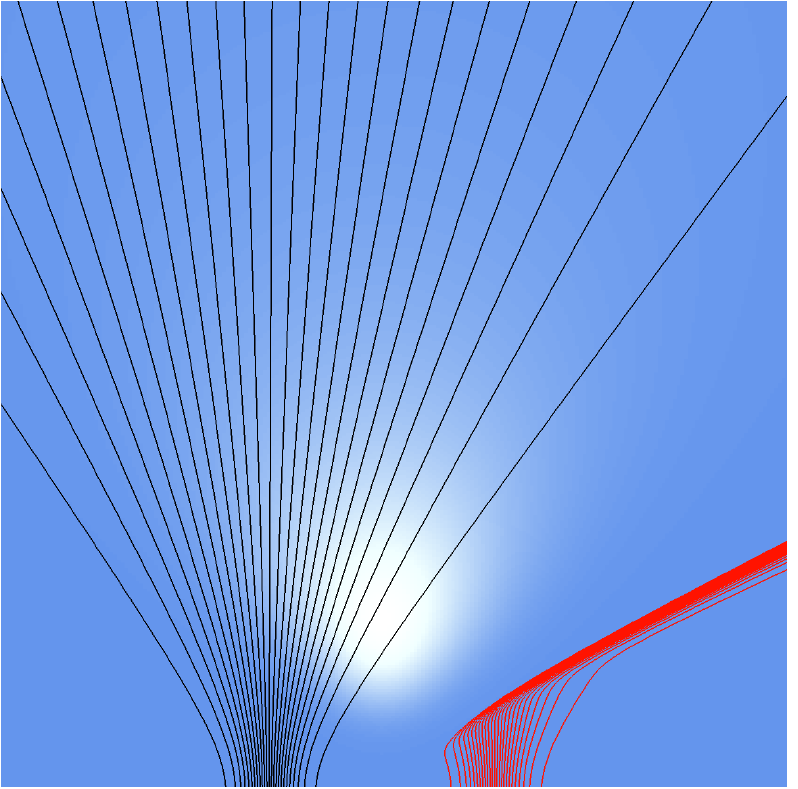}}

\caption{Double-slit experiment with completely incoherent channels. The right
hand side beam is weakened by factor $a$.\label{fig:sw.3}}
\end{figure*}

\begin{sidewaysfigure*}
\begin{centering}
\includegraphics[width=0.85\textwidth]{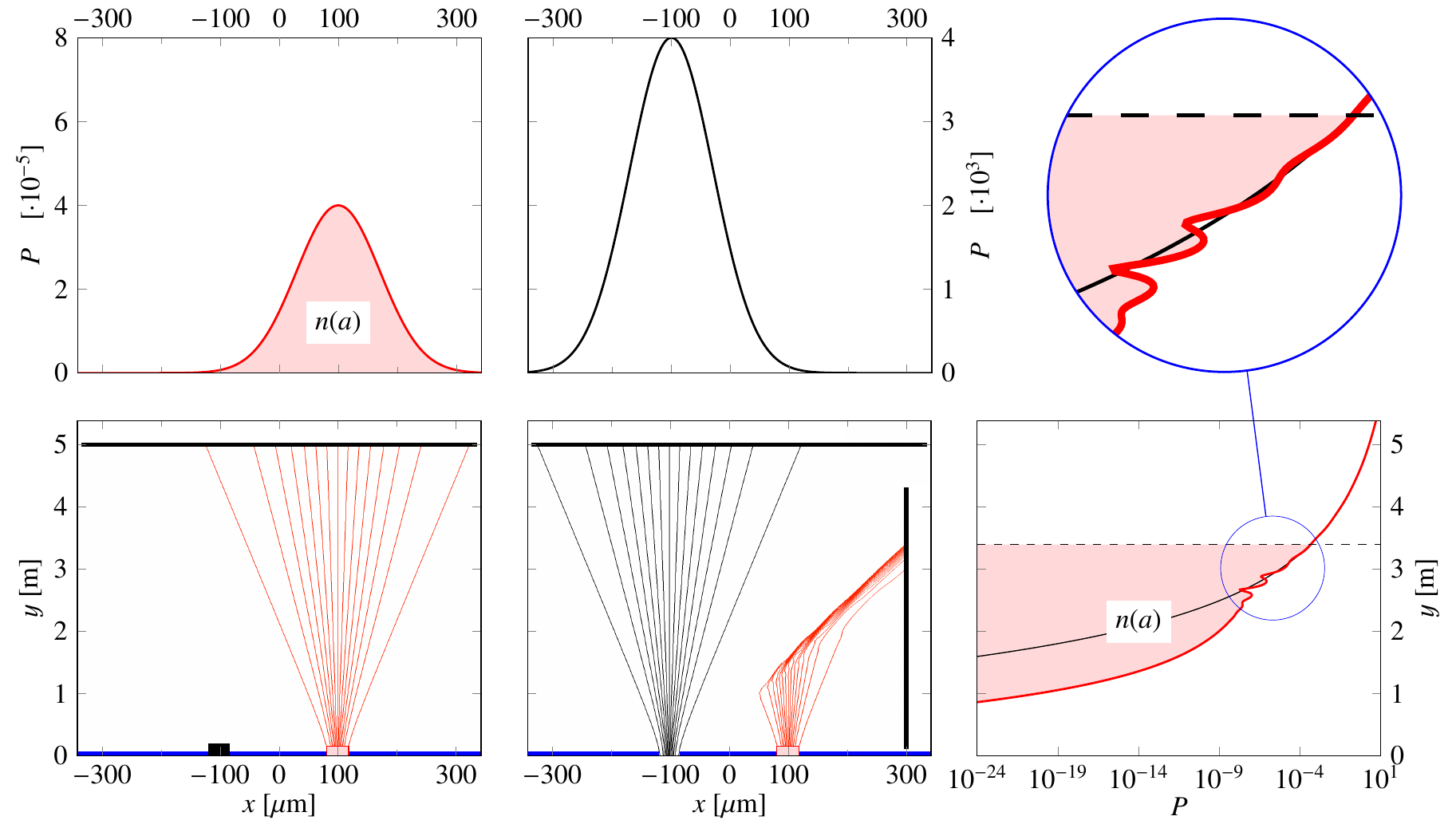}
\par\end{centering}
\begin{centering}
\hfill{}\subfloat[\label{fig:sw.4a}]{}\hfill{}\subfloat[\label{fig:sw.4b}]{}\hfill{}\subfloat[\label{fig:sw.4c}]{}\hfill{}
\par\end{centering}
\caption{Registration of particles during extreme beam attenuation, $a=10^{-8}$.
(a) If slit~1 is closed, a small number $n\left(a\right)$ of particles
coming from slit~2 is registered on the forward screen. (b) If, then,
slit~1 is fully opened (i.e.\ with $a=1$), one registers a much
higher number of particles for slit~1, but apparently none for slit~2.
Instead, $n\left(a\right)$ particles are then registered on the sideways
screen parallel to the $y$-axis. (c) The probability density distribution
for the latter exhibits marked signs of interference effects due to
the compressed wave superpositions within the bunching area caused
by the sweeper effect.\label{fig:sw.4}}
\end{sidewaysfigure*}

Let us now turn to the case of incoherent beams. For, although we
shall refrain from constructing a concrete model of incoherence and
implementing it in our schema, we already have the tools of an effective
theory, i.e.\ to describe incoherence without the need of a specified
mechanism for it. Namely, as full incoherence between two (Gaussian
or other) beams is characterized by the complete absence of the interference
term in the overall probability distribution of the system, this means
that $P_{\mathrm{tot}}=R_{1}^{2}+R_{2}^{2}$, since the interference
term
\begin{equation}
R_{1}R_{2}\left(\VEC v_{\mathrm{1}}+\VEC v_{2}\right)\cos\varphi=0\label{eq:sw.15}
\end{equation}
vanishes. For the case $\cos\varphi=0$, i.e.\ with $\varphi=\frac{\pi}{2}$,
Eq.~\eqref{eq:sw.15} vanishes which effectively describes the situation
of two incoherent beams in the double-slit system. What about the
two interference terms in the probability current $\VEC J_{\mathrm{tot}}$~\eqref{eq:Jfinal},
then? Well, the first term is identical with the vanishing~\eqref{eq:sw.15},
but for the second term we obtain from entangling current~\eqref{eq:Je}
with $\varphi=\frac{\pi}{2}$ 
\begin{equation}
\frac{\hbar}{m}R_{1}R_{2}\left(\frac{\nabla R_{2}}{R_{2}}-\frac{\nabla R_{1}}{R_{1}}\right)=\frac{\hbar}{m}\left(R_{1}\nabla R_{2}-R_{2}\nabla R_{1}\right).\label{eq:sw.16}
\end{equation}
As the distributions $R_{i}$ may have long wiggly tails \textendash{}
summing up, after many identical runs, to a Gaussian with no cut-off,
but spreading throughout the whole domain of the experimental setup
(cf.\ section~\ref{sec:The-constituting-setup} and \cite{Groessing.2013dice})
\textendash{} the expression~\eqref{eq:sw.16} is not at all guaranteed
to vanish. In fact, a look at Fig.~\ref{fig:sw.3} shows that there
is an effect even for incoherent beams: Although the product $R_{1}R_{2}$
is negligible and therefore leads to no interference fringes on the
screen, nevertheless expression~\eqref{eq:sw.16} has the effect
of bending average trajectories so as to obey the no-crossing rule
well known from our model as well as from Bohmian theory.

As was already pointed out in~\cite{Sanz.2009context}, or more recently,
in~\cite{Luis.2015what}, the resulting trajectories of Fig.~\ref{fig:sw.3a}
can be understood as a nonlinear effect that is not usually considered
in standard quantum mechanics, but explainable in the Bohmian picture.
There, it is the structure of the velocity field which is genuinely
nonlinear and therefore allows for the emergence of the type of trajectory
behaviour. However, also in our approach, the emergence of the trajectories
of Fig.~\ref{fig:sw.3} is completely understandable as it can be
traced back to the non-vanishing of expression~\eqref{eq:sw.16}:
The average trajectories never cross the central symmetry line in
Fig.~\ref{fig:sw.3a}, a fact due to the diffusion related hot spot
indicated in red-to-yellow-to-white (depicting both interference terms
of Eq.~\eqref{eq:sw.16}), which represents a kinetic energy reservoir
that effectively gives particles a push in the forward direction;
The intensity of Eq.~~\eqref{eq:sw.16} is weakened by the factor
$a=10^{-8}$ in Fig.~\ref{fig:sw.3b}, which is why it does not affect
the strong beam. However, it is sufficient for the attenuated beam
to become deflected. 

In sum, then, performing a double-slit experiment with incoherent
beams leads to an emergent behaviour of particle propagation which
can be explained by the effectiveness of diffusion waves with velocities
$\VEC u_{i}$ interacting with each other, thereby creating a hot
spot where the intensity of the diffusive currents is highest and
leads to a deflection into the forward direction such that no-crossing
of the average velocities beyond the symmetry line is made possible
(Fig.~\ref{fig:sw.3a}). This is therefore in clear contradiction
to the scenario where only one slit is open for the particle to go
through. If the slits are not open simultaneously, the particles could
propagate to locations beyond the symmetry line, i.e.\ to locations
forbidden in the case of the second slit being open.~\cite{Sanz.2009context}

As our velocity fields $\VEC v_{i}$~\eqref{eq:diss.4.40} and $\VEC u_{i}$~\eqref{eq:diss.4.41}
are identical with the Bohmian and the osmotic momentum, respectively,
one can relate them also to the technique of weak measurements. The
latter have turned out~\cite{Leavens.2005weak,Wiseman.2007grounding,Hiley.2012weak,Hiley.2016quantum,deGosson.2016weak,deGosson.2016observing}
to provide said velocities as \textit{weak values}, which are just
given by the real and complex parts of the quantum mechanical expression
\begin{equation}
\frac{\left\langle \mathbf{r}\mid\hat{p}\mid\varPsi\left(\mathbf{\mathit{t}}\right)\right\rangle }{\left\langle \mathbf{r}\mid\varPsi\left(\mathbf{\mathit{t}}\right)\right\rangle }\,,
\end{equation}
i.e.\ the weak values associated with a weak measurement of the momentum
operator $\hat{p}$ followed by the usual (``strong'') measurement
of the position operator $\hat{r}$ whose outcome is $\mathbf{r}$.
In other words, in principle the trajectories for intensity hybrids
generally, and for the quantum sweeper in particular, are therefore
accessible to experimental confirmation.

In the standard quantum mechanical description of double-slit experiments
with intensity hybrids one is usually only concerned with the gradual
fading out of wave-like properties like interference fringes. However,
in our  model we are dealing with diffusion-based wave-like properties
throughout all magnitudes of attenuation of, e.g., slit~2, even in
the case of incoherent beams. For here, if we observe particles coming
through slit~2 characterized by a very low intensity such as $a=10^{-8}$,
one faces the sweeper effect (Fig.~\ref{fig:sw.4}).

The number $n\left(a\right)$ of particles which we do see come through
slit~2 and which produces the distribution (red) in Fig.~\ref{fig:sw.4a}
actually is deflected from the forward screen when slit~1 is opened,
but the same number $n\left(a\right)$ can easily be detected on the
sideways screen to the right in Fig.~\ref{fig:sw.4b}. Although the
particles would eventually also be detected on a more elongated forward
screen as in Fig.\ref{fig:gupferl-P}, the effect would be much smaller
simply due to the geometry, whereas the sideways screen setup allows
the registration with maximal contrast. In principle, for beam attenuation
as schematized in Fig.~\ref{fig:sw.4}, if one employs a sideways
screen, one thus obtains a different outcome than the one expected
due to standard quantum mechanical lore. According to the latter,
the beam from slit~2 should be unaffected by the situation at slit~1.
This would mean that in the unaffected scenario less than a number
of $\frac{n\left(a\right)}{2}$ particles could eventually be registered
on any sideways screen parallel to the $y$-axis along a wide spatial
extension, whereas our result predicts that the totality of the number
$n\left(a\right)$ of particles can be registered within a comparatively
narrow spatial domain. In Fig.~\ref{fig:sw.4c}, the vertical screen
setup reveals interesting features of the probability density distribution,
accounting both for the interference and the sweeper effects. The
black line indicates the continuation of the probability density distribution
for the one-slit case, which is of course being modified once the
interference effect in the coherent case of adding an attenuated beam
is allowed for. However, even in the incoherent scenario not showing
the comparatively small interference effects, one still obtains the
full sweeper effect, with a smooth transition between the two curves
in the upper and the lower parts of Fig.~\ref{fig:sw.4c}, respectively.
This is due to the non-vanishing of~\eqref{eq:sw.16}, i.e.\ a significant
contribution from the diffusive terms despite the smallness of the
transmission factor $a$.

\section{The quantum mechanical description of the sweeper effect\label{sec:sweeper quantum}}

\begin{figure}
\begin{centering}
\subfloat[The two cases of the attenuation factor at the right slit of a double
slit system, i.e. $a=10^{-4}$ and $a=10^{-8}$, respectively, essentially
provide the same distribution at moderate resolution. \label{fig:gupferl-a}]{\centering{}\includegraphics{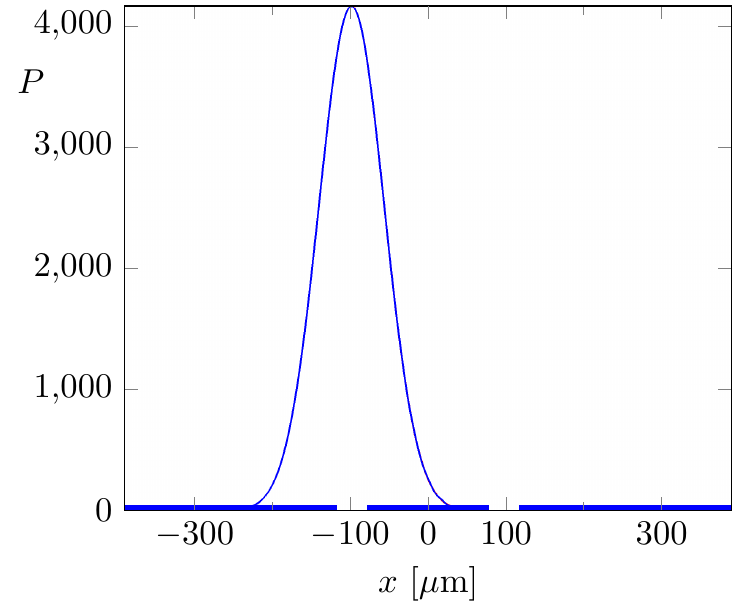}}\hfill{}\subfloat[Same as in (a); by zooming in with a factor of 1,000 two cases are
discernible: interference phenomena for $a=10^{-4}$ (blue), \textit{vs.\ }apparently
smooth behaviour for $a=10^{-8}$ (red). \label{fig:gupferl-b}]{\centering{}\includegraphics{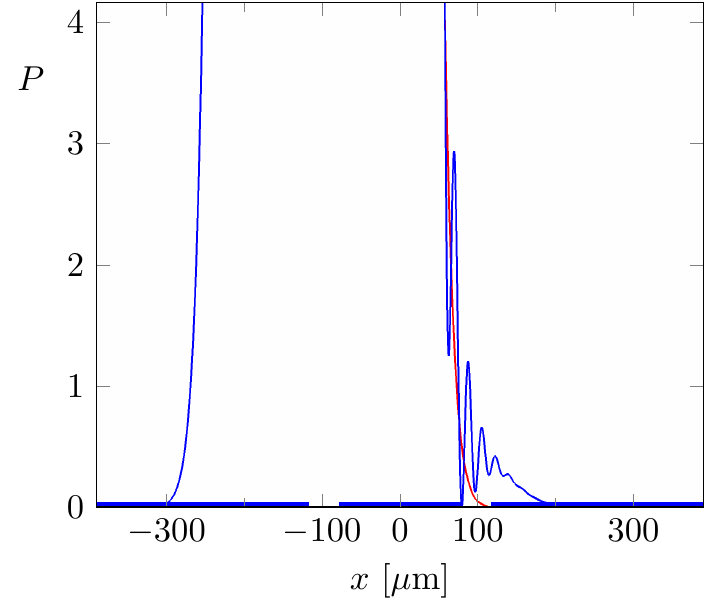}}
\par\end{centering}
\begin{centering}
\vspace{-0.2cm}
\subfloat[Same as in (a) on a logarithmic scale. Dotted initial distributions
for the cases of $a=10^{-4}$ (blue) and $a=10^{-8}$ (red), respectively,
evolve into distributions clearly showing interference phenomena which
have been ``swept aside'' far to the right.\label{fig:gupferl-c}]{\centering{}\includegraphics{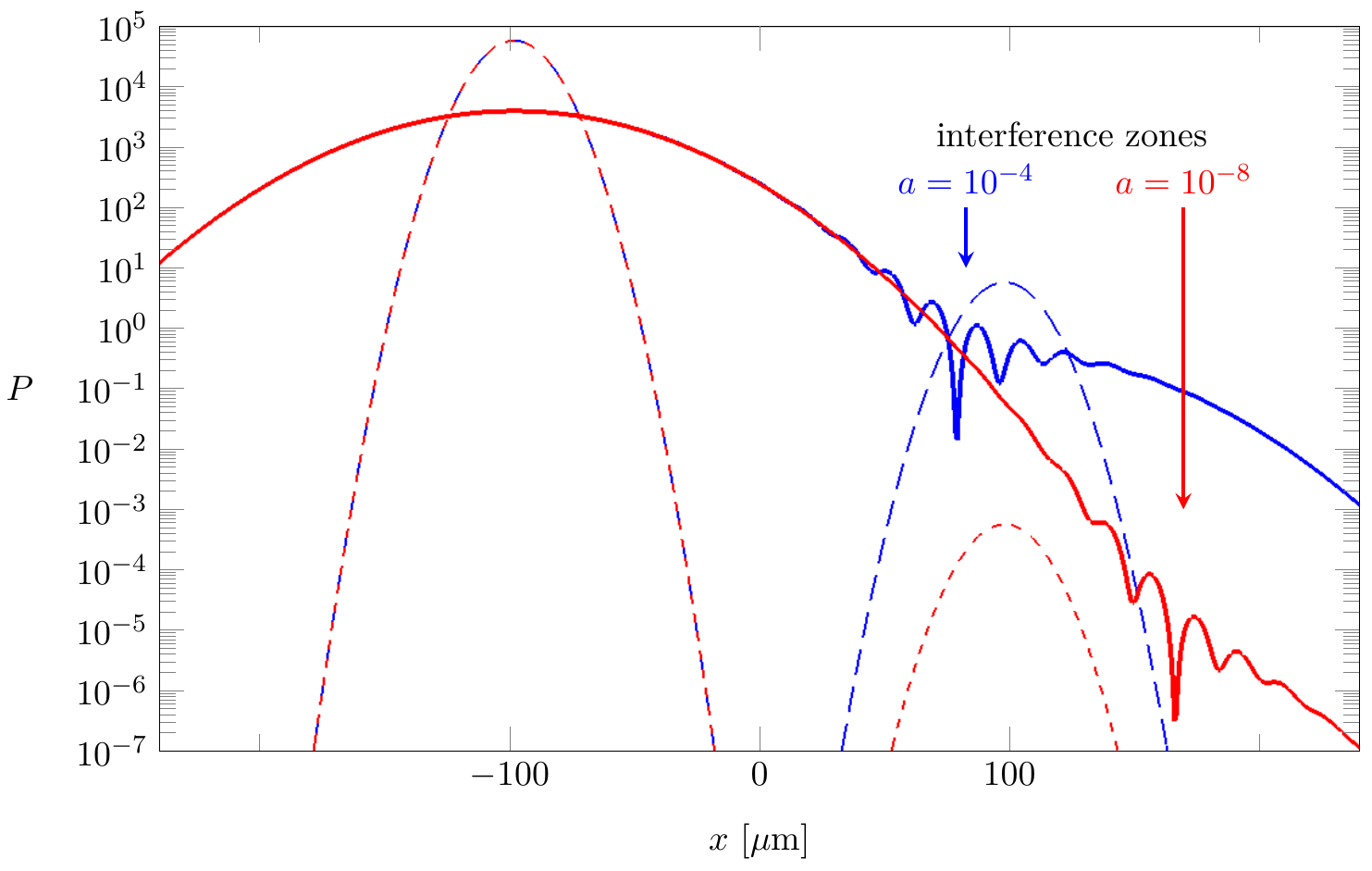}}
\par\end{centering}
\centering{}\caption{The sweeper effect as described by quantum mechanics. Probability
density distribution $P$ in a distance of 5\,m from the double slit.
\label{fig:gupferl}}
\end{figure}
\begin{figure}
\noindent \centering{}\includegraphics{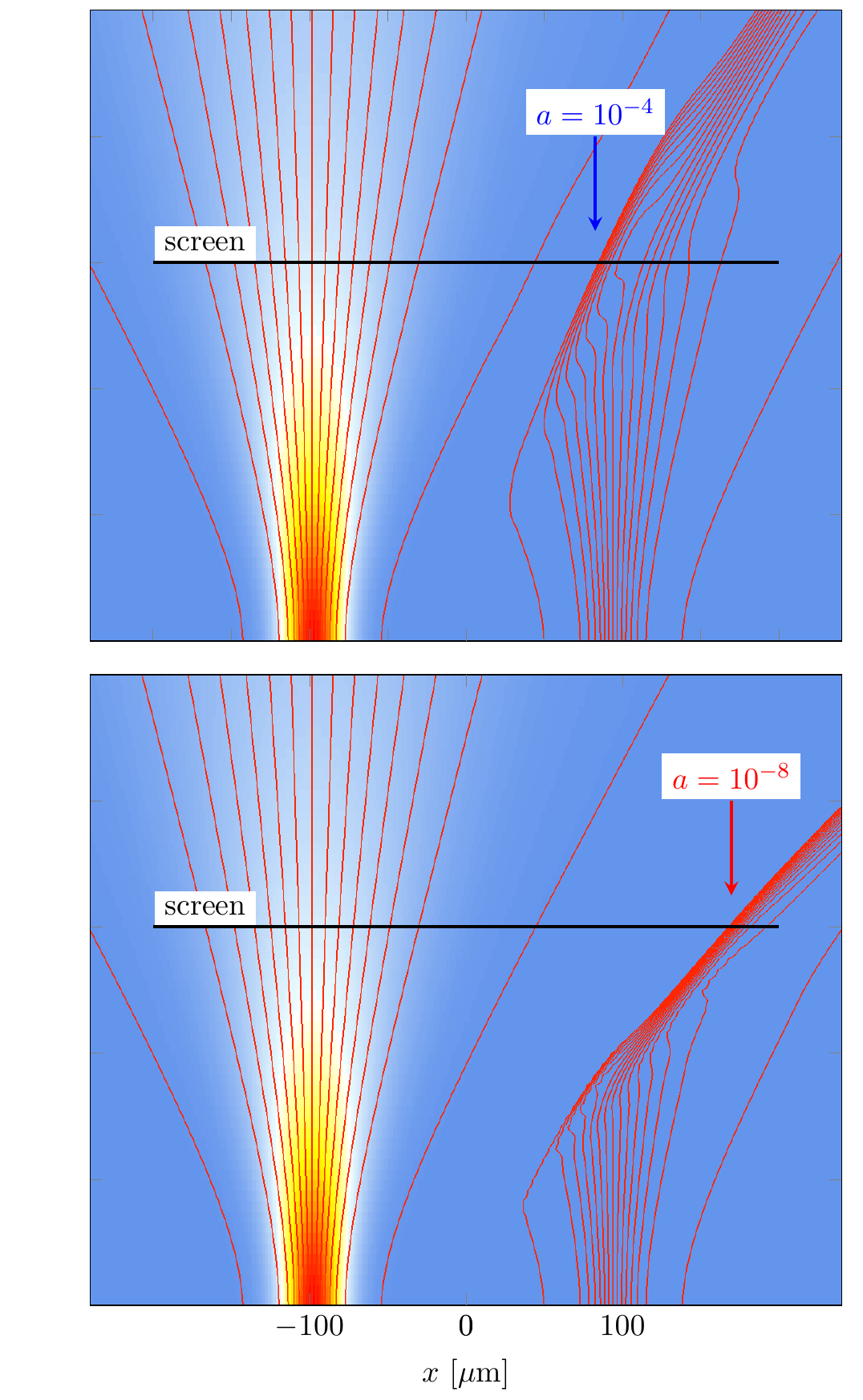}\caption{{\small{}Probability density distributions $P$ emanating from the
double slit with transmission factor $a=10^{-4}$ (top) and $a=10^{-8}$
(bottom) according to the red and blue distributions in Fig.~\ref{fig:gupferl},
respectively.} The arrows indicate the position of the interference
zones as measured at the screen.\label{fig:gupferl-P}}
\end{figure}
Let us now consider the stochastic attenuation discussed above in
purely quantum mechanical terms. As already mentioned, the probability
density distribution is given by Equation \eqref{eq:sw.2}. A graphic
representation of this distribution in a distance of 5m from the double
slit is shown in Fig.~\ref{fig:gupferl}. Two cases of the attenuation
factor at one of the two slits of a double slit system are shown,
i.e. $a=10^{-4}$ and $a=10^{-8}$ affecting the right slit, respectively.
As is to be expected, on a linear scale the distribution will appear
as if practically the whole intensity goes through the left un-attenuated
slit (Fig.~\ref{fig:gupferl-a}). Zooming in with a factor of 1000
as shown in Fig.~\ref{fig:gupferl-b}, one can see the faint rest
of interference phenomena for the case of $a=10^{-4}$ (blue), whereas
for $a=10^{-8}$ (red) apparently smooth behaviour is seen. Still,
the full effect is best visible on the logarithmic scale shown in
Fig.~\ref{fig:gupferl-c}. Compared to the dotted initial distributions
for the cases of $a=10^{-4}$ (blue) and $a=10^{-8}$ (red), respectively,
the whole distribution clearly shows interference phenomena which
have been ``swept aside'' far to the right. The probability distribution
for latter is shown in Fig.~\ref{fig:gupferl-P} in which the relative
positions of the red and blue arrow are the same as in Fig.~\ref{fig:gupferl-c}
indicating the positions of the detected interference zones. Thus,
the quantum sweeper effect is confirmed also via orthodox language. 

The bunching together of low counting rate particles within a very
narrow spatial domain, or channel, respectively, counters naive expectations
that with ever higher beam attenuation nothing interesting may be
seen any more. The reason why these expectations are not met is given
by the explicit appearance of the nonlinear structure of the probability
current $\VEC J_{\mathrm{tot}}$~\eqref{eq:Jfinal} in these domains
for very low values of $a$. 

\section{Implications\label{sec:conclusion-1}}

We have shown that for transmission factors below $a\lesssim10^{-4}$
in intensity hybrids, new effects appear which are not taken into
account in a naive, i.e.\ linear, extrapolation of expectations based
on higher-valued transmission factors. We have described the phenomenology
of these \textit{quantum sweeper} effects, including the bunching
together of low counting rate particles within a very narrow spatial
domain, or channel, respectively. However, we also stress that these
results are in accordance with standard quantum mechanics, since we
just used a re-labelling and re-drawing of the constituent parts of
the usual quantum mechanical probability currents. However, concerning
the explicit phenomenological appearances due to the nonlinear structure
of the probability current in the respective domains for very low
values of $a$, our sub-quantum model is better equipped to deal with
these appearances explicitly.

With the discovery of the quantum sweeper effect on the basis of a
 causal approach to quantum mechanics, we claim to have presented
a first example as it was demanded by Rabi\footnote{In his criticism of David Bohm's causal interpretation of the quantum
mechanical formalism, Isidor Rabi made the following statement in
the 1950ies which is still shared by quite some researchers today:
``I do not see how the causal interpretation gives us any line to
work on other than the use of the concepts of quantum theory. Every
time a concept of quantum theory comes along, you can say yes, it
would do the same thing as this in the causal interpretation. But
I would like to see a situation where the thing turns around, when
you predict something and we say, yes, the quantum theory can do it
too.'' \cite{FreireJr..2005science}}. We are optimistic that through further developments, both in theory
employing sub-quantum mechanics and in weak measurement techniques
capable of probing the latter regime, more unexpected new effects
can be predicted and eventually be confirmed in experiment. 

\section{Conclusion and outlook\label{sec:conclusion-2}}

Summarizing, it has been shown that for transmission factors below
$a\lesssim10^{-4}$ in intensity hybrids, new effects appear which
are not taken into account in a naive, i.e.\ linear, extrapolation
of expectations based on higher-valued transmission factors. One describes
the phenomenology of these quantum sweeper effects, including the
bunching together of low counting rate particles within a very narrow
spatial domain. It has also been stressed that these results are in
accordance with standard quantum mechanics, since just a re-labelling
and re-drawing of the constituent parts of the usual quantum mechanical
probability currents has been used. The reason why the above-mentioned
naive expectations are not met is given by the explicit appearance
of the nonlinear structure of the probability current in these domains
for very low values of $a$. In this regard, the presented sub-quantum
model is better equipped to deal with these appearances explicitly.

\chapter{Numerical methods\label{chap:Numerical-methods}}
\begin{quote}
In chapters~\ref{sec:3.probability-distributions} to \ref{sec:6.extreme-beam-attenuation}
we have already used numerical methods to produce distribution pictures.
We shall here give a detailed explanation on how the results have
been computed. As the mathematics of said numerical methods is rather
extensive and would thus be misplaced in between the physically oriented
explanations of the last chapters, an overview on the procedure of
simulation shall be provided here. A quick overview on the simulation
setup already provided in chapters~\ref{sec:The-constituting-setup}
and \ref{sec:path-1} will be given. Then the practical handling of
action and phase will be introduced as well as a note on the usage
of diffusion coefficients as an addition to chapter~\ref{sec:3.3.The-derivation-of-Dt}.

A bigger part will be dedicated to finite difference procedures, especially
the two particular ones we used throughout the work, coupled map lattices
and the Crank\textendash Nicolson's method, as well as the respective
stability criteria.

We shall present the construction of trajectories whose representation
is not quite clear, especially in the sweeper figures of chapter~\ref{sec:6.extreme-beam-attenuation}.
Finally, we shall show how to calibrate our tools by using measurement
data of neutron double-slit experiments.
\end{quote}

\section{Preliminaries}

\subsection{The simulation setup\label{sec:The-simulation-setup}}

\begin{figure}[th]
\centering{}\includegraphics{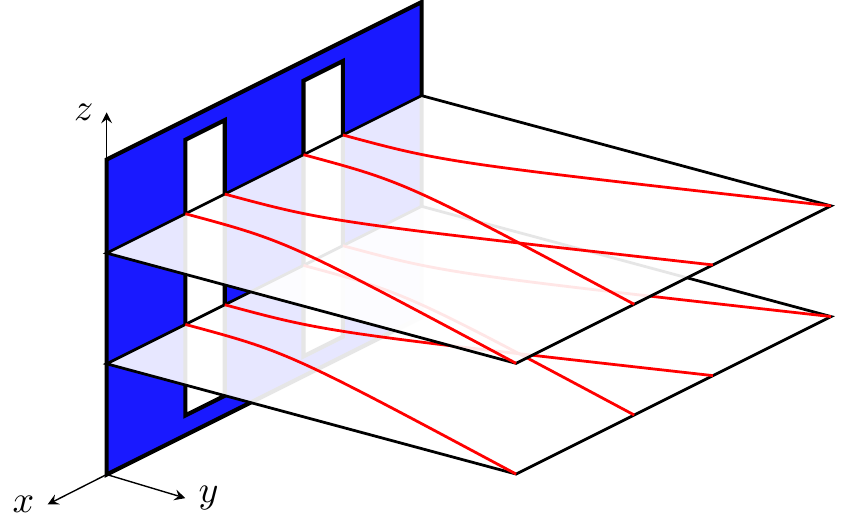}\caption{Setting of a double-slit experiment in three dimensions with Bohm-type
trajectories sketched on different layers\label{fig:if2d.1}}
\end{figure}

In section~\ref{sec:The-constituting-setup} the setting of a single-slit
experiment has been sketched, which has been further extended to at
least a double-slit in chapter~\ref{sec:path-1} as shown in Fig.~\ref{fig:if2d.1},
comprising a three-dimensional problem with slits in the $xz$-plane
elongated in $z$-direction. Consider a Gaussian entering a slit propagating
in the positive $y$-direction. Its spreading in the $xy$-plane is
essentially independent of the $z$-position as any spreading into
the $z$-direction is compensated by the spreading of a neighboured
plane. For simplicity, we neglect the impact of the slit's top and
bottom edges, thus assuming sufficiently large slits.

The dispersion is assumed to feature an ideal Gaussian shape not being
refracted at the slit's side edges. Furthermore, the Gaussians extends
along the whole $x$-direction, i.e.\ the Gaussian function is not
cut by the slit it runs through, as indicated in Fig.~\ref{fig:single-slit-with-P}
by the left most shape not cut by the slit. Then, one does not need
to consider about phase-free spaces along any light-cone-like structures
which would arise otherwise. Right after the slit the initial probability
density at a given slit centre position $x_{0}$ reads as
\begin{equation}
P(x,0)=\frac{1}{\sqrt{2\pi}\sigma_{0}}\;\e^{\nicefrac{-(x-x_{0})^{2}}{2\sigma_{0}^{2}}}\label{eq:Pi(x,0)}
\end{equation}
at initial time $t=0$ and initial standard deviation $\sigma_{0}$.
$P(x,0)$~(\ref{eq:Pi(x,0)}) is the distribution to start with in
every single simulation. The connection between the $y$-axis and
time $t$ is given by Eq.~(\ref{eq:y=00003Dv_yt}), 
\begin{equation}
y\left(t\right)=\frac{\hbar k_{y}t}{m}\thinspace.\label{eq:diss.3.48-chap.7.1}
\end{equation}

\subsection{Action and phase\label{sec:The-role-of-phase}}

As soon as more slits are available in a given setup, the phase relations
between the distributions after the slits have to be considered. In
order to derive the phase relations of coherent beams, we recall the
definition of the phase~\cite{Mesa.2013variable}
\begin{equation}
\varphi(x,t)=S(x,t)/\hbar\label{eq:td.38}
\end{equation}
with the classical action function $S(x,t)$ as defined in chapter~\ref{sec:orthogonality}.
We identifying the total velocity $v_{\mathrm{tot}}(x,t)$ of Eq.~(\ref{eq:td.20})
along a trajectory with 
\begin{equation}
v_{\mathrm{tot}}(x,t)=\frac{\nabla S(x,t)}{m}\thinspace.\label{eq:td39}
\end{equation}
We assume that there is no potential and the paths described by $v_{\mathrm{tot}}(x,t)$,
as sketched in Fig.~\ref{fig:surface_S}, correspond to particle
trajectories of free particles and thus the energy is constant, $E=\mathrm{const.}$ 

These presumptions then lead to the action
\begin{equation}
S(x,t)=\intop_{vt}^{x}mv_{\mathrm{tot}}\d x'-\intop_{0}^{t}E\d t'=m\intop_{vt}^{x}\left[v+\frac{u_{0}^{2}t}{\sigma_{0}^{2}+u_{0}^{2}t^{2}}\,\xi(t)\right]\d x'-\intop_{0}^{t}E\d t'\label{eq:td.40}
\end{equation}
with $E$ being the system's total energy and $m$ the mass of the
particle involved. According to Fig.~\ref{fig:td.1}, the lower bound
of the integral in Eq.~(\ref{eq:td.40}) is set to $vt$ being the
starting point of the diffusion which is different from zero due to
velocity $v$ causing an angle of inclination of the incident plane
wave. According to the motion in $t$-direction, there is the constant
component $mv_{y}^{2}t$ to be added to $S(x,t)$ in Eq.~(\ref{eq:td.40})
which we put into $Et$.

As $v=\mathrm{const.}$ as well as $E=\mathrm{const.}$ we can solve
both integrals, providing
\begin{equation}
S(x,t)=mvx+\frac{mu_{0}^{2}}{2}\left[\frac{\xi(t)}{\sigma(t)}\right]^{2}t-mv^{2}t-Et.\label{eq:S(x,t)_1}
\end{equation}
In Eq.~(\ref{eq:S(x,t)_1}), the right most term depends on $t$
only and will cancel out later.

Finally, we write the phase defined by Eq.~(\ref{eq:td.38}) as
\begin{equation}
\begin{aligned}\varphi(x,t) & =\frac{1}{\hbar}\left[mvx+\frac{mu_{0}^{2}}{2}\left[\frac{x-vt}{\sigma(t)}\right]^{2}t-mv^{2}t-Et\right].\end{aligned}
\begin{gathered}\end{gathered}
\label{eq:td.42-2}
\end{equation}
Expression~(\ref{eq:td.42-2}) sticks to the coordinate system and
will turn out to be very helpful for  interference calculations on
a grid.

Now, if we extend the setup to a double-slit system, as sketched in
Figs.~\ref{fig:if2d.1} or \ref{fig:td.2}, we need the Gaussian
shaped probability densities coming out from each slit as well as
the overall phase which is a combination of the single phases $\varphi(x,t)\ \eqref{eq:td.42-2}$.
Since each Gaussian has its own phase~(\ref{eq:td.42-2}) we are
free to add a phase shifter $\Delta\varphi(x,t)$ for one of the slits
of the two-slit experiment, say slit~1, which modifies $\varphi_{1}(x,t)$
to
\begin{equation}
\varphi'_{1}(x,t)=\frac{S_{1}(x,t)}{\hbar}+\Delta\varphi(x,t)\label{eq:td.46}
\end{equation}
which yields for the phase difference
\begin{equation}
\begin{aligned}\varphi_{12}(x,t) & =\vphantom{\frac{}{{\displaystyle \intop}}}\varphi_{2}(x,t)-\varphi'_{1}(x,t)\\
 & =\vphantom{\frac{}{{\displaystyle \intop}}}\frac{m}{\hbar}\left[\vphantom{\intop}v_{2}(x-x_{02})-v_{1}(x-x_{01})-(v_{2}^{2}-v_{1}^{2})t\right]-\Delta\varphi(x,t)\\
 & \quad\quad+\frac{mt}{2\hbar}\left[\frac{u_{02}^{2}(x-x_{02}-v_{2}t)^{2}}{\sigma_{2}^{2}(t)}-\frac{u_{01}^{2}(x-x_{01}-v_{1}t)^{2}}{\sigma_{1}^{2}(t)}\right].
\end{aligned}
\begin{gathered}\end{gathered}
\label{eq:td.44}
\end{equation}
Even though the phase shifter $\Delta\varphi(x,t)$ allows for modification
of $x$ and $t$ independently, in this work we only provide simulations
with the phase shifter $\Delta\varphi(t)$ being a function of time
only, e.g.\ as is clearly indicated in Fig.~(\ref{fig:dice-phase-shift}).

\subsection{The diffusion coefficient in computations}

\begin{figure}[!ht]
\centering{}\includegraphics{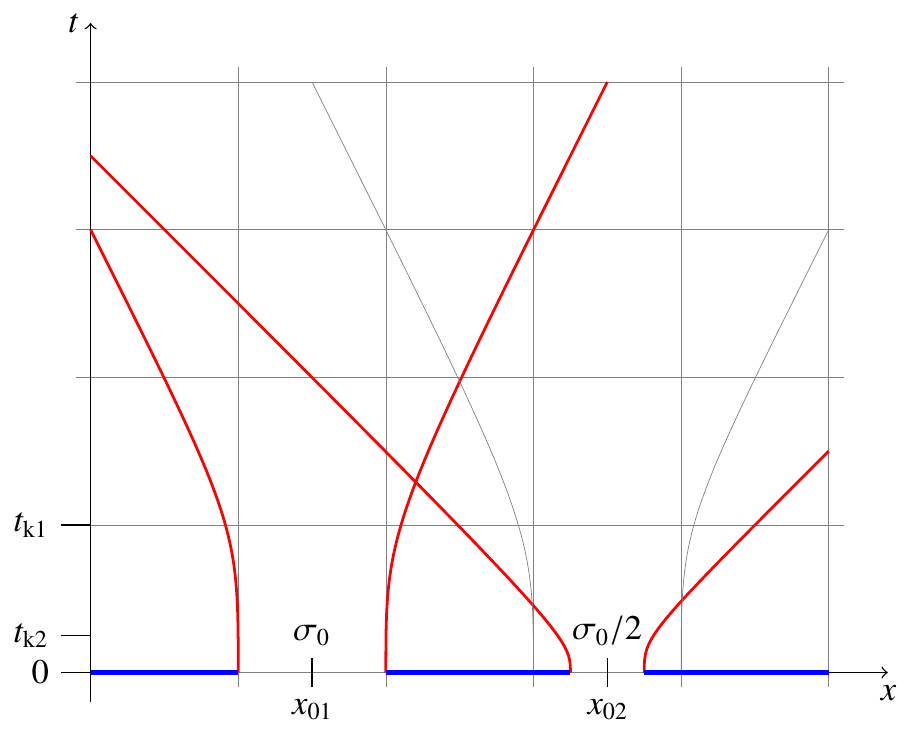}\caption{Sketch of a double-slit with two different widths and Bohm-type trajectories
(and same-widths scenario indicated by grey lines)\label{fig:td.2}}
\end{figure}

The two slits at positions $x_{01}$ and $x_{02}$ could also have
different slit widths and hence different parameters, $\sigma_{01}$,
$\sigma_{1}(t)$, $u_{01}$ and $\sigma_{02}$, $\sigma_{2}(t)$,
$u_{02}$, respectively, as sketched in Fig.~\ref{fig:td.2}. 
Note, the phase difference $\varphi_{12}$~(\ref{eq:td.44}) is at
any time defined for the whole domain as already pointed out in section~\ref{sec:The-simulation-setup}.

Now we take a closer look at the time $t=t_{\mathrm{k}}$ of the kink
(see Fig.~\ref{fig:td.2}), i.e.\ the time when the wave packet
changes its spreading behaviour. According to Eq.~(\ref{eq:td.11}),
\begin{equation}
\frac{\sigma(t)}{\sigma_{0}}=\sqrt{1+\frac{D^{2}t^{2}}{\sigma_{0}^{4}}}\thinspace,\label{eq:td.11-chap.7.2}
\end{equation}
this is obviously when the two terms under the square root become
of equal value, which yields 
\begin{equation}
1=\frac{D^{2}t_{\mathrm{k}}^{2}}{\sigma_{0}^{4}}\thinspace,
\end{equation}
hence the number under the square root becomes $2$, thus we get $\sigma(t_{\mathrm{k}})=\sqrt{2}\sigma_{0}$.
With the help of Eq.~(\ref{eq:td.22}) we find that
\begin{equation}
D_{\mathrm{t}}=\frac{t}{t_{\mathrm{k}}}D.\label{eq:td.23}
\end{equation}
At time $t=t_{\mathrm{k}}$ the diffusion coefficient\footnote{Note that the diffusivity $D=\hbar/2m$ is constant for all times
$t$ and has to be distinguished from the diffusion coefficient $D_{\mathrm{t}}$.
See also section~\ref{sec:3.3.The-derivation-of-Dt}} becomes $D_{\mathrm{t}}=D$. For an exemplary picture, consider the
scenario depicted in Fig.~\ref{fig:td.2} comprising two slits of
different widths. We assume the initial Gaussians passing the slits
have a standard deviation according to the respective slit widths,
e.g., $\sigma_{01}=\sigma_{0}$ and $\sigma_{02}=\sigma_{0}/2$, respectively.
The resulting Bohm-type trajectories of the two decaying Gaussians
have the properties that the time at the kink quadruples while the
spreading only doubles, 
\begin{equation}
\sigma_{01}=2\sigma_{02}\qquad\Longrightarrow\qquad t_{\mathrm{k}1}=4t_{\mathrm{k}2},
\end{equation}
$t_{\mathrm{k}i}$ being the time at which the kink arises at the
respective slit $i$, as indicated in Fig.~\ref{fig:td.2} by red
lines compared with the greyed-out lines for the spreading of slit~2
for the case both slits would have standard deviation $\sigma_{0}$.
According to Eq.~(\ref{eq:td.22}), the diffusion coefficients of
the two slits, now different from each other and thus indicated by
$D_{\mathrm{t},i}$ corresponding to slit $i$, yield 
\begin{equation}
D_{\mathrm{t},1}(t)=\frac{D^{2}t}{\sigma_{02}^{2}}\quad\neq\quad D_{\mathrm{t},2}(t)=\frac{D^{2}t}{\sigma_{01}^{2}},\qquad\forall t>0,\label{eq:diss.Dt1-Dt2}
\end{equation}
which implies that one cannot compute both spread distributions in
a single step, as the associated diffusivities evolve different in
time. Instead, one has to compute each single probability distribution
and combine them afterwards  according to Eq.~(\ref{eq:Ptot2slit}).

As an example of a double-slit setup with different slit widths which
considers also Eqs.~(\ref{eq:td.44}) and (\ref{eq:diss.Dt1-Dt2})
in comparison to a double-slit experiment with equal widths.~\cite{Groessing.2012doubleslit,Mesa.2013variable}

\begin{sidewaysfigure}
\subfloat[$\sigma_{01}=\sigma_{02}$\label{fig:td.3a}]{\centering{}\includegraphics[width=0.48\textwidth]{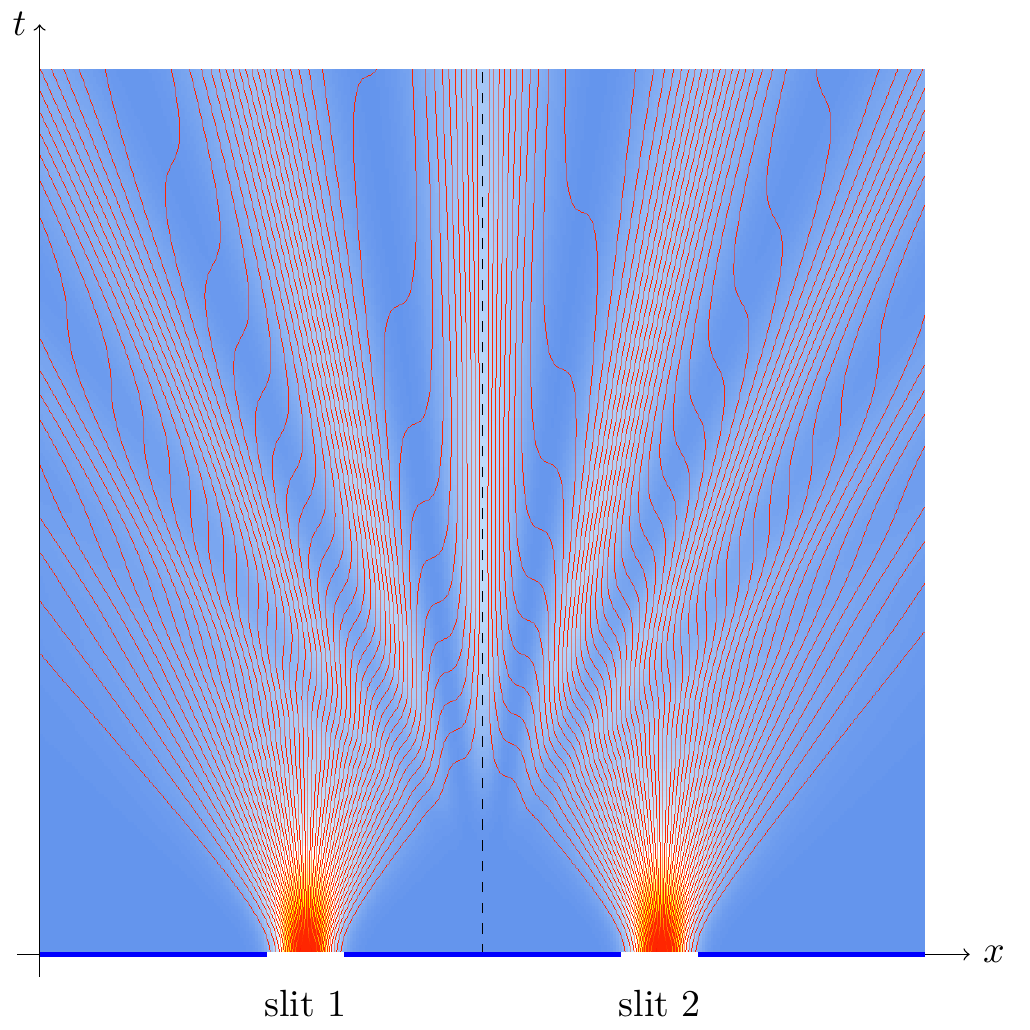}}\hfill{}\subfloat[$\sigma_{01}=2\sigma_{02}$\label{fig:td.3b}]{\centering{}\includegraphics[width=0.48\textwidth]{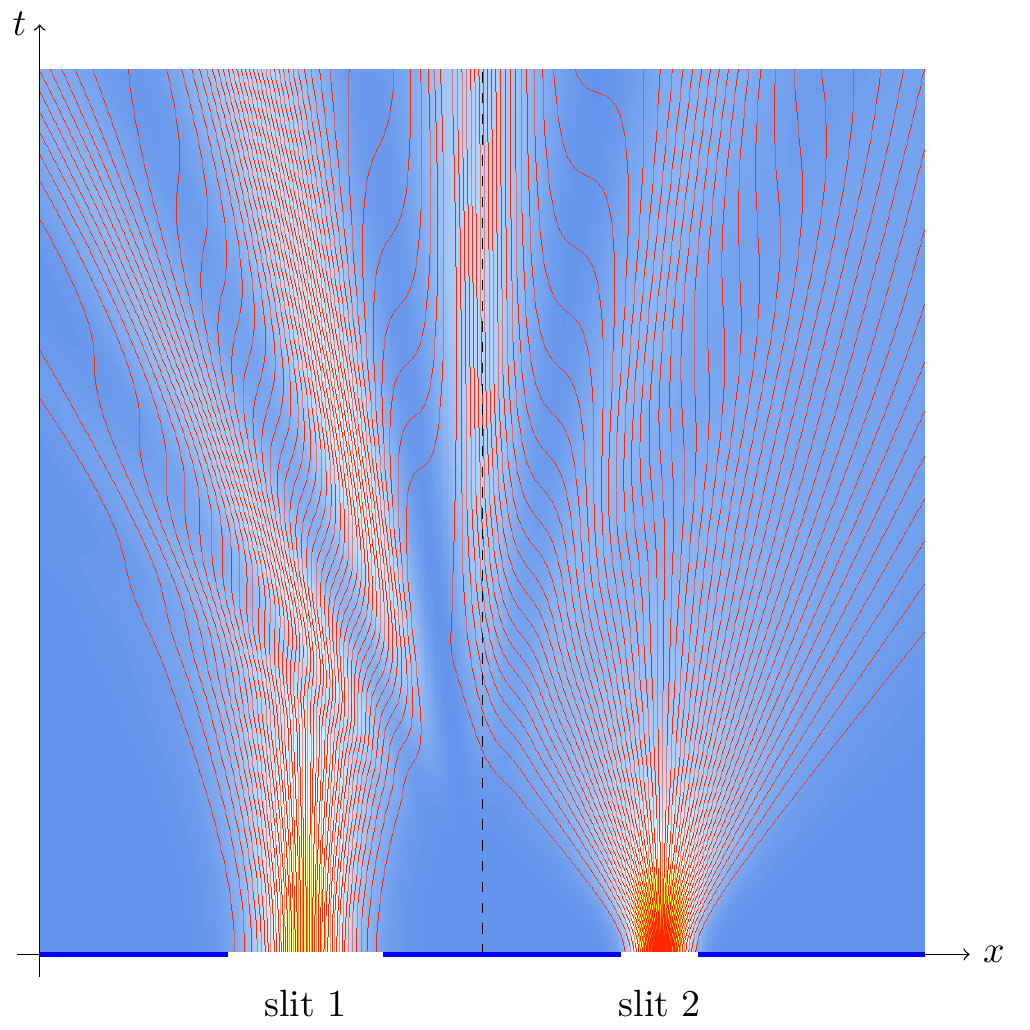}}\caption{Classical computer simulation of the interference pattern with different
slit widths: intensity distribution with increasing intensity from
white through yellow and orange, with trajectories (red) for two Gaussian
slits, and with large dispersion (evolution from bottom to top; $v_{x,1}=v_{x,2}=0$).\label{fig:td.3}}
\end{sidewaysfigure}
The graphical results providing the interference patterns thereto
are shown in Fig.~\ref{fig:td.3}. In Fig.~\ref{fig:td.3a} the
maximum of the intensity is distributed along the symmetry line exactly
in the middle between the two slits, as well as in Fig.~\ref{fig:td.3b},
though slit~2 has doubled width. In the exemplary figures, trajectories
according to Eq.~(\ref{eq:Ptot2slit}) for the two Gaussian slits
are shown. For an explanation on the meaning of these patterns, see
chapter~\ref{sec:Double-slit-interference}.

\section{The finite difference method}

In section~\ref{sec:3.3.The-derivation-of-Dt} we have formulated
the ballistic diffusion equation~(\ref{eq:ballisticDE}),
\begin{equation}
\frac{\partial P}{\partial t}=\underbrace{\frac{D^{2}t}{\sigma_{0}^{2}}}_{D_{\mathrm{t}}}\,\frac{\partial^{2}P}{\partial x^{2}}\thinspace,\label{eq:ballisticDE-fdm}
\end{equation}
with diffusion constant $D=\hbar/2m$. Eq.~(\ref{eq:ballisticDE-fdm})
is valid per slit of width $\sigma_{0}$. In a multi-slit system Eq.~(\ref{eq:ballisticDE-fdm})
has to be evaluated once per slit and combined with phases~(\ref{eq:td.44}).

In this section we describe the evaluation procedure of $P(x,t)$
in order to solve Eq.~(\ref{eq:ballisticDE-fdm}) with initial value
$P(x,0)$ given by Eq.~(\ref{eq:Pi(x,0)}) by means of finite difference
methods (FDM). FDMs are numerical methods for solving differential
equations by approximation with difference equations. Here, the ballistic
diffusion equation~(\ref{eq:ballisticDE-fdm}) is solved per slit
on a discretised grid. As first relations, we define 
\begin{equation}
t=T\Delta t,\qquad x=X\Delta x,\label{eq:diss.7.1}
\end{equation}
with $t$ and $x$ denoting time and position in the physical domain
while $T$ and $X$ denote time and position of the simulation domain,
respectively. Then we have for the step widths 
\begin{equation}
\Delta t=\frac{t}{T}\;,\qquad\Delta x=\frac{x}{X}\:,\label{eq:diss.7.2}
\end{equation}
thereby defining the scaling between the physical domain and the numerical
discretisation.

Now, we take a closer look at two different numerical procedures to
solve Eq.~(\ref{eq:ballisticDE-fdm}) and the stability conditions
of these procedures.

\subsection{Coupled map lattices\label{subsec:Coupled-map-lattices}}

\begin{figure}[!h]
\begin{centering}
\includegraphics{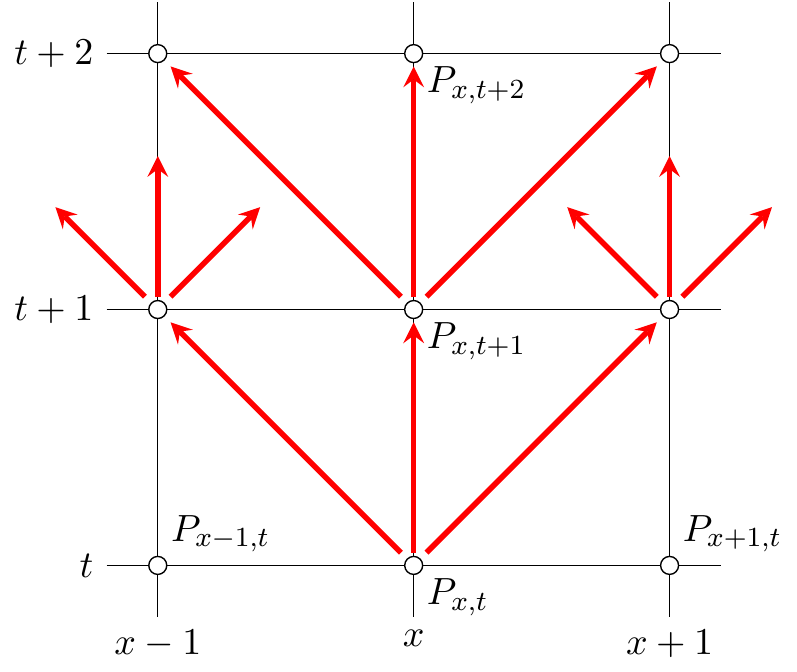}
\par\end{centering}
\caption{Neighbourhood in coupled map lattices\label{fig:diss.7.01-neighborhood}}
\end{figure}

Coupled map lattices, or short CML, are equivalent to cellular automata,
though each cell\footnote{We shall use the terms ``cell'' and ``node'' synonymously. However,
for CMLs or cellular automata the term cell is more common which is
associated with the idea of a space filled with some entities.} is represented by real values instead of integers (see Fig.~\ref{fig:diss.7.01-neighborhood}).
CMLs allow to inquire into dynamical processes of emergent processes
and could model not only general phenomenological aspects of our world
but also directly the laws of physics themselves. CMLs then could
be a powerful tool to get a deeper understanding on what is going
on because they are information-preserving and thus retain one of
the most fundamental features of microscopic physics \textendash{}
namely reversibility.~\cite{Toffoli.1987cellular}

Coupled map lattices reduce macroscopic phenomena to precisely defined
microscopic processes which make them of prime methodological interest,
but in order to obtain such features, in general one has no choice
but to implement an explicit finite difference forward scheme, a so-called
\textit{Euler scheme}, respectively. 

An explicit \textit{forward} scheme is characterized by the fact that
solely solutions of already elapsed time steps are sufficient to calculate
the solution of the next time step. In a coupled map lattice, then,
all values of the next time step of the whole domain are computable
within a single iteration only out of values already calculated before.
The crucial point of this definition is that these upcoming values
are \textit{computable} in the same iteration, these values must therefore
not be part of a condition that is itself subject to be solved before,
otherwise the scheme were \textit{implicit}. In this sense coupled
map lattices are completely specified, discrete dynamical systems
of a local relation, i.e.\ neighbourhood rules, as is the case for
continuous dynamical systems defined by partial differential equations.
And hence coupled map lattices are the discrete physicist's concept
of \textit{fields}.

In order to derive the coupled map lattices' relations we replace
the differential terms of Eq.~(\ref{eq:ballisticDE-fdm}) by discrete
differences,\footnote{To make things easily readable and taking into account that the indexed
variables are only used in this chapter, we leave the naming of the
variables untouched, even though the indexed variables define a grid
comprising only natural number, i.e.\ $x,t\in\mathbb{N}$, whereas
non-index variables represent physical quantities. In this chapter
let us define: A variable being an index ($\cdot_{x}$, $\cdot_{t}$)
pertains to the grid, else ($x,$$t$) it represents a physical quantity.\label{fn:on-the-index-notation}}
\begin{align}
\frac{\partial P}{\partial t} & \rightarrow\frac{P_{x,t+1}-P_{x,t}}{\Delta t}+O(\Delta t^{2}),\label{eq:diss.7.5explicit}\\
\frac{\partial^{2\vphantom{\vphantom{\int^{8}}}}P}{\partial x^{2}} & \rightarrow\frac{P_{x+1,t}-2P_{x,t}+P_{x-1,t}}{\Delta x^{2}}+O(\Delta x^{2}),\label{eq:diss.7.6explicit}
\end{align}
using two-dimensional cells $P_{x,t}$ for each value $P(x,t)$ on
a discrete lattice. $\Delta x$ and $\Delta t$ being the step width
in space and time, respectively. The Landau notation $O$ describes
the limiting behaviour of the functions, both of which are here of
order $2$.

The resulting finite difference equation is obtained by simply substituting
Eqs.~(\ref{eq:diss.7.5explicit}) and (\ref{eq:diss.7.6explicit})
into (\ref{eq:ballisticDE-fdm}), thereby omitting the Landau notation
$O$,
\begin{equation}
\frac{P_{x,t+1}-P_{x,t}}{\Delta t}=\frac{D_{x,t+1}}{\Delta x^{2}}\Bigl(P_{x+1,t}-2P_{x,t}+P_{x-1,t}\Bigr),
\end{equation}
and in case $D_{\mathrm{t}}(x,t)$, or its pendant on the lattice
$D_{x,t}$, is independent of $x$, then, after reordering the equation
reads as
\begin{equation}
P_{x,t+1}=P_{x,t}+\frac{D_{t+1}\Delta t}{\Delta x^{2}}\Bigl(P_{x+1,t}-2P_{x,t}+P_{x-1,t}\Bigr)\label{eq:FDMexplicit}
\end{equation}
with spatial variable $x$, time $t$, and initial Gaussian distribution
$P(x,0)$ having standard deviation $\sigma_{0}$ at $t=0$. The calculation
of a cell's value $P_{x,t+1}$ (at time $t+\Delta t$) only depends
on cell values at the previous time step $t$, which fulfils the necessary
condition for coupled map lattices as stated above. In Eq.~(\ref{eq:FDMexplicit})
the time-dependent diffusion coefficient $D_{t+1}$ can be calculated
without any knowledge of neighbouring cells because it only depends
on time. As this diffusion coefficient represents the underlying physical
process at a given cell it is calculated in Eq.~(\ref{eq:FDMexplicit})
for the evaluated time step $t+\Delta t$ at which $P_{x,t+1}$ is
evaluated, hence $D_{t+1}$ instead of $D_{t}$.

Concerning the neighbourhood rules as local relations, a cell's value
affects only itself and its direct neighbours in the next time step
thereby defining a light-cone-like $45^{\circ}$ line in the unity-sized
grid of the coupled map lattice as shown in Fig.~\ref{fig:diss.7.01-neighborhood}.
However, this is an impact of the construction of derivatives in the
finite different scheme as is obvious from Eqs.~(\ref{eq:diss.7.5explicit})
and (\ref{eq:diss.7.6explicit}).

\subsubsection*{Stability of coupled map lattices}

The solutions of finite difference schemes may provide instabilities
which are related to high-frequency oscillations. Instability is essentially
a local phenomenon as at the points where the oscillations arise the
derivative of the solution is discontinuous. Even though, the oscillations
caused by the instability propagate to other regions, which can eventually
make the disturbance seem to be global in extent. Here, we examine
the conditions to be taken into consideration under which and when
the system is stable.

The stability condition for the scheme~(\ref{eq:FDMexplicit}) is
that 
\begin{equation}
\left|\frac{D_{\mathrm{t}}\Delta t}{\Delta x^{2}}\right|\leq\frac{1}{2}\label{eq:tdde.1.4.1}
\end{equation}
be satisfied for all values of the cells in the domain of computation.
The general procedure is that one considers each of the \textit{frozen
coefficient problems} arising from the scheme. The frozen coefficient
problems are the constant coefficient problems obtained by fixing
the coefficients at their values attained at each point in the domain
of the computation (cf.\ Strikwerda~\cite{Strikwerda.2004finite}). 

To fix the coefficients in Eq.~(\ref{eq:tdde.1.4.1}) the variables
$\Delta x$ and $\Delta t$ are kept constant during the whole computation,
whereas the value of $D_{\mathrm{t}}(t)$ grows with increasing time.
In order to obtain the best possible estimate with Eq.~(\ref{eq:tdde.1.4.1})
we substitute the maximum possible value of $D_{t+1}$, i.e.\ 
\begin{equation}
D_{\mathrm{t}}(t)\to\max(D_{t+1})=D_{t_{\mathrm{max}}+1}
\end{equation}
to be kept up for the sake of derivation of the stability conditions
only.

Substituting $D_{\mathrm{t}}(t)$ of Eq.~(\ref{eq:ballisticDE-fdm})
into (\ref{eq:tdde.1.4.1}) leads to 
\begin{align}
\Delta t & \leq\frac{\Delta x^{2}\sigma_{0}^{2}}{2D^{2}t}\label{eq:tdde.1.4.2-1}
\end{align}
which reaches its minimum value in the domain at $t=t_{\mathrm{max}}$,
thereby defining the largest allowed step width $\Delta t_{\mathrm{max}}$
to ensure stability. Using these limits, i.e.\ $t_{\mathrm{max}}=T_{\mathrm{max}}\Delta t_{\mathrm{max}}$,
yields
\begin{align}
\Delta t_{\mathrm{max}} & \leq\frac{\Delta x^{2}\sigma_{0}^{2}}{2D^{2}t_{\mathrm{max}}}=\frac{\Delta x^{2}\sigma_{0}^{2}}{2D^{2}T_{\mathrm{max}}\Delta t_{\mathrm{max}}}\thinspace,\label{eq:tdde.1.4.3-1}\\
\Delta t_{\mathrm{max}}^{2} & \leq\frac{\Delta x^{2\vphantom{\vphantom{\int^{8}}}}\sigma_{0}^{2}}{2D^{2}T_{\mathrm{max}}}\;,\label{eq:tdde.1.4.4-1}
\end{align}
 and eventually leads to the\textit{ stability condition} 
\begin{equation}
\ensuremath{{\displaystyle {\displaystyle \Delta t_{\mathrm{max}}\leq\frac{\Delta x\sigma_{0}}{D\sqrt{2T_{\mathrm{max}}}}\thinspace.}}}\label{eq:stability-1}
\end{equation}
While the numerator's variables, $\Delta x$ and $\sigma_{0}$, are
solely determined by the setup in $x$-direction, the denominator's
variables, $D=\hbar/2m=\mathrm{const.}$ and $T_{\mathrm{max}}$,
the latter is solely determined by the $t$-direction. If, for example,
one extends the time development, i.e.\ by setting 
\begin{equation}
T_{\mathrm{max}}\to aT_{\mathrm{max}},\qquad a>1,
\end{equation}
one then has to shrink 
\begin{equation}
\Delta t_{\mathrm{max}}\to\Delta t_{\mathrm{max}}/\sqrt{a}
\end{equation}
 simultaneously to ensure stability.

The stability condition~(\ref{eq:stability-1}) turns out to be a
problematic restriction on computability. However, in cases with moderate
spreading we obtain pretty good results and proved the method of coupled
map lattices to work fine. Nevertheless, coupled map lattices demand
explicit methods, as already stated above. As there are also examples
in this thesis where this method does not work economically usefully,
we then must employ other methods (see chapter~\ref{subsec:Crank-Nicolson's-method}).

Finally, we want to point out that, in cases where coupled map lattices
are stable, the approximation follows the exact solution at least
linearly with $x$ and $t$. The complete proves can be found in textbooks,
e.g.\ from~Toffoli \textit{et al.}~\cite{Toffoli.1987cellular},
Schwarz and Köckler~\cite{Schwarz.2009numerische}, or Haas~\cite{Haas.1999numerische}.

\subsection{Crank\textendash Nicolson's method\label{subsec:Crank-Nicolson's-method}}

\begin{figure}[!h]
\begin{centering}
\includegraphics{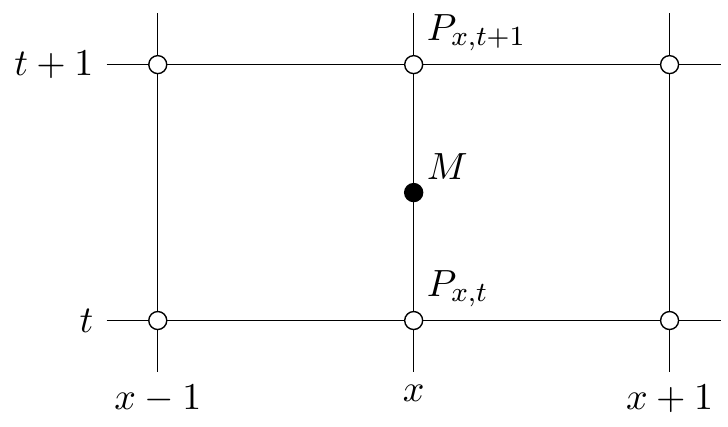}
\par\end{centering}
\caption{Lattice according to Crank\textendash Nicolson's method\label{fig:Crank-lattice}}

\end{figure}

Now we investigate Crank\textendash Nicolson's method as an example
of an \textit{implicit} method. From the viewpoint of the difference
approximation the disadvantage with the derivative of an explicit
method is that the used difference quotients approximate their associated
derivatives at different positions of the domain. In order to enhance
the approximations the second derivative is replaced in the following
way: Instead of using rule~(\ref{eq:diss.7.6explicit}) we approximate
$\partial^{2}P/\partial x^{2}$ by the arithmetic mean of the two
difference quotients at nodes $[x,t]$ and $[x,t+1]$ at two consecutive
time steps, as shown in Fig.~\ref{fig:Crank-lattice},
 and obtain for the approximations\footnote{We repeat here the statement of footnote~\ref{fn:on-the-index-notation}
\vpageref{fn:on-the-index-notation}} with respect to $M$~\cite{Schwarz.2009numerische} 
\begin{align}
\frac{\partial^{2}P}{\partial x^{2}} & =\frac{P_{x+1,t}-2P_{x,t}+P_{x-1,t}}{2\Delta x^{2}}+\frac{P_{x+1,t+1}-2P_{x,t+1}+P_{x-1,t+1}}{2\Delta x^{2}}+O(\Delta x^{2}),\label{eq:diss.7.17implicit}\\
\frac{\partial P}{\partial t} & =\frac{P_{x,t+1}-P_{x,t}}{\Delta t}+O(\Delta t^{2}).\label{eq:diss.7.18implicit}
\end{align}
The limiting behaviour of each function is of order $2$ as indicated
by the Landau notation $O$. As $D_{\mathrm{t}}$ is independent of
$x$, the resulting finite difference equation is obtained by substitution
of Eqs.~(\ref{eq:diss.7.17implicit}) and (\ref{eq:diss.7.18implicit})
into (\ref{eq:ballisticDE-fdm}), thereby omitting the Landau notation
$O$, which reads after reordering as
\begin{equation}
\begin{array}{clclcl}
 & -d_{t}P_{x-1,t+1} & + & \Bigl(2+2d_{t}\Bigr)P_{x,t+1} & - & d_{t}P_{x+1,t+1}\\
= & \hphantom{-}d_{t}P_{x-1,t} & + & \Bigl(2-2d_{t}\Bigr)^{\vphantom{\vphantom{\int^{8}}}}P_{x,t} & - & d_{t}P_{x+1,t}
\end{array}\begin{gathered}\end{gathered}
\label{eq:FDMimplicit}
\end{equation}
with
\begin{equation}
d_{t}:=\frac{D_{t}\Delta t}{\Delta x^{2}}\label{eq:d_t}
\end{equation}
thereby assuming the value 
\begin{equation}
D_{t}:=D_{t+1/2}=D_{\mathrm{t}}(t+\Delta t/2)
\end{equation}
 at $M$.

A quick look at Eq.~(\ref{eq:FDMimplicit}) illuminates why the scheme
is implicit: The values of the next time step cannot be calculated
directly out of the former ones. Instead, a linear equation system
has first to be solved to obtain the solution. In comparison with
coupled map lattice, then, one has to put more effort into computer
programs.

\subsubsection*{Stability of Crank-Nicolson's method}

We set $d_{t}$~(\ref{eq:d_t}) being constant in a first step, i.e.\ $d_{t}\rightarrow d$,
and rewrite Eq.~(\ref{eq:FDMimplicit}) as

\begin{equation}
\begin{aligned} & \left(\begin{array}{cccc}
2+2d & -d\\
-d & 2+2d & -d\\
\ddots & \ddots & \ddots\\
 & -d & 2+2d & -d\\
 &  & -2d & 2+2d
\end{array}\right)\mathbf{p}_{t+1}\\
= & \left(\begin{array}{cccc}
2-2d & d\\
d & 2-2d & d\\
\ddots & \ddots & \ddots\\
 & d & 2-2d & d\\
 &  & 2d & 2-2d
\end{array}\right)\mathbf{p}_{t}
\end{aligned}
\end{equation}
or short
\begin{equation}
(2\mathbf{I}+d\mathbf{J})\mathbf{p}_{t+1}=(2\mathbf{I}-d\mathbf{J})\mathbf{p}_{t}
\end{equation}
with $\mathbf{I}$ being the identity matrix and
\begin{align}
\mathbf{J} & :=\left(\begin{array}{cccccc}
2 & -1\\
-1 & 2 & -1\\
 & -1 & 2 & -1\\
 &  & \ddots & \ddots & \ddots\\
 &  &  & -1 & 2 & -1\\
 &  &  &  & -2 & 2
\end{array}\right)\qquad\in\mathbb{R}^{n,n},\\
\mathbf{p}_{t} & :=\left(\begin{array}{c}
P_{1,t}\\
P_{2,t}\\
P_{3,t}\\
\vdots\\
P_{n-1,t}\\
P_{n,t}
\end{array}\right)
\end{align}
with $n$ being the number of nodes in $x$-direction. Because of
$d>0$ the matrix $(2\mathbf{I}+d\mathbf{J})$ is diagonal dominant
and regular, thus we obtain formally
\begin{equation}
\mathbf{p}_{t+1}=(2\mathbf{I}+d\mathbf{J})^{-1}(2\mathbf{I}-d\mathbf{J})\mathbf{p}_{t}.
\end{equation}
This method is absolutely stable if the absolute values of the eigenvalues
$\lambda_{i}$ of the matrix $(2\mathbf{I}+d\mathbf{J})^{-1}(2\mathbf{I}-d\mathbf{J})$
are less than one. Because of the form of $\mathbf{J}$ the eigenvalues
$\mu_{i}$ are real and $0<\mu_{i}<4$~\cite{Schwarz.2009numerische}
and hence
\begin{equation}
-1<\lambda_{i}=\frac{2-d\mu_{i}}{2+d\mu_{i}}<1.
\end{equation}
This proves Crank\textendash Nicolson's method absolutely stable because
the value $d$ is not restricted. For we allow any positive values
for $D_{\mathrm{t}}$ and hence any arbitrary values $d\rightarrow d_{t}=D_{\mathrm{t}}\Delta t/(\Delta x)^{2}$~(\ref{eq:d_t})
without loss of stability.

The approximation follows the exact solution at least with $O(\Delta x^{2})$
and $O(\Delta t^{2})$, respectively, and converges thus $10$ times
faster than coupled map lattices. However, the iteration steps must
not be chosen too big because, though stability is given, the approximation
error increases~$\propto O(\Delta x^{2}+\Delta t^{2})$. The proves
can be found in textbooks, e.g.\ from Toffoli \textit{et al.}~\cite{Toffoli.1987cellular},
Schwarz and Köckler~\cite{Schwarz.2009numerische}, or Haas~\cite{Haas.1999numerische}.

\subsection{Comparison of the finite difference schemes}

We compared two finite difference schemes and provided a short overview
on advantages and restrictions in both cases. The coupled map lattices,
as an example of an explicit scheme~(\ref{eq:FDMexplicit}), has
its most advantageous feature definitely in its quick and easy implementation
at the cost of problematic restrictions on the step width. For the
Crank-Nicolson method, as an example of an implicit scheme~~(\ref{eq:FDMimplicit}),
the implementation task is rather on the expensive side because of
the equations solvers needed for, while its advantage lies in its
convergence behaviour for any step width. For an overview see Tab.~(\ref{tab:FDMschemes}).
\begin{table}[!h]
\selectlanguage{english}%
\begin{centering}
\renewcommand*\arraystretch{2.5}\foreignlanguage{british}{}%
\begin{tabular}{|c|c|c|c|}
\hline 
\selectlanguage{british}%
Scheme\selectlanguage{british}%
 & \selectlanguage{british}%
Stable\selectlanguage{british}%
 & \selectlanguage{british}%
Error\selectlanguage{british}%
 & \selectlanguage{british}%
Comment\selectlanguage{british}%
\tabularnewline
\hline 
\hline 
\selectlanguage{british}%
Coupled map lattices\selectlanguage{british}%
 & \selectlanguage{british}%
$\ensuremath{{\displaystyle {\displaystyle \Delta t_{\mathrm{max}}\leq\frac{\Delta x\sigma_{0}}{D\sqrt{2T_{\mathrm{max}}}}}}}$\selectlanguage{british}%
 & \selectlanguage{british}%
$O(\Delta x+\Delta t)$\selectlanguage{british}%
 & \selectlanguage{british}%
easy to implement\selectlanguage{british}%
\tabularnewline
\hline 
\selectlanguage{british}%
Crank\textendash Nicolson\selectlanguage{british}%
 & \selectlanguage{british}%
yes\selectlanguage{british}%
 & \selectlanguage{british}%
$O(\Delta x^{2}+\Delta t^{2})$\selectlanguage{british}%
 & \selectlanguage{british}%
converges always\selectlanguage{british}%
\tabularnewline
\hline 
\end{tabular}\bigskip{}
\par\end{centering}
\selectlanguage{british}%
\begin{centering}
\caption{Overview on the two compared finite difference schemes.\label{tab:FDMschemes}}
\par\end{centering}
\end{table}

For our simulations within this thesis we employed both coupled map
lattices as well as Crank\textendash Nicolson's method. For both of
which we developed on a standard personal computer using \textit{Scilab}~\cite{Campbell.2010modeling}
and recently also \textit{Julia language}~\cite{Bezanson.2014julia:},
two open source packets for numerical computation.

\section{The simulation procedure\label{sec:8.2.Simulation-procedure}}

The simulation procedure, which is schematically shown in Fig.~\ref{fig:Diss.simulation-procedure},
comprises the following steps to simulate solutions according to the
ballistic diffusion equation~(\ref{eq:ballisticDE}):
\begin{sidewaysfigure}
\noindent \begin{centering}
\includegraphics[width=1\textwidth]{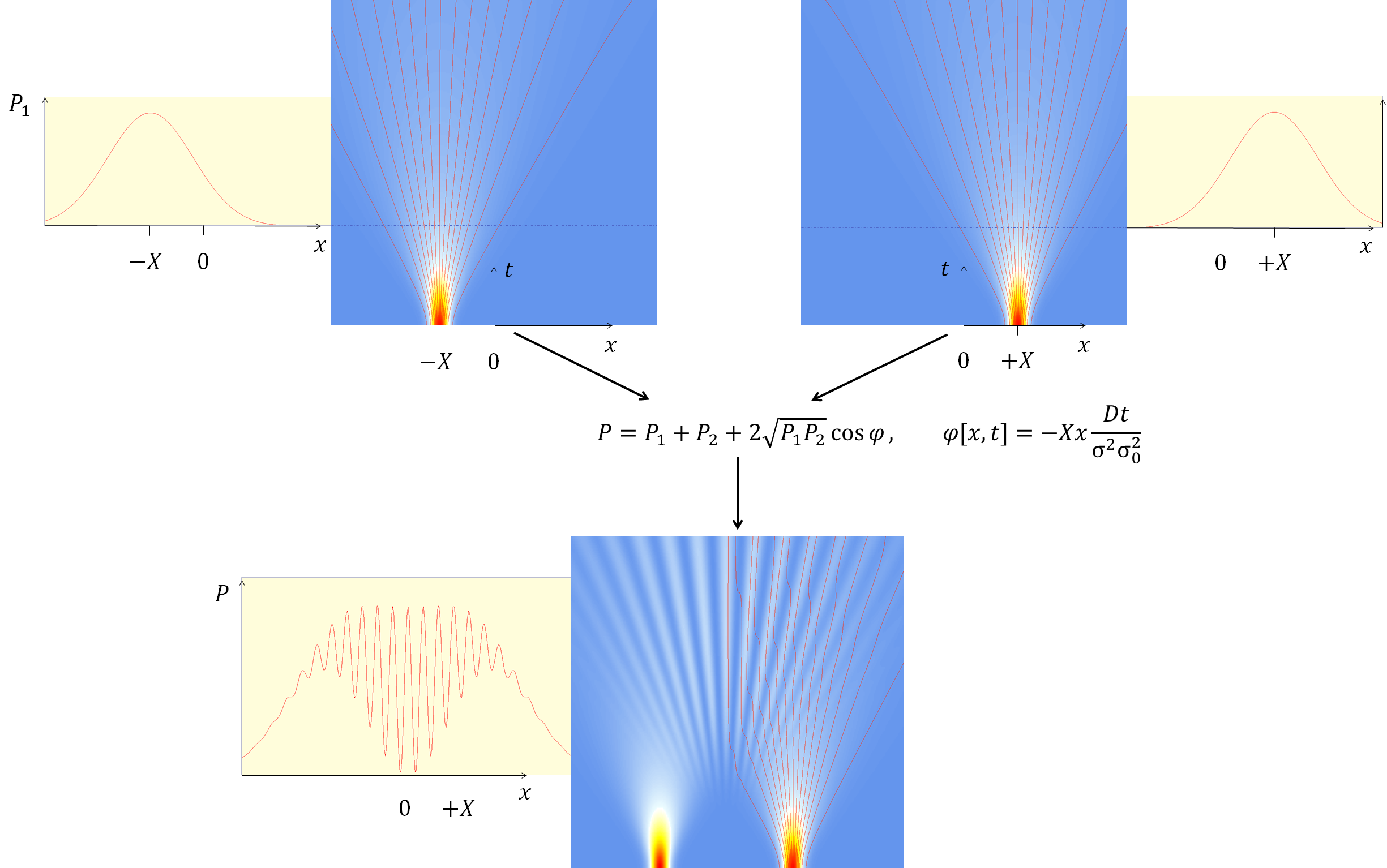}
\par\end{centering}
\caption{Simulation procedure steps, exemplarily shown for a probability density
$P$. Averaged trajectories are shown in red\label{fig:Diss.simulation-procedure}}
\end{sidewaysfigure}

\begin{enumerate}
\item Define an initial probability distribution $P$ as in Eq.~(\ref{eq:Pi(x,0)}),
\item Compute the spreading:~(\ref{eq:FDMexplicit}) or (\ref{eq:FDMimplicit}),
\item Calculate the associated phase $\varphi$ according to Eqs.~(\ref{eq:td.40})
and (\ref{eq:td.46})
\item Combine to

\begin{enumerate}
\item either a total probability distribution~(\ref{eq:Ptot2slit}),
\item or a total probability current~(\ref{eq:Jfinal}).
\end{enumerate}
\end{enumerate}
Accordingly, with this procedure we simulate intensity probabilities
as well as current distributions. 

\section{Trajectories\label{sec:Trajectories}}

If one considers a particle as a walker obeying a Brownian-type motion
including the zitterbewegung, then the resulting trajectory would
be erratic and thus of little usefulness for the purpose of repeated
experiments (see section~\ref{sec:The-constituting-setup} for further
explanation). Therefore, the particle's trajectories in the pictures
within this thesis are the results of averaging of a huge number of
such walkers, in the mean obeying a Bohmian-type trajectory which
is sufficiently smooth to explain repeated experiments then. The emerging
trajectories are in full accordance with those obtained from the Bohmian
approach, as can be seen by comparison with references~\cite{Holland.1993,Bohm.1993undivided,Sanz.2009context,Sanz.2012trajectory,Davidovic.2015description,deGosson.2016observing},
for example.

Accordingly, trajectories of Bohmian-type are shown, which are always
computed from the underlying probability distribution $P$. On some
occasions the distances between two single, adjacent trajectories
differ for didactic reasons, 
\begin{itemize}
\item as in Fig.~\ref{fig:1}, for example, where each two single trajectories
are chosen equally spaced, and hence the trajectories are initially
equidistant,
\item as in Fig.~\ref{fig:td.3}, for example, where the flux, i.e.\ every
value $\Delta P$, between any two adjacent trajectories is equal
and kept constant, hence the trajectories reach their highest density
around the maximum of the intensity distribution.
\end{itemize}
While the former method of displaying trajectories is mostly used
for comparison reasons with older pictures in literature, the latter
one gives a better idea of properties.

In most of the pictures the same number of trajectories for each Gaussian
is used thereby resulting in well proportioned figures as long as
the distributions possess about the same intensities. However, if
the relation of the intensities differ considerable, this easy recipe
fails. For example, in figures in chapter~\ref{sec:6.extreme-beam-attenuation}
the number of trajectories are chosen to be equal for each slit thereby
resulting in sweeper effects comprising trajectories that do not provide
the correct physical proportions: Thus, if one maintained the trajectories
of said sweeper-figures to enclose the same amount of flux for both
beams at the same time, and thus for the whole picture, then either
the low-intensity beam had no visible trajectories or the high-intensity
beam had too many trajectories so that one couldn't distinguish between
the single lines.

\section{Calibrating the simulation tools\label{sec:Calibrating-the-simulation}}

The double-slit experiment is of particular interest and therefore
there is a bunch of measured data available.
\begin{figure}
\begin{centering}
\includegraphics[width=1\textwidth]{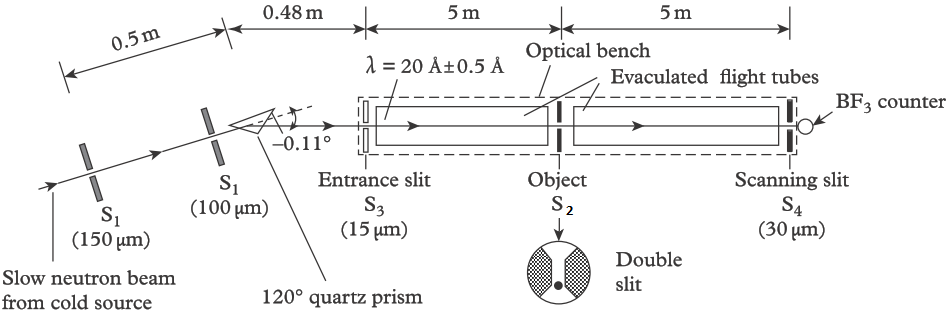}
\par\end{centering}
\caption{Experimental setup. From~\cite{Zeilinger.1988single-,Rauch.2015neutron}\label{fig:8.4.Experimental-setup}}
\end{figure}
\begin{figure}[!htb]
\begin{centering}
\subfloat[\foreignlanguage{english}{Probability distribution\label{fig:diss.8.4a}}]{\centering{}\includegraphics[width=0.45\textheight]{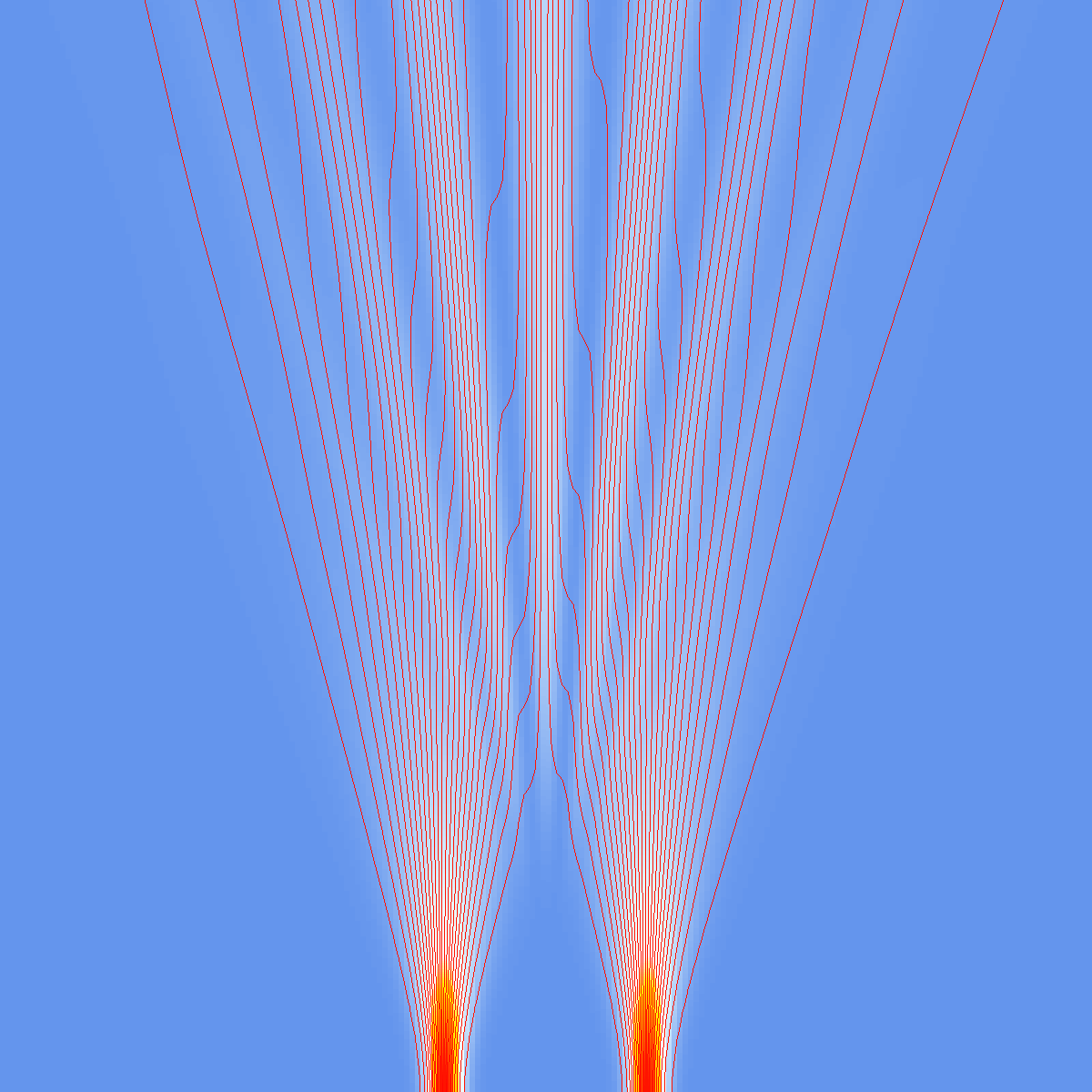}}
\par\end{centering}
\centering{}\subfloat[Intensity recorded at $y=5\thinspace\mathrm{m}$ comprising the measured
curve (black) from~\cite{Zeilinger.1988single-}, and the simulation's
result (red)\label{fig:diss.8.4b}]{\centering{}\includegraphics[width=0.6\textheight]{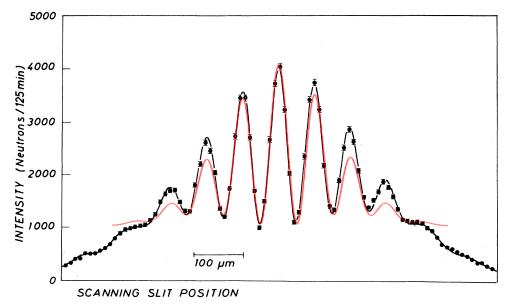}}\caption{Classical computer simulation of the interference pattern for $\lambda=1.8\thinspace\mathrm{nm}$,
the slit width is $22\thinspace\mu\mathrm{m}$ each, with their centres
being $200\thinspace\mu\mbox{m}$ apart\label{fig:diss.8.4}}
\end{figure}
In an actual experiment as sketched in Fig.~\ref{fig:8.4.Experimental-setup},
the double-slit diffraction of neutrons has been measured~\cite{Rauch.2000neutron,Zeilinger.1988single-}.
The typical wavelength used is $\lambda=1.845\thinspace\mathrm{nm}$.
The Gaussian slit width is $21.9\thinspace\mu\mathrm{m}$ and $22.5\thinspace\mu\mathrm{m}$,
respectively with their centres being $126.3\thinspace\mu\mbox{m}$
apart, and the intensity distributions are recorded on a screen $S_{4}$
located in the forward direction at a distance of $5\thinspace\mbox{m}$
from the object slit $S_{5}$.

In Fig.~\ref{fig:diss.8.4} we show the results of our computer simulations
for the probability density distributions of a neutron beam using
the parameters of this experiment.

Comparing these simulations' results with actual measurement data
in Rauch and Werner~\cite{Rauch.2000neutron} as well as from Zeilinger~et\,al.~\cite{Zeilinger.1988single-}
enabled us to adjust the parameters. It turned out that there is a
certain ratio between the slit width and $\sigma_{0}$ to be maintained
that is around
\begin{equation}
\sigma_{0}\approx\frac{\mathrm{slit\,width}}{3}
\end{equation}
such that the correct shape of the intensity recorded at $y=5\thinspace\mathrm{m}$
can be ensured.

Even though the curves do not perfectly fit, the result is sufficiently
accurate, taking into account, that the actual measurement did not
have taken place with idealized Gaussians but with real neutron beams.
Zeilinger~et\,al.~\cite{Zeilinger.1988single-} carried out in
their paper how they compared the measured data with theory. In fact,
they integrated of course the whole length of the optical bench, i.e.\ in
Fig.~\ref{fig:8.4.Experimental-setup} this corresponds to the paths
from $S_{3}$ to $S_{4}$. In our model this is not possible as we
do not yet allow objects in the path. Thus, our path contains the
second half of the optical bench, i.e.\ in Fig.~\ref{fig:8.4.Experimental-setup}
corresponding to the paths from $S_{2}$ to $S_{4}$, thereby assuming
an idealized Gaussian behind $S_{5}$. The scope of this thesis is
to simulate a Gaussian beam in one dimension without diffraction,
therefore, modelling diffracted Gaussians would need further investigation.

\section{Conclusions and outlook}

In this chapter, the simulation means for obtaining probability distributions
as well as density currents has been provided. Preliminarily, the
setup, the phase conditions and the diffusion coefficient for different
slit widths has been discussed.

As a numerical means to solve the ballistic diffusion equation two
finite different schemes have been introduced. The first one, coupled
map lattices as an example for an explicit scheme has been shown to
be beneficial for exploration of the dynamical behaviour bringing
in the advantage of easy implementation. The second one, the implicit
scheme of Crank\textendash Nicolson has been proven to be unconditionally
stable for the cost of much more computational effort, however, it
allows obtaining solutions independent of the domain even in situations
where coupled map lattices collapse.

The method of constructing trajectories which are part of the numerical
procedure, has been explained.

Finally, a calibration procedure has been provided: The comparison
of measured results from a neutron-experiment with simulated results
of the same setup has yielded the relation between the slit width
and the initial half-width of the Gaussian $\sigma_{0}$.

\clearpage{}

\appendix

\chapter{Classical mechanics and Boltzmann's relation\label{chap:boltz}}
\begin{quote}
The equations of mechanics can be deduced from a least action principle\index{action!least action principle},
where usually the varied path in configuration space always terminates
at end points representing the system configuration at the same times,
$t_{0}$ and $t_{1}$, as the natural path.

In the following one starts with the derivation of a less constrained
$\vardelta$\textendash variation with a varied path over which an
integral is evaluated that may end at other times than the natural
path, and there may be a variation in the coordinates at the end points.
By defining a relation between heat and mechanical work one follows
the thoughts of Brillouin~\cite[Chapter 11]{Brillouin.1964tensors},
and to some extent of Goldstein~\cite{Goldstein.2002classical},
Scheck~\cite{Scheck.2010mechanics}, Hamel~\cite[pp.312-314]{Hamel.1967theoretische},
and Hand~\cite[pp.230ff]{Hand.1998analytical} leading directly to
the Boltzmann relation of periodic motion.
\end{quote}

\section{The principle of least action\label{sec:boltz.principle}}

We consider a general problem with time-dependent holonomic constraints.
With kinetic energy $T$\nomenclature[Tq]{$T(q^k, \dot{q}^k, t)$}{kinetic energy},
potential energy $V$\nomenclature[Vq]{$V(q^k, t)$}{potential energy},
time $t$ \nomenclature[t]{$t$}{time}, generalized coordinates $q^{k}$
and velocities $\dot{q}^{k}$\nomenclature[qkdot]{$\dot{q}^k$}{velocity},
$k=1,\ldots,r$ ($r$ being the remaining coordinates), we form then
the Lagrangian function\nomenclature[Langrangian]{$L(q^k,\dot{q} ^k,t)$}{Lagrangian function}
\begin{equation}
L(q^{k},\dot{q}^{k},t)=T(q^{k},\dot{q}^{k},t)-V(q^{k},t).\label{eq:boltz.1}
\end{equation}
We have further the momentum\nomenclature[pk]{$p_k$}{momentum} $p_{k}$
conjugate to the coordinate $q^{k}$ given by
\begin{equation}
p_{k}=\frac{\p T}{\p\dot{q}^{k}}=\frac{\p L}{\p\dot{q}^{k}}\,,\label{eq:boltz.2}
\end{equation}
and Lagrange's equation\index{Lagrange's equation} takes the form
\begin{equation}
\frac{\d p_{k}}{\d t}=\frac{\p L}{\p q^{k}}\,.\label{eq:boltz.3}
\end{equation}

We will study the value of the action integral\index{action!integral}
\begin{equation}
S=\intop L\d t\label{eq:boltz.4}
\end{equation}
during the evolution of the system.

For the $\delta$-variation\index{delta-variation@$\delta$-variation}\nomenclature[delta]{$\delta$}{variation with fixed end points and fixed times}
the varied path always terminates at end points representing the system
configuration at the same times, $t_{0}$ and $t_{1}$, as the natural
path. To obtain Lagrange's equations of motion, it is also required
that the varied path returns to the same end points in configuration
space, i.e.\ $\delta q^{k}(t_{0})=\delta q^{k}(t_{1})=0$. \cite{Brillouin.1964tensors,Goldstein.2002classical}

\begin{figure}[!htb]
\centering{}%
\begin{minipage}[c][1\totalheight][t]{0.6\columnwidth}%
\begin{center}
\includegraphics{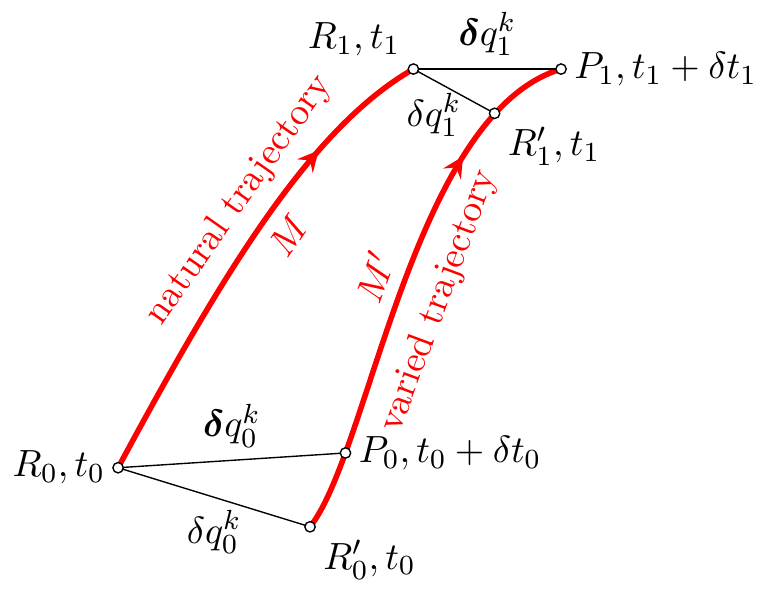}
\par\end{center}%
\end{minipage}\hspace*{\fill}%
\begin{minipage}[c][1\totalheight][t]{0.35\columnwidth}%
\begin{center}
\caption{The $\protect\vardelta$-variation in configuration space, composed
of a variation in space, $\delta q^{k}$, and, additionally, of a
variation in time, $\delta t$.\label{fig:boltz.1}}
\par\end{center}%
\end{minipage}
\end{figure}

Now, we define a less constrained $\vardelta$-variation\nomenclature[deltabold]{$\vardelta$}{variation with possible variation in end points as well as in times}
(note the bold $\delta$-symbol) according to Fig.~\ref{fig:boltz.1}
with a varied path over which an integral is evaluated that may end
at other times than the natural path, i.e.\ the paths have different
throughput times, and there may be an additional variation in the
coordinates at the end points.

As shown in Fig.~\ref{fig:boltz.1} the two usual variations, $\delta q^{k}$
in space and $\delta t$ in time, lead to the $\vardelta$-variation
of the space variable by the relation
\begin{equation}
\vardelta q^{k}=\delta q^{k}+(\dot{q}^{k}+\delta\dot{q}^{k})\delta t\approx\delta q^{k}+\dot{q}^{k}\delta t\label{eq:boltz.5}
\end{equation}
with negligible second order correction $\delta\dot{q}^{k}\delta t$.
The variation of action integral $S$\nomenclature[S]{$S$}{action integral},
i.e.
\begin{equation}
\vardelta S=\vardelta\intop_{t_{0}}^{t_{1}}L\d t,\label{eq:boltz.6}
\end{equation}
is the difference of the action of the natural trajectory $M$ from
$R_{0}$ to $R_{1}$ and the action of the varied trajectory $M'$
from $R_{0}$ to $R_{1}$. Thus we can rewrite Eq.~\eqref{eq:boltz.6}
as
\begin{equation}
\vardelta S=\intop_{t_{0}+\delta t_{0}}^{t_{1}+\delta t_{1}}L(q^{k}+\vardelta q^{k})\d t-\intop_{t_{0}}^{t_{1}}L(q^{k})\d t
\end{equation}
with the Lagrangian function $L$ along the varied trajectory in the
first integral and $L$ along the natural trajectory in the second
integral. We separate the integration over the terminal segments $R_{0}'P_{0}$
and $R_{1}'P_{1}$ and obtain
\begin{align}
\vardelta S= & \underbrace{L(q_{1}^{k}+\delta q_{1}^{k})\delta t_{1}}_{Q_{1}'P_{1}}-\underbrace{L(q_{0}^{k}+\delta q_{0}^{k})\delta t_{0}}_{Q_{0}'P_{0}}+\underbrace{\intop_{t_{0}}^{t_{1}}L(q^{k}+\delta q^{k})\d t}_{M'}-\underbrace{\intop_{t_{0}}^{t_{1}}L(q^{k})\d t}_{M}\nonumber \\
= & L(q_{1}^{k}+\delta q_{1}^{k})\delta t_{1}-L(q_{0}^{k}+\delta q_{0}^{k})\delta t_{0}+\intop_{t_{0}}^{t_{1}}\delta L\d t.\label{eq:boltz.8}
\end{align}
Here the variation in the last integral can be carried out through
a parametrization of the varied path, 
\begin{equation}
\delta L=L(q^{k}+\delta q^{k})-L(q^{k})=\sum_{k}\left(\frac{\p L}{\p q^{k}}\delta q^{k}+\frac{\p L}{\p\dot{q}^{k}}\delta\dot{q}^{k}+\frac{\p L}{\p t}\delta t\right),\label{eq:boltz.9}
\end{equation}
where the last term in the bracket vanishes because we have chosen
two simultaneous positions and hence $\delta t=0$. We integrate the
second term by parts, using the exchange relation $\delta\left(\frac{\d\cdot}{\d t}\right)=\frac{\d}{\d t}\left(\delta\cdot\right)$
for $\dot{q}^{k}$and obtain
\begin{equation}
\frac{\p L}{\p\dot{q}^{k}}\frac{\d}{\d t}\left(\delta q^{k}\right)=\frac{\d}{\d t}\left(\frac{\p L}{\p\dot{q}^{k}}\delta q^{k}\right)-\delta q^{k}\frac{\d}{\d t}\left(\frac{\p L}{\p\dot{q}^{k}}\right)\,.
\end{equation}
Substitution of these expressions into Eq.~\eqref{eq:boltz.9} leads
to
\begin{equation}
\intop_{t_{0}}^{t_{1}}\delta L\d t=\intop_{t_{0}}^{t_{1}}\sum_{k}\left[\frac{\p L}{\p q^{k}}-\frac{\d}{\d t}\left(\frac{\p L}{\p\dot{q}^{k}}\right)\right]\delta q^{k}\d t+\sum_{k}\left.\frac{\p L}{\p\dot{q}^{k}}\delta q^{k}\right|_{R_{0}}^{R_{1}}.\label{eq:boltz.11}
\end{equation}
On account of Eqs.~\eqref{eq:boltz.2} and \eqref{eq:boltz.3}, the
equation within the square brackets of the integral disappears entirely.
Now we substitute Eqs.~\eqref{eq:boltz.9} and \eqref{eq:boltz.11}
into~\eqref{eq:boltz.8} and find 
\begin{equation}
\vardelta S=\left[L(q_{1}^{k})+\delta L\right]\delta t_{1}-\left[L(q_{0}^{k})+\delta L\right]\delta t_{0}+\sum_{k}\left.\frac{\p L}{\p\dot{q}^{k}}\left(\vardelta q^{k}-\dot{q}^{k}\delta t\right)\right|_{R_{0}}^{R_{1}}.
\end{equation}
We identify $L(q_{i}^{k})=L_{i}$, neglect the second order terms
and reorder to find our final result as
\begin{equation}
{\displaystyle \vardelta S=H_{0}\delta t_{0}-H_{1}\delta t_{1}-\sum_{k}p_{0k}\vardelta q_{0}^{k}+\sum_{k}p_{1k}\vardelta q_{1}^{k}}.\label{eq:boltz.13}
\end{equation}
Here, we substituted Eq.~\eqref{eq:boltz.2} and introduced the Hamiltonian\nomenclature[Hamiltonian]{$H(q^k,\dot{q} ^k,t)$}{Hamiltonian}
\begin{equation}
H_{i}=\sum_{k}p_{ik}\dot{q_{i}}^{k}-L(q_{i}^{k},\dot{q}_{i}^{k},t_{i}).
\end{equation}

\subsection{The conservative case}

Along with the integral $S$ defined in Eq.~\eqref{eq:boltz.4} we
shall consider the \textit{abbreviated action\index{action!abbreviated}}\nomenclature[F]{$F$}{abbreviated action integral}
\begin{align}
F= & 2\intop T\d t=\intop\sum_{k}p_{k}\dot{q}^{k}\d t=\intop\sum_{k}p_{k}\d q^{k}.\label{eq:boltz.15}
\end{align}
Taking into account
\begin{equation}
H=\sum_{k}p_{k}\dot{q}^{k}-L=2T-L=T+V=E.\label{eq:boltz.16}
\end{equation}
For conservative systems\index{conservative systems} the total energy
$E$ remains constant, $H_{0}=H_{1}=E$, and from \eqref{eq:boltz.13}
we find
\begin{equation}
\vardelta S=E(\delta t_{0}-\delta t_{1})-\sum_{k}p_{0k}\vardelta q_{0}^{k}+\sum_{k}p_{1k}\vardelta q_{1}^{k}.\label{eq:boltz.18}
\end{equation}
We reconsider Eq.~\eqref{eq:boltz.4} and set up the equation connecting
$F$ with action $S$ by
\begin{equation}
S=\intop L\d t=\int(T-V)\d t=\intop(2T-E)\d t=F-\intop E\d t.\label{eq:boltz.17}
\end{equation}

We compare the values of the integrals $F$ taken along two neighbouring
trajectories, the natural and a nearby entirely arbitrary trajectory.
On the natural trajectory, the total energy $E$ remains constant,
but this is not so on the varied trajectory. We obtain then 
\begin{align}
\vardelta F= & \vardelta S+\vardelta\intop E\d t=\vardelta S+\intop\delta E\d t+E(\delta t_{1}-\delta t_{0}),\label{eq:boltz.19}
\end{align}
where the last term in~\eqref{eq:boltz.19} is an expression for
the variation at the endpoints of the trajectory from 0 to 1. Substitution
of Eq.~\eqref{eq:boltz.18} into \eqref{eq:boltz.19} yields
\begin{equation}
\vardelta F=\intop\delta E\d t+\sum_{k}p_{1k}\vardelta q_{1}^{k}-\sum_{k}p_{0k}\vardelta q_{0}^{k}\label{eq:boltz.20}
\end{equation}
as a general result. A nearby trajectory, although entirely arbitrary,
is only subject to the conditions of respecting constraints~\cite{Brillouin.1964tensors,Hamel.1967theoretische}.
Now, we investigate the influence of a modification of such constraints.

\subsection{Reduced constraints\label{sec:boltz.reduced-constraints}}

We consider, again, a system of $N$ mass points defined by their
$3N$ position coordinates. We further suppose that there exist $l$
holonomic constraints among these points so that there remains only
\begin{equation}
r=3N-l
\end{equation}
independent degrees of freedom. We assume a conservative system characterized
by time-independent holonomic constraints, hence we can define a total
energy $E$ remaining constant in time during the natural evolution
of the system.

We find
\begin{equation}
E=T+V=\textrm{const.},\qquad L=T-V=2T-E\label{eq:boltz.22}
\end{equation}
for the natural motion of the conservative system, the usual Lagrangian
$L$ referring only to the visible coordinates $q^{1},\ldots,q^{r}$.
Now, we allow for a variation of constraints and we will use the asterisk
$^{*}$ to indicate the overall quantities containing the independent
coordinates $q^{r+1},q^{r+2},\ldots,q^{3N}$. So as not to give useless
complication to the equations, we take it, that the forces guaranteeing
the constraints are derived from a potential energy $V^{*}$ by 
\begin{equation}
V^{*}=\sum_{k=r+1}^{3N}A_{k}(q^{k})^{2}
\end{equation}
with very large positive coefficients $A_{k}$ thereby guaranteeing
very small $q^{k}$s. This form corresponds to the hypothesis that
the constraints are realized by very rigid elastic systems. A small
displacement $q^{k}$ brings into action a very great force $-2A_{k}q^{k}$
which opposes this change. The coordinates $q^{r+1},q^{r+2},\ldots,q^{3N}$
then remain practically constant, their corresponding velocities $\dot{q}^{k}$
vanish; the corresponding momenta $p_{k}$ however will not always
vanish due to their dependence on $\dot{q}_{i}^{k}$, $i=1,\ldots,l=3N-r$.

The kinetic energy $T$ is unchanged in the natural motion, for, all
the velocities $q^{r+1},q^{r+2},\ldots,q^{3N}$ of the hidden coordinates
are practically constant (and zero) for this trajectory and hence
$p_{k}\dot{q}^{k}\approx0$ for $k=r+1,\ldots,2N$. With these definitions
we find the total energy as
\begin{equation}
E^{*}=E+V^{*},\label{eq:boltz.energy24}
\end{equation}
including a new term coming from the new potential energy $V^{*}$.
The complete Lagrangian function reads
\begin{equation}
L^{*}=T-V-V^{*}=L-V^{*}\label{eq:boltz.26}
\end{equation}
with $L$ being the usual Lagrangian referring only to the visible
coordinates $q^{1},\ldots,q^{r}$. Note that $E^{*}$and $L^{*}$
in Eqs.~\eqref{eq:boltz.energy24} and \eqref{eq:boltz.26} are related
to the natural trajectory.

On a varied trajectory, the kinetic energy $T^{*}$ changes. In this
case we write
\begin{equation}
2T^{*}=2T+\sum_{k=r+1}^{3N}p_{k}\dot{q}^{k},\label{eq:boltz.24}
\end{equation}
and
\begin{equation}
F^{*}=F+\sum_{k=r+1}^{3N}\intop p_{k}\dot{q}^{k}\d t=F+\sum_{k=r+1}^{3N}\intop p_{k}\d q^{k}.\label{eq:boltz.27}
\end{equation}
We can thus apply Eq.~\eqref{eq:boltz.20} to our system with the
quantities marked with asterisks and we obtain 
\begin{align}
\vardelta F^{*}= & \intop\delta E^{*}\d t+\sum_{k=1}^{3N}p_{1k}\vardelta q_{1}^{k}-\sum_{k=1}^{3N}p_{0k}\vardelta q_{0}^{k},\label{eq:boltz.28}
\end{align}
indicating visible and hidden coordinates, whereas Eq.~\eqref{eq:boltz.20}
contained only the $r$ visible coordinates. Equation~\eqref{eq:boltz.15}
must also hold for the constraints, thus
\begin{equation}
\vardelta F^{*}=\vardelta\intop2T^{*}\d t=\vardelta F+\vardelta\intop\sum_{k=r+1}^{3N}p_{k}\d q^{k}=\vardelta F+\intop\delta\sum_{k=r+1}^{3N}p_{k}\d q^{k}+\left.\sum_{k=r+1}^{3N}p_{k}\vardelta q^{k}\right|_{0}^{1}.\label{eq:boltz.29}
\end{equation}
Returning now to the quantities without asterisks we get by substituting
Eq.~\eqref{eq:boltz.29} into \eqref{eq:boltz.28} that
\begin{equation}
\vardelta F=\intop\left(\delta E+\delta V^{*}-\delta\sum_{k=r+1}^{3N}p_{k}\dot{q}^{k}\right)\d t+\sum_{k=1}^{r}p_{1k}\vardelta q_{1}^{k}-\sum_{k=1}^{r}p_{0k}\vardelta q_{0}^{k}\label{eq:boltz.30}
\end{equation}
 because the term $\sum_{k=r+1}^{3N}p_{1k}\vardelta q_{1}^{k}-\sum_{k=r+1}^{3N}p_{0k}\vardelta q_{0}^{k}$
cancels.

\section{A thermodynamical analogy\label{sec:boltz.thermodynamical}}

\begin{figure}[!htb]
\centering{}%
\begin{minipage}[c][1\totalheight][t]{0.6\columnwidth}%
\begin{center}
\includegraphics{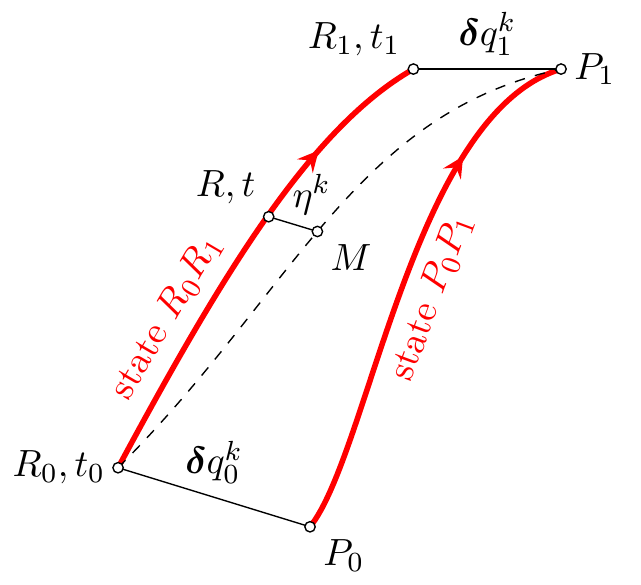}
\par\end{center}%
\end{minipage}\hspace*{\fill}%
\begin{minipage}[c][1\totalheight][t]{0.35\columnwidth}%
\begin{center}
\caption{A very slow transformation from a physical state $R_{0}R_{1}$ to
the physical state $P_{0}P_{1}$ leads to Boltzmann's formula.\label{fig:boltz.2}}
\par\end{center}%
\end{minipage}
\end{figure}

To carve out the thermodynamical analogy we suppose a given physical
state $R_{0}R_{1}$ represented by a first trajectory as shown in
Fig.~\ref{fig:boltz.2}. Suppose we wish to make a transition of
state $R_{0}R_{1}$, characterized by different pressure, volume,
and temperature, into state $P_{0}P_{1}$, represented by another
trajectory. These two trajectories correspond to different constant
values of the coordinates $q^{r+1},\ldots,q^{3N}$, called the macroscopic
coordinates in the thermodynamical sense.

We must have special forces that are capable of acting on all the
molecules, and these forces supply work. The work supplied by these
forces will be equivalent to the heat supplied to the system. If in
this transition the volume is changed, external work will be done
against the forces which cause the constraints $\delta W=\delta V^{*}$.

The heat supplied to the system will, on the one hand, increase the
total internal energy $E$ and, on the other hand, furnish the external
work $W$. We have then
\begin{equation}
\delta Q=\delta E+\delta W=\delta E+\delta V^{*}\label{eq:A.31}
\end{equation}
which is the heat supplied to the system ($\delta Q$), the increase
of disordered internal energy ($\delta E$) and the ordered work furnished
by the system against constraint mechanism ($\delta W=\delta V^{*}$)
according to Boltzmann~\cite{Boltzmann.1866uber} (see also~\cite{Groessing.2008vacuum,Groessing.2009origin}).

We assume a continuous and gradual transition from the trajectory
$R_{0}R_{1}$, characterized by certain constant values of $q^{r+1},\ldots,q^{3N}$,
to the trajectory $P_{0}P_{1}$, characterized by values of the coordinates
$q^{k}+\vardelta q^{k}$, that starts at time $t_{0}$ and ends at
time $t_{1}$, represented by the path $R_{0}MP_{1}$. At time $t$
the ratio of change between the two states is represented by the segment
$RM$ given as
\begin{equation}
\eta^{k}=\frac{t-t_{0}}{t_{1}-t_{0}}\vardelta q^{k}.
\end{equation}
In a time $\d t,$ the change is
\begin{equation}
\d\eta^{k}=\frac{\d t}{t_{1}-t_{0}}\vardelta q^{k}.
\end{equation}

The heat furnished to the system in the time $\d t$ to bring about
the change is
\begin{equation}
\d(\delta Q)=\frac{\d t}{t_{1}-t_{0}}\delta Q.
\end{equation}
The work done by the system is then
\begin{equation}
\d(\delta V^{*})=\frac{\d t}{t_{1}-t_{0}}\delta V^{*}
\end{equation}
where the definitions are exactly those used before, $\delta Q$ and
$\delta V^{*}$ being the quantities defined for a sudden jump and
$\d(\delta Q)$, $\d(\delta V^{*})$ being the same quantities for
an infinitesimal transformation. 

The total supply of heat $\Delta Q$ given to the system during time
$t_{1}-t_{0}$ of the transformation with the use of \eqref{eq:A.31}
reads
\begin{equation}
\Delta Q=\intop_{t_{0}}^{t_{1}}\d(\delta Q)=\frac{1}{t_{1}-t_{0}}\intop_{t_{0}}^{t_{1}}\delta Q\d t=\frac{1}{t_{1}-t_{0}}\intop_{t_{0}}^{t_{1}}(\delta E+\delta V^{*})\d t.\label{eq:boltz.34}
\end{equation}
To compare this result with integral~\eqref{eq:boltz.30} we make
the hypothesis that \textit{the varied motion keeps the values of
the hidden coordinates $q^{r+1}\ldots q^{3N}$ constant and very small}.
Under those circumstances the velocities $\dot{q}^{k}$, $k=r+1,\ldots,3N$,
would vanish in the varied motion as it does in the natural motion,
and the term
\begin{equation}
\delta\sum_{k=r+1}^{3N}p_{k}\dot{q}^{k}=0
\end{equation}
disappears in Eq.~\eqref{eq:boltz.30}. Therefore, we find a general
form of Boltzmann's formula by substitution of Eq.~\eqref{eq:boltz.30}
into \eqref{eq:boltz.34} as
\begin{align}
\Delta Q= & \frac{1}{t_{1}-t_{0}}\left(\vardelta F-\left.\sum_{k=1}^{r}p_{k}\vardelta q^{k}\right|_{t_{0}+\delta t_{0}}^{t_{1}+\delta t_{1}}\:\right).\label{eq:boltz.36}
\end{align}

\subsection{Periodic motions}

\begin{figure}[H]
\centering{}%
\begin{minipage}[c][1\totalheight][t]{0.6\columnwidth}%
\begin{center}
\includegraphics{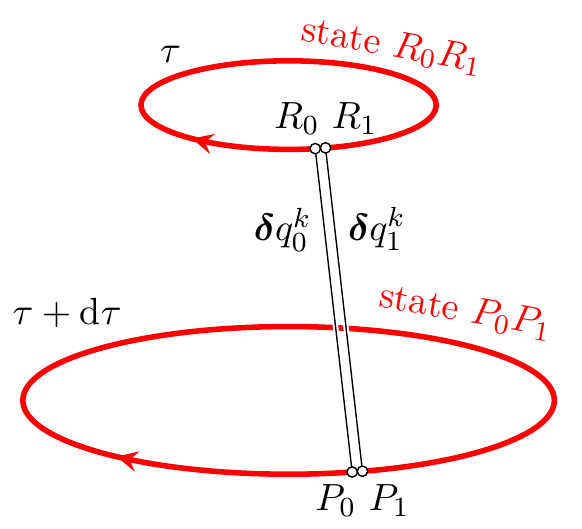}
\par\end{center}%
\end{minipage}\hspace*{\fill}%
\begin{minipage}[c][1\totalheight][t]{0.35\columnwidth}%
\begin{center}
\caption{The $\protect\vardelta$\textendash variation adapted to periodic
motion with each start point, $R_{0}$ and $P_{0}$, connected to
its corresponding end point, $R_{1}$ and $P_{1}$, respectively.\label{fig:boltz.3}}
\par\end{center}%
\end{minipage}
\end{figure}

At that point, we move one step further and close the trajectories
of Fig.~\ref{fig:boltz.2} which yields a periodic configuration
as provided in Fig.~\ref{fig:boltz.3}. In this special case the
two points $R_{0}$ and $R_{1}$ coincide, as well as the two points
$P_{0}$, $P_{1}$ of the varied trajectory.
The $\vardelta q_{i}^{k}$ are equal, and also the momenta $p_{ik}$,
\begin{equation}
\vardelta q_{0}^{k}=\vardelta q_{1}^{k},\qquad p_{0k}=p_{1k}.
\end{equation}
We apply these relations to Eq.~\eqref{eq:boltz.36} where the sum
disappears. We then find for a cyclic motion with period $\tau=t_{1}-t_{0}=2\pi/\omega$
that 
\begin{equation}
\Delta Q=\frac{1}{\tau}2\vardelta\intop_{0}^{\tau}T\d t.\label{eq:boltz.38}
\end{equation}

Suppose that we could make a canonical transformation (c.f.~\cite{Hand.1998analytical,Goldstein.2002classical})
from variables $p$, $q$ to a different, but still canonical, set
of variables $I$, $\psi$, whereby in terms of the new variables
the new Hamiltonian lacks any dependence on $\psi,$ i.e.\ $H=H(I)$.
Because $H$ is constant in this periodic system and depends only
on $I$, $I$ itself must be a constant of the motion, thus
\begin{equation}
\dot{I}=-\frac{\p H}{\p\psi}=0,\qquad\dot{\psi}=\frac{\p H}{\p I}=\mathrm{const.}
\end{equation}
The variable $\psi$ must increase linearly with the time
\begin{equation}
\omega(I)\equiv\dot{\psi}=\frac{\p H}{\p I},\qquad\psi=\omega(t-t_{0}).
\end{equation}
Here, $I$ is the action variable which plays the role of a momentum,
while $\psi$ is the coordinate conjugate to $I$ and is called the
angle variable.

With the use of an appropriate type-$F_{1}$ generating function,
$\tilde{W}(q,\psi)$, which is a function of both old and new coordinate
variables. Since the motion is periodic in $p$, $q$, then the motion
must also be periodic in $\psi$, so $\tilde{W}(q,\psi)$ is a periodic
function of $\psi$. We have then
\begin{equation}
\d\tilde{W}=p\d q-I\d\psi
\end{equation}
and integration over a single period of the motion, q returns to its
original value, while $\psi$ advances by the amount of one period,
$2\pi$,
\begin{equation}
\oint\d\tilde{W}=0=\oint p\d q-\oint I\d\psi.
\end{equation}
Because $I$ is a constant, it can be taken out of the integral. With
the integral $\oint\d\psi=2\pi$ we get
\begin{equation}
I=\frac{1}{2\pi}\oint p\d q.\label{eq:boltz.39}
\end{equation}

Comparing this result with Eq.~\eqref{eq:boltz.15}, one recognizes
immediately the identity of $I$ with the abbreviated action $F$,
since
\begin{equation}
\oint p\d q=\intop_{0}^{\tau}p\dot{q}\d t=\intop_{0}^{\tau}2T\d t=F
\end{equation}
and hence
\begin{equation}
\vardelta F=2\intop_{0}^{\tau}\vardelta T\d t.
\end{equation}
On the other hand, for the special case that the period $\tau$ remains
constant\footnote{Due to the definition of the bold-faced $\vardelta$-variation~\eqref{eq:boltz.5},
$\vardelta q^{k}\approx\delta q^{k}+\dot{q}^{k}\delta t$, the rightmost
term containing $\delta t=0$ vanishes and reduces the variation to
a standard variation, $\vardelta\to\delta$.} during the transition, i.e.\ $\delta\tau=0$, Boltzmann~\cite{Boltzmann.1866uber}
has shown that the heat supplied to the system then splits up into
two equal parts, heat and work energy, respectively, expressed by
\begin{equation}
\delta Q=2\delta E,\qquad\delta L=\delta E.
\end{equation}
Now, this formulation is equivalent with the vanishing of the variation
of the potential energy $V$, i.e.\ $\delta V=0$, and hence the
variation of the action due to the change from the natural to the
varied trajectory reads as 
\begin{equation}
\delta S=\delta\intop_{0}^{\tau}(T-V)\d t=\intop_{0}^{\tau}\delta T\d t,
\end{equation}
which leads by substitution of Eq.~\eqref{eq:boltz.39} into \eqref{eq:boltz.38}
to~\cite{Groessing.2008vacuum,Groessing.2009origin}
\begin{equation}
\Delta Q=\omega\delta F=2\omega\delta S.\label{eq:boltz.40}
\end{equation}

\selectlanguage{english}%

\chapter{Mathematical relations}

\section{Random variables\label{sec:aceq.sub.0}}

Let $X$ be a random variable. If the values $x$ which $X$ can assume
are continuously distributed, we define the probability density of
the random variable to be $P(x)$. This means that $P(x)\d x$ is
the probability that $X$ assumes a value in the interval $[x,x+\d x]$.
The total probability must be one, i.e.\ $P(x)$ is normalized to
one: 
\begin{equation}
\intopinfty P(x)\d x=1.\label{eq:aceq.0.1}
\end{equation}
The mean value of $X$ is defined by 
\begin{equation}
\meanx X=\intopinfty xP(x)\d x.\label{eq:aceq.0.2}
\end{equation}
Now let $F(X)$ be a function of the random variable $X$; we call
$F(X)$ a random function. Its mean value is defined corresponding
to Eq.~\eqref{eq:aceq.0.2} by 
\begin{equation}
\meanx{F(X)}=\intopinfty F(x)P(x)\d x.\label{eq:aceq.0.3}
\end{equation}

By default, we shall use different symbols for mean values over space
$\meanx x$, and mean values over time $\meant x$, if not otherwise
noted (see any good textbook, e.g.~\cite{Schwabl.2006en,Wachter.2006compendium}).

Let us consider continuous probability densities on the real line,
i.e.\ in one dimension, with or without explicit time dependence:
$P\in L^{1}(R);\,\int_{R}P(x)\mathrm{d}x=1$. Then we can define the
expectation value (mean value) by 
\begin{equation}
\mu:=\meanx x=\intop xP(x)\mathrm{d}x,\label{meandef}
\end{equation}
and the variance by 
\begin{equation}
\sigma^{2}:=\meanx{(x-\meanx x)^{2}}=\intop(x-\mu)^{2}P(x)\mathrm{d}x.\label{deviationdef}
\end{equation}
The standard deviation $\sigma$ equals the square root of the variance.

\section{Vectors and fields\label{sec:ef.vectors}}

The following is an overview on often used identities in Cartesian
vector calculus (see any good Textbook, e.g.,~\cite{Prechtl.2010vorlesungen}).

Let's start with the nabla operator 
\begin{equation}
\nabla=\VEC e_{x}\frac{\partial}{\partial x}+\VEC e_{y}\frac{\partial}{\partial y}+\VEC e_{z}\frac{\partial}{\partial z}\thinspace,\label{eq:appa.1.1}
\end{equation}
which is of vector type. $\VEC e_{u}$ denotes the unit vector in
$u$-direction. If needed, brackets have to be set in order to define
the scope of the operator, e.g., 
\begin{eqnarray}
\nabla\left(fg\right) & = & \left(\nabla f\right)g+f\nabla g,\label{eq:appa.1.2}\\
\nabla\left(\VEC f\cdot\VEC g\right) & = & \left(\nabla\otimes\VEC f\right)\cdot\VEC g+\left(\nabla\otimes\VEC g\right)\cdot\VEC f,\label{eq:appa.1.3}\\
\nabla\otimes\left(f\VEC g\right) & = & \left(\nabla f\right)\otimes\VEC g+f\nabla\otimes\VEC g.\label{eq:appa.1.4}
\end{eqnarray}
In Eqs.~\eqref{eq:appa.1.2}\textendash \eqref{eq:appa.1.4} the
nabla operator applies on a product of terms, the result is of vector
type or tensor type as in Eq.~\eqref{eq:appa.1.4}, respectively.

Tab.~\ref{tab:ef.1} shows elementary nabla operations in Cartesian
coordinates.

\begin{table}[!htb]
\noindent \begin{centering}
\renewcommand*\arraystretch{1.5}%
\begin{tabular}{|l|}
\hline 
$\nabla f=\VEC e_{x}\partial_{x}f+\VEC e_{y}\partial_{y}f+\VEC e_{z}\partial_{z}f$\tabularnewline
$\nabla\cdot\VEC f=\partial_{x}f_{x}+\partial_{y}f_{y}+\partial_{z}f_{z}$\tabularnewline
$\nabla\times\VEC f=\VEC e_{x}(\partial_{y}f_{z}-\partial_{z}f_{y})+\VEC e_{y}(\partial_{z}f_{x}-\partial_{x}f_{z})+\VEC e_{z}(\partial_{x}f_{y}-\partial_{y}f_{x})$\tabularnewline
$\nabla^{2}f=\partial_{x}^{2}f+\partial_{y}^{2}f+\partial_{z}^{2}f$\tabularnewline
\hline 
$\nabla(f+g)=\nabla f+\nabla g$\tabularnewline
$\nabla\cdot(\VEC f+\VEC g)=\nabla\cdot\VEC f+\nabla\cdot\VEC g$\tabularnewline
$\nabla\times(\VEC f+\VEC g)=\nabla\times\VEC f+\nabla\times\VEC g$\tabularnewline
\hline 
$\nabla(fg)=f\nabla g+g\nabla f$\tabularnewline
\hline 
$\nabla\cdot(f\VEC g)=f\nabla\cdot\VEC g+\VEC g\cdot\nabla f$\tabularnewline
$\nabla\times(f\VEC g)=f\nabla\times\VEC g+(\nabla f)\times\VEC g$\tabularnewline
\hline 
$\nabla(\VEC f\cdot\VEC g)=\VEC f\cdot\nabla\VEC g+\VEC g\cdot\nabla\VEC f+\VEC f\times(\nabla\times\VEC g)+\VEC g\times(\nabla\times\VEC f)$\tabularnewline
\hline 
$\nabla\cdot(\VEC f\times\VEC g)=\VEC g\cdot(\nabla\times\VEC f)-\VEC f\cdot(\nabla\times\VEC g)$\tabularnewline
$\nabla\times(\VEC f\times\VEC g)=\VEC f\,\nabla\cdot\VEC g-\VEC g\,\nabla\cdot\VEC f+\VEC g\cdot\nabla\VEC f-\VEC f\cdot\nabla\VEC g$\tabularnewline
\hline 
$\nabla\cdot(\nabla f)=\nabla^{2}f=\Delta f$\tabularnewline
$\nabla\times(\nabla f)=\VEC 0$\tabularnewline
$\nabla\cdot(\nabla\times\VEC f)=0$\tabularnewline
$\nabla\times(\nabla\times\VEC f)=\nabla(\nabla\cdot\VEC f)-\nabla^{2}\VEC f$\tabularnewline
\hline 
\end{tabular}
\par\end{centering}
\bigskip{}
\caption{\label{tab:ef.1} Elementary nabla calculus; Cartesian coordinates
$x$, $y$, $z$; $f$ and $g$ denote scalar fields, $\protect\VEC f$
and $\protect\VEC g$ denote vector fields.}
\end{table}

Finally, we mention some frequently used rules in one variable, denoting
the derivative by prime: 
\begin{eqnarray}
\left(fg\right)' & = & f'g+fg'\label{productrule}\\
\left(\frac{f}{g}\right)' & = & \frac{f'g-fg'}{g^{2}}\label{quotientrule}\\
\left(f(g)\right)' & = & f'(g)g'\label{chainrule}\\
\int f'g & = & fg-\int fg'\label{intudotv}\\
\int\frac{f'}{f} & = & \ln|f|\label{intufracv}
\end{eqnarray}
 The result of integration over~\eqref{productrule} yields Eq.~\eqref{intudotv}.
Eq.~\eqref{intufracv} can also be achieved by substituing $f\rightarrow\ln f$
and $g\rightarrow f$ into~\eqref{chainrule}.

\section{Entropic functionals\label{sec:ef.entropic}}

Now we derive some practical identities. These identities hold true
on general information theoretic grounds and are thus not bound to
quantum mechanical issues. Symbols being used are scalars $f$ and
vectors $\VEC f$ in Cartesian coordinates (cf.~\cite{Garbaczewski.2008information}).
Starting with 
\begin{equation}
\nabla\ln f=\frac{\nabla f}{f}\label{nablalnu}
\end{equation}
 which is a special case of Eq.~\eqref{chainrule} and hence analogue
to~\eqref{intufracv}. Next, 
\begin{align}
\nabla^{2}\ln f & =\nabla\cdot\left(\nabla\ln f\right)\\
 & =\nabla\cdot\left(\frac{\nabla f}{f}\right)\\
 & =\frac{f\nabla^{2}f-(\nabla f)^{2}}{f^{2}}\label{nabla2lnu1}\\
 & =\frac{\nabla^{2}f}{f}-\left(\frac{\nabla f}{f}\right)^{2},
\end{align}
 and thus, reordered, we obtain 
\begin{equation}
\ensuremath{{\displaystyle \frac{\nabla^{2}f}{f}=\nabla^{2}\ln f+\left(\frac{\nabla f}{f}\right)^{2}}}.\label{nabla2udivu}
\end{equation}
 Obviously, Eq.~\eqref{quotientrule} has been used in~\eqref{nabla2lnu1}.
Substitution of $f=\sqrt{g}$ into~\eqref{nabla2udivu} leads to
\begin{eqnarray}
\frac{\nabla^{2}\sqrt{g}}{\sqrt{g}} & = & \frac{\nabla^{2}g^{1/2}}{g^{1/2}}\nonumber \\
 & = & \nabla^{2}\ln g^{1/2}+\left(\nabla\ln g^{1/2}\right)^{2}\label{eq:nabla2sqrtdivsqrt}\\
 & = & \frac{1}{2}\nabla^{2}\ln g+\frac{1}{4}\left(\nabla\ln g\right)^{2}\,,\nonumber 
\end{eqnarray}
and
\begin{equation}
\begin{gathered}\begin{aligned}\frac{\nabla f}{f} & =\frac{\nabla\sqrt{g}}{\sqrt{g}}=\nabla\ln\sqrt{g}=\frac{1}{2}\nabla\ln g\\
 & =\frac{1}{2}\thinspace\frac{\nabla g}{g}\thinspace.
\end{aligned}
\end{gathered}
\label{eq:nablap2p=00003D1_2nablar2r}
\end{equation}

For further calculations we make use of the above introduced probability
function $P$ which leads us from~\eqref{nablalnu} to
\begin{equation}
\begin{gathered}\begin{aligned}\meanx{\nabla\ln P} & =\meanx{\frac{\nabla P}{P}}=\intop P\frac{\nabla P}{P}\d x\\
 & =\intop\nabla P\d x=P\bigg|_{-\infty}^{\infty}=0\,.
\end{aligned}
\end{gathered}
\end{equation}
 Note that $P(-\infty)=0$ and $P(\infty)=0$ must hold since the
integral over $R$ equals a finite value, namely 1. Further, from~\eqref{nabla2udivu}
we obtain 
\begin{align}
\meanx{\frac{\nabla^{2}P}{P}} & =\meanx{\nabla^{2}\ln P}+\meanx{\left(\frac{\nabla P}{P}\right)^{2}}\label{meannabla2udivu1}\\
 & =\intop P\frac{\nabla^{2}P}{P}\d x=\intop\nabla^{2}P\d x\label{eq:meannabla2udivu}\\
 & =\intop\nabla\cdot\nabla P\d x=\nabla P\bigg|_{-\infty}^{\infty}=0.\label{meannabla2udivu2}
\end{align}
 For $P$ we must demand that any derivative of $P$ must vanish at
its limits, i.e.\ $\lim\limits _{x\to\pm\infty}\nabla^{n}P(x)=0$,
$n\ge0$, otherwise we have $\int_{-\infty}^{\infty}\nabla^{n}P(x)\d x\neq0$
that leads us to at least one further integral from $-\infty$ to
$-\infty$ which therefore cannot be finite.

As~\eqref{meannabla2udivu2} is the value of the l.h.s.\ of~\eqref{meannabla2udivu1},
the r.h.s.\ of~\eqref{meannabla2udivu1} must also vanish. By considering
Eq.~\eqref{nablalnu} we obtain 
\begin{equation}
-\meanx{\nabla^{2}\ln P}=\meanx{\left(\nabla\ln P\right)^{2}}=\meanx{\left(\frac{\nabla P}{P}\right)^{2}}\thinspace.\label{eq:nabla2lnu}
\end{equation}
The mean value of Eq.~\eqref{eq:nabla2sqrtdivsqrt} can hence be
obtained by 
\begin{eqnarray}
\meanx{\frac{\nabla^{2}\sqrt{P}}{\sqrt{P}}} & = & \frac{1}{2}\meanx{\nabla^{2}\ln P}+\frac{1}{4}\meanx{(\nabla\ln P)^{2}}\\
 & = & \frac{1}{4}\meanx{\nabla^{2}\ln P}=-\frac{1}{4}\meanx{\left(\frac{\nabla P}{P}\right)^{2}}\thinspace.
\end{eqnarray}
\selectlanguage{british}%

\raggedright

\printbibliography[title=Bibliography,heading=bibintoc]

\clearpage{}

\chapter*{Publication list}

\cehead[]{Publication list}

\cohead[]{Publication list}

\addcontentsline{toc}{chapter}{Publication List} 

\newrefsection

\nocite{%
Groessing.2010emergence,%
Groessing.2011explan,%
Groessing.2011dice,%
Groessing.2012doubleslit,%
Schwabl.2012quantum,%
Mesa.2012classical,%
Grossing.2012quantum,%
Groessing.2012vaxjo,%
Groessing.2013dice,%
Mesa.2013variable,%
Fussy.2014multislit,%
Groessing.2014emqm13-book,%
Groessing.2014relational,%
Groessing.2014attenuation,%
Groessing.2015implications,%
Groessing.2015dice,%
Groessing.2016emqm,%
Mesa.2016emqm,%
Groessing.2016emqm15-book}

\newrefcontext[sorting=none]
\printbibliography[env=nolabelbib,title={Journals},keyword=mypaperj,heading=subbibliography,resetnumbers=true]
\printbibliography[env=nolabelbib,title={Conference Proceedings},keyword=mypaperc,heading=subbibliography,resetnumbers=true]
\printbibliography[env=nolabelbib,title={Books},keyword=mypaperb,heading=subbibliography,resetnumbers=true] 

\end{document}